\documentclass[sn-basic]{sn-jnl}

\usepackage{graphicx}%
\usepackage{multirow}%
\usepackage{amsmath,amssymb,amsfonts}%
\usepackage{amsthm}%
\usepackage{mathrsfs}%
\usepackage[title]{appendix}%
\usepackage{xcolor}%
\usepackage{textcomp}%
\usepackage{manyfoot}%
\usepackage{booktabs}%
\usepackage{algorithm}%
\usepackage{algorithmicx}%
\usepackage{algpseudocode}%
\usepackage{listings}%
\usepackage{xspace}%

\def\kms{{\text{km\,s}$^{-1}$}}

\def\Lsun{{\rm L$_{\odot}$}}
\def\Rsun{{\rm R$_{\odot}$}}
\def\Msun{{\rm M$_{\odot}$}}

\def\sun{\odot}
\def\halpha{{\rm H$\alpha$}\xspace}
\def\hbeta{{\rm H$\beta$}\xspace}

\newcommand{\io}[2]{#1\,{\textsc{#2}}}

\def\he2{{He~{\small II}}\xspace}

\begin{document}

\title[Red novae, their progenitors, and remnants]{Red novae, their progenitors, and remnants}
\author*[1]{\fnm{Tomasz} \sur{Kamiński}}\email{tomkam@ncac.torun.pl}

\author*[2,3,4]{\fnm{Nadejda} \sur{Blagorodnova}}\email{nblago@fqa.ub.edu}

\affil*[1]{\orgdiv{Laboratory for Astrophysics I}, \orgname{Nicolaus Copernicus Astronomical Center}, \orgaddress{\street{ul. Rabia\'{n}ska 8}, \city{Toruń}, \postcode{87-100}, \state{kujawsko-pomorskie}, \country{Poland}}}

\affil[2]{\orgdiv{Institut de Ciències del Cosmos (ICCUB)}, \orgname{Universitat de Barcelona (UB)}, \orgaddress{\street{c. Martí i Franquès, 1}, \city{Barcelona}, \postcode{08028},  \country{Spain}}}

\affil[3]{\orgdiv{Departament de Física Quàntica i Astrofísica (FQA)}, \orgname{Universitat de Barcelona (UB)}, \orgaddress{\street{c. Martí i Franquès, 1}, \city{Barcelona}, \postcode{08028},  \country{Spain}}}

\affil[4]{ \orgname{Institut d'Estudis Espacials de Catalunya (IEEC)}, \orgaddress{\street{Edifici RDIT, Campus UPC}, \city{Castelldefels}, \postcode{08860},  \country{Spain}}}


\abstract{Red novae or luminous red novae are a class of optical transients that have emerged over the past two decades. They occupy an intermediate luminosity regime between classical novae and supernovae and are characterized by cool, slowly expanding ejecta and a pronounced evolution toward red, dust-enshrouded remnants. These events are now widely interpreted as the outcome of binary coalescence involving non-compact stars, providing a rare opportunity to directly observe the dynamical phases of stellar mergers and their immediate aftermath. Observational studies of red novae provide a glimpse into the still poorly understood physics of unstable mass transfer and common-envelope evolution in binary stars, responsible for the formation of high-energy astrophysical phenomena, compact binary systems, and gravitational wave sources.
In this review, we synthesize current observational knowledge of red novae, including their outburst properties, population characteristics, and long-term remnants. Observations of light curves, spectra, and circumstellar environments reveal a complex interplay between mass ejection, collisions, radiative processes, and dust formation. Archival detections of red novae progenitors show a diversity of systems, ranging from low-mass contact binaries to massive evolved stars, with a notable representation of post-main-sequence stars.
We also examine current efforts to predict red nova outbursts and establish robust event rates, both of which remain challenging. The growing sample of extragalactic transients suggests that the brightest red novae may be even more frequent than core-collapse supernovae in the local Universe, underscoring their importance for binary evolution and stellar population studies. Finally, we outline future prospects, including the impact of large-scale time-domain surveys and the potential connection between stellar mergers and gravitational-wave sources.
%
}
\keywords{Transients, Stellar mergers, Common envelope evolution, Binary evolution, Circumstellar matter}

\maketitle
\tableofcontents

\section*{Abbreviations}
\begin{tabular}{ll}
 AGB & asymptotic giant branch  \\ 
 ALMA & Atacama Large Millimeter Array\\ 
 CE, CEE & common envelope, common envelope evolution\\ 
 CSE & circumstellar envelope\\ 
 DDI & delayed dynamical instability\\ 
 FIR & far-infrared\\ 
 FWHM & full width at half-maximum\\ 
 GEE & grazing envelope evolution\\ 
 HR & Hertzsprung--Russell (diagram)\\ 
 ILOT & intermediate-luminosity optical transients \\ 
 ILRT & intermediate-luminosity red transients \\ 
 IR & infrared \\ 
 ISM & interstellar medium \\
 JWST & James Webb Space Telescope\\ 
 L2, L3 & the second and third Lagrange points, respectively\\ 
 LBV & luminous blue variable\\ 
 LC & light curve\\ 
 LRN & luminous red nova (=\,red nova)\\ 
 LTE & local thermodynamic equilibrium\\ 
 LSST & Large Synoptic Survey Telescope\\ 
 MIR & mid-infrared\\ 
 MS & main sequence\\ 
 MW & Milky Way\\ 
 NIR & near-infrared\\ 
 RLOF & Roche-lobe overflow \\ 
 RGB & red giant branch \\ 
 RN & red nova\\ 
 RSG & red supergiant\\ 
 SED & spectral energy distribution\\ 
 SN & supernova\\ 
 S/N & signal-to-noise ratio\\ 
 SPRITE & eSPecially Red Intermediate-luminosity Transient Event \\ 
 UV & ultraviolet\\ 
 WD & white dwarf\\ 
 YHG & yellow hypergiant\\ 
 YSG & yellow supergiant\\ 
 YSO & young stellar object\\ 
 ZAMS & zero-age main sequence
\end{tabular}

\section{Introduction} 
Red novae (RNe) and luminous red novae (LRNe) are a class of optical transients recognized over the past two decades as a distinct subgroup of intermediate-luminosity events. They typically span peak magnitudes of $-3 \gtrsim M_r \gtrsim -16$ mag and evolve on timescales of weeks to years. Despite this diversity, they share several defining characteristics: low effective temperatures at maximum light, outflow velocities of a few hundred \kms, and rapid post-peak cooling that produces a red, often dust-enshrouded remnant. These properties distinguish them from other transients in a similar luminosity range.

Red nova eruptions are non-terminal and are widely interpreted as the observable outcome of stellar mergers in binaries composed of non-compact stars, from main-sequence (MS) objects to evolved giants and supergiants. This interpretation was firmly established with V1309 Sco, whose progenitor was observed as a contact binary prior to merger. Red novae therefore provide a direct window into the physics of extreme binary interaction, including common-envelope and grazing-envelope evolution---processes that remain central, yet poorly understood, in stellar astrophysics and population synthesis. Although only $\simeq$29 events are currently known, the volumetric rate of the more luminous events in the Local Group is comparable to that of core-collapse supernovae (and even higher for low-mass red nova events), and their detection rate is expected to increase rapidly with ongoing and upcoming time-domain surveys.

In this review, we summarize observational progress in the study of red novae, building on earlier works focused on their phenomenology \citep[e.g.,][]{KamiSubmm,Pastorello2019review}. We emphasize the connection between observed properties and the underlying physics of stellar mergers, while referring the reader to dedicated reviews for theoretical aspects of common-envelope evolution and merger modeling \citep[e.g.,][]{Ivanova2013Rev,RoepkeDeMarco2023,SchneiderAnnRev}.

This paper is organized as follows. Section~\ref{sect-phonomenon} introduces the red nova phenomenon and its historical context. Section~\ref{sect-outburst} describes their photometric and spectroscopic evolution during outburst. Section~\ref{sect-progenitors} reviews the properties of progenitor systems and methods of their identification. Section~\ref{sect-population} discusses population trends and event rates. Section~\ref{section-remnants} summarizes the properties of red nova remnants. Finally, Section~\ref{sect-future} outlines future directions, including prospects for detecting gravitational waves from selected systems.

\section{The red nova phenomenon}\label{sect-phonomenon}
\subsection{Historical perspective and milestones}\label{sect-history}
While unusually red transients have been sporadically observed in the 20th century, the modern field of red novae began with the enigmatic eruption of V838~Mon in early 2002. Initially considered a classical nova, V838~Mon defied expectations for all known nova-like objects and transients at the time \citep{MunariConf}. Additionally, a spectacular light echo from the eruption was captured by the Hubble Space Telescope \citep{BondEcho}, making V838~Mon an especially attractive target for study. It was noted early on that the unusual properties of V838~Mon's outburst matched well with two earlier outbursts, that is, of the Milky Way object V4332 Sgr, which erupted in 1994 \citep{Martini1999}, and of the Red Variable, which erupted in the Andromeda Galaxy in 1988 \citep[M31-RV;][]{Mould1990}. The latter was particularly perplexing for the contemporary observers, but was assumed to be an atypical outburst of a red giant star \citep{Rich1989}. Only the renewed interest, triggered by studies of V838~Mon in the early 2000s, changed this view. V838~Mon, M31-RV, and V4332 Sgr were quickly associated with each other, and it was realized that they belong to an entirely new type of astrophysical eruption, most commonly known today as luminous red novae (LRNe). It took another two decades to link red novae to an even older eruption, CK Vul, which was observed by European astronomers in 1670--1672 \citep{Hevelius,Shara1985,Kato2003,KamiNat}. While the observations of red novae can be dated back to the late 1980s or even the 17th century, the current interpretation of the phenomenon---as a manifestation of stellar coalescence observed in real-time---dates to the fundamental work of \citet{TylendaSoker2006}. They demonstrated that collisions between non-compact stars can trigger optical outbursts of relatively high luminosity and that such events would be similar to V838\,Mon-like eruptions.  

\begin{figure*}
    \centering
    \includegraphics[width=1.0\textwidth]{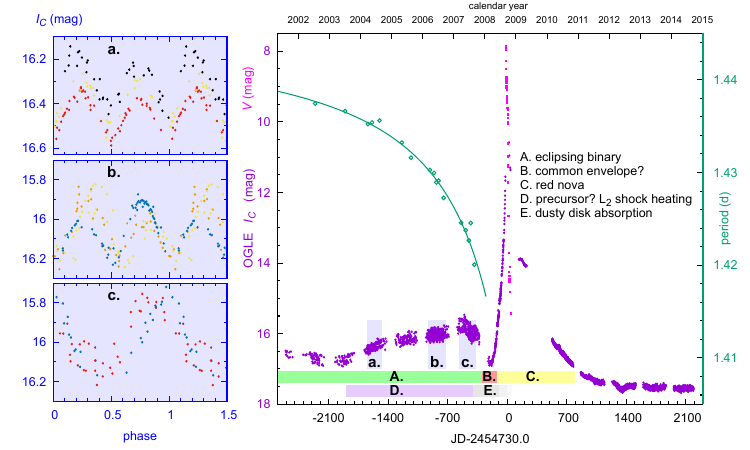}
    \caption{The light curve of V1309 Sco before and during the outburst. The main panel shows photometric measurements (purple and magenta points for $I_C$ and $V$, respectively) and the derived period (green points). The main phases in V1309 Sco photometric evolution are indicated by color bars at the bottom of the plot. Sample phased light curves are shown in the three left panels and correspond to the shaded areas a, b, and c in the main panel. Points of different colors in the smaller panels show individually-phased curves for selected time ranges. The reference epoch was arbitrarily selected but is common to all presented curves. Based on \citet{Tylenda2011}  and \citet {Pejcha2017} (and references therein). Data from OGLE ($V$ and $I_C$), ASAS ($V$), and AAVSO ($V$).}
    \label{fig-v1309lc}
\end{figure*}

A major breakthrough in the investigations of red novae was the discovery of Nova Sco 2008, now known as V1309 Sco, which underwent an outburst in late 2008 \citep{Mason2010}. The progenitor system was photometrically monitored by the OGLE project \citep{ogle}, whose observations showed an eclipsing contact binary with a decaying orbital period \citep[][see Fig.~\ref{fig-v1309lc}]{Tylenda2011}. This was the first time the spiraling-in of two stars into a common envelope phase was observed, predating the first detection of gravitational waves from in-spiraling black holes by a few years. V1309 Sco provides the most direct and compelling evidence that collisions between non-compact stars produce moderately energetic eruptions visible at optical and infrared (IR) wavelengths.     

Subsequent discoveries of more luminous extragalactic red novae have significantly expanded the known sample. For events within a few Mpc, progenitor systems have been identified in archival data--including M31‑LRN‑2015, M101‑2015‑OT, and NGC4490‑2011OT1--  extending the population to higher‑mass progenitors. In addition, several previously classified {\it supernova impostor}–like events have been reinterpreted as red novae. The current sample discussed in the literature is summarized in Table \ref{tab-main}.

\subsection{Naming conventions}
The first time the term `red nova' (RN) was used in the literature to describe the phenomenon known today under this name was by \citet{GoranskijBarsukova2007} and in multiple conference proceedings in \citet{2007proceedings}. These early studies referred to Galactic (V838~Mon and V4332 Sgr) and extragalactic objects (M31-RV). The extended term `luminous red nova', coined later, was used first to reference a particularly luminous extragalactic red nova in M85 \citep{Kulkarni2007Natur,Rau2007}. However, the terms RN and LRN later became interchangeable, regardless of the actual luminosity of the transient. Additionally, the recent recognition of planetary engulfment events as red novae led to the introduction of the term `underluminous red nova' \citep{DeNature}. It is important to note that distances and true bolometric luminosities of Galactic objects are often poorly constrained, so the quantification in the naming may be confusing or incorrect.

\begin{figure}[ht]
    \centering
    \includegraphics[trim=0 0 20 0,width=0.95\textwidth]{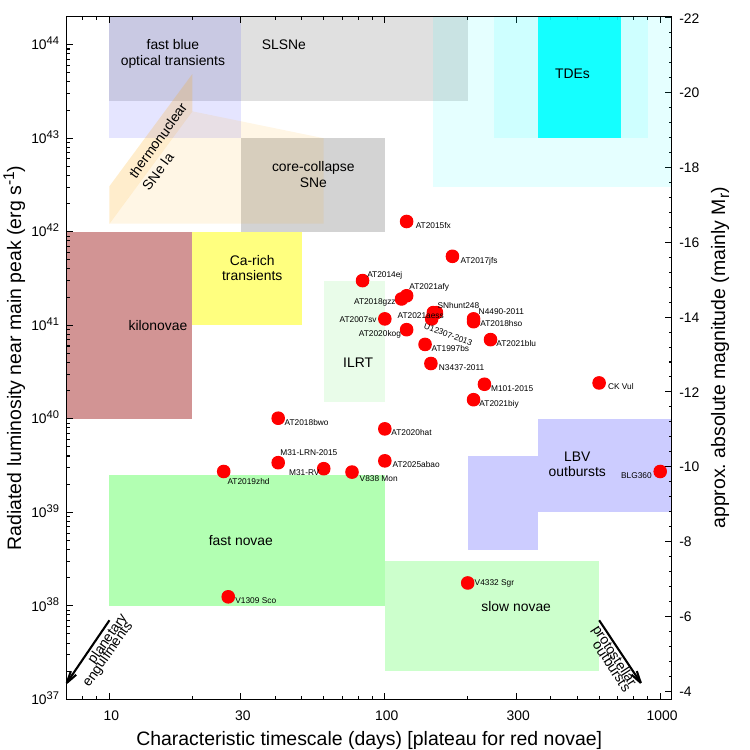}
    \caption{Peak luminosities vs. duration parameter space for most known red novae (red circles) in the context of other astrophysical transients (shaded regions). For red novae, the plotted timescale is the plateau time span from Table \ref{tab-main}. For other types of transients, this timescale is obtained rather arbitrarily, usually as the duration of the plateau or the decay time from peak by 2--3 magnitudes. Note that the distances to most Galactic red novae are poorly constrained, making their luminosities especially uncertain. For a variation of the diagram where total energy is considered (i.e., including kinetic energy), see, for instance, \cite{Kashi2010arXiv}.}
    \label{fig-diagram}
\end{figure}

Several groups have used alternative naming conventions, such as red transients, mergebursts, and tylendars. Within the extragalactic community, red novae are members of the so-called `interacting gap transients' or simply `gap transients' \citep{SokerKashi2012, Kasliwal2012PASA,Cai2022Universe}. This nomenclature denotes the broad class of transients with peak luminosities in the `gap' between classical novae and supernovae (SNe). Following \cite{Kasliwal2012PASA}, our Fig.\,\ref{fig-diagram} illustrates the relation of red novae to other (gap) transients, in terms of their radiative luminosity and characteristic duration of the outburst. 

Initially, some authors termed these outbursts `intermediate luminosity optical transients' (ILOTs) or `intermediate-luminosity \emph{red} transients' (ILRTs) \citep{PastorelloFraser2019}. However, more recently, this name has been reserved for NGC300-OT-like and SN2008S-like transients \citep{Thompson2009ApJ,Valerin2025AA_ILRTphot,Valerin2025AA_ILRT_spec}. Whenever detected, their progenitors appeared to be dust-enshrouded stars that faded below progenitor level years after the burst \citep{Prieto2008,Berger2009ApJ,Adams2016MNRAS}, supporting a terminal explosion scenario. Ironically, the transient in M85 that originally gave rise to the term \textit{luminous red nova} was later reclassified as an ILOT, based on the identification of an IR-bright progenitor \citep{Ofek2008ApJ}. Nowadays, the terms ILOT and ILRT are almost exclusively used to refer to these weak explosions, which are likely from electron-capture SNe \citep[e.g.,][]{Botticella2009MNRAS,Kochanek2011ApJ,Cai2021,Cai2022Universe}. However, in some works this name is still used as a synonym of `gap transient' \citep[e.g.,][]{Soker2016MNRAS}.

`Supernova impostors' are also commonly placed within the gap transient population \citep{VanDyk2000_1997bs,VanDyk2002PASP,Kochanek2012ApJ} and partially overlap with the parameter space of red novae. Observationally, SN impostors are characterized by narrow hydrogen emission lines which are narrower than in genuine SNe, i.e., with full width at half maximum, FWHM, $<2000$\,\kms.  
Early phases closely resemble type IIn SNe \citep{Filippenko1997ARAA}, which led to initial misclassifications as genuine supernovae. Impostor spectra also resemble the most energetic red novae at early stages, making initial classification challenging and leading to misclassifications and confusion among these three classes. Indeed, some SN impostors \citep[e.g., SN\,1997bs, AT\,2007sv, AT\,2015fx;][]{VanDyk2000_1997bs,Tartaglia2015_2007sv,Tartaglia2016ApJ_AT20215fx} were reclassified as red novae based on spectroscopic and photometric evolution at optical and IR wavelengths \citep{Pastorello2019review,ReguittiIR}. 

Although the class of supernova impostors defines objects with broadly similar observational signatures, it actually represents a heterogeneous group of transients whose outbursts may be powered by distinct physical mechanisms \citep{Smith2011_LBVs,VanDyk2012}. In some cases, supernova impostors have been observed to precede a genuine supernova explosion, such as SN~2009ip \citep{Mauerhan2013_SN2009ip,Pastorello2013ApJ_SN2009ip,Margutti2014_SN2009ip,Soker2013ApJ_Sn2009ip}, leading to their interpretation as supernova precursors. Other supernova impostors were claimed to be giant eruptions of luminous blue variables (LBVs) originating from very massive progenitors (M$>$40\,M$_{\odot}$), such as P Cyg and $\eta$ Car \citep{DavidsonHumphreys1997ARAA,Weis2020Galax_LBV}. One observational property that helps set them apart from other classes, such as red novae, is their significantly longer timescale: their outbursts typically persist for several hundred days \citep[e.g., $\eta$ Car or UGC~2773-OT;][]{Smith2010AJ_precursorLBV} and, in some cases, recur \citep[e.g., P Cyg or SN~2000ch;][]{Wagner2004PASP_SN2000ch}.

%

One final variation of red novae is `especially red intermediate-luminosity transient events' (SPRITEs) \citep{Kasliwal2017,Jencson2019}, which are events detected at IR wavelengths, with only weak or non-detected visual counterparts. Unfortunately, the different names and classifications are not used consistently in the field, and the re-classification of historic events is still ongoing. 

We use here the simplified but more generic name \emph{red novae} in reference to all transients of this type, regardless of their luminosity, location (Galactic and extragalactic), or whether they were observed solely in the optical or IR (i.e., SPRITEs are included). In this naming scheme, `red' reflects the property that the immediate product of the outburst is an intrinsically cool star, often of spectral type M, which is additionally reddened by circumstellar dust formed from the ejecta. As will be discussed later, dust obscuration is so high that it would be more adequate to call them `infrared' novae. `Nova' may be thought of as having the original ancient meaning, that is, `a new star in the sky', which relates to the transient nature of the objects. `Nova' also seems to be an adequate term since, in the merger interpretation, a rejuvenated star with new properties is formed.   

\subsection{The merger interpretation of the red novae} 
Soon after the eruption of V838~Mon, a discussion on the nature of this and similar transients began. Multiple scenarios were proposed, including: an atypical nova-like outburst \citep[e.g.,][]{Rauch2002_V838Nova,Shara2010}; helium-shell flashes in an AGB star or post-AGB star \citep[including a born-again AGB star][]{2005MNRAS.361..695L}; and a consecutive engulfment of several planets by the parent star \citep{Retter}. \citet{TylendaSoker2006} \citep[see also][]{SokerTylenda2006, Tylenda2005} critically examined these scenarios and showed that they fail to consistently account for the observed characteristics of V838~Mon and similar objects. Some proposed scenarios fail to reproduce the observed outburst luminosities and progenitor brightnesses, fail to predict chemical abundances consistent with observations, or do not match the inferred timescales, ejecta velocities, and ejecta masses based on the best estimates available at the time. Most characteristically, many of the considered eruption mechanisms are primarily powered by thermonuclear burning, which requires very high temperatures and which consequently results in a very hot remnant. For instance, classical novae reach temperatures above 10$^5$ K after the eruption. As discussed later, red novae evolve to very low photospheric temperatures, reaching values as low as $\sim$2000 K, and are embedded in circumstellar material spanning temperatures of $\sim$10–1000 K. These properties are inconsistent with expectations for thermonuclear outburst models.

The merger model proposed by \citet{TylendaSoker2006} remains the most successful framework for explaining V838~Mon and other red novae discovered since then. The eruption model developed for V838~Mon envisioned a collision in a binary consisting of a B3 main-sequence star (6--8 M$_{\sun}$) and a protostar of $\approx$0.4 M$_{\sun}$ \citep[][see also Sect. \ref{sect-progenitors-gal}]{TylendaProgenitor}. Later on, it was also found that the binary had a tertiary B-type companion at a wide orbit. The collision and subsequent coalescence in V838~Mon released the gravitational energy of the inner binary system. Part of the energy was used for heating the gas in shocks and powering the electromagnetic outburst, whereas an approximately similar energy fraction was used for unbinding matter from the system. Their interpretation was that the event was thus powered mainly by accretion and did not require sustained thermonuclear burning at high temperatures. 

In the framework proposed by \citet{TylendaProgenitor}, after the secondary had plunged into the envelope of the primary, the energy released under the surface of the primary caused rapid expansion of the star, and optically-thick winds developed \citep{Lipunov,Quataert}. This formed an enormously bloated object with a low surface temperature. The scenario explains very well the observed properties of V838~Mon during its 3-month outburst and the following and ongoing relaxation phase. Later studies of red novae and similar systems added another layer of complexity to the energetics of collisions by considering the release of recombination energy \citep[e.g.,][]{IvanovaScience,Reichardt2020,MatsumotoMetzger,ChenIvanova2024} or a short activation of nucleosynthesis \citep{TylendaCKfinal}. However, the overall energy budget in these scenarios is still dominated by the release of gravitational (orbital) energy from the binary. 

The low-mass companion consumed by the V838~Mon primary had a too low luminosity to significantly contribute to the progenitor photometry, so its mass was calculated to account for the total power of the eruption. This mass had to be accreted on a timescale of at most a few days. If thermonuclear sources are excluded, as advocated above, V838~Mon's eruption must have been primarily powered by the release of gravitational energy. To account for the short timescale of the event, the infalling mass likely required a relatively compact configuration, such as a star, rather than loosely bound circumstellar matter or a gas cloud. It was thus different from powerful protostellar eruptions, such as FU Ori-type events. 

The stellar merger model originally proposed for V838~Mon has since been widely applied to other red novae. Proposed progenitor systems include mergers involving main-sequence stars, red giants, and yellow (super)giants. While V838~Mon represents a merger in a very young triple system, most red novae are thought to arise from mergers in more evolutionarily advanced binaries. This is especially true for post-MS stars that are rapidly expanding and are more prone to mass transfer and orbital instabilities leading to a merger. Yellow supergiants, which are objects in the Hertzsprung-Russell gap, are particularly well represented in the extragalactic sample of known red novae \citep{Tranin} (Sect.~\ref{sect-progenitors-ex}). Mergers involving very compact stars, such as neutron stars, black holes, and their pairs, produce transients of very different characteristics than red novae (e.g., kilonovae) and have been widely discussed in the literature \citep{Metzger2019}. They are omitted in this review.

Observations of red novae have often been linked to the long-standing problems of unstable mass transfer leading to common-envelope evolution (CEE) in binary stars \citep[e.g.,][]{PaczynskiCE,IvanovaScience,Ivanova2013Rev,Ivanova2020ceebook}.
During this stage, orbital tightening combined with the loss of co-rotation between the donor's envelope and the binary orbit causes the secondary star to move inside the expanded donor's envelope. During this dynamical phase, the binary's orbital angular momentum is transferred to the envelope, leading to its partial or total ejection \citep{Webbink1984ApJ,Ivanova2013Rev}. 

The formation of a common envelope is the culmination of a mass transfer episode, which can be initiated either by the expansion of the donor (primary star in this case) filling its Roche lobe, or the loss of orbital angular momentum from the binary, leading to the shrinkage of the orbit, or a combination of both. This loss of orbital angular momentum can occur via various mechanisms, such as magnetic braking, emission of gravitational waves, spin-orbit exchange between the stars \citep[commonly referred to as the Darwin instability;][]{Darwin1879_instability}, or mass loss from the binary system \citep{Pols2011,Henneco2024}. Finally, in the context of triple systems, there is also a possibility of a secular instability \citep{Eggleton2001ApJ}. None of the observed events so far has been associated with high-velocity collisions in globular clusters (i.e., not strictly in a long-lasting binary), but they too may manifest in the sky as red novae \citep{ChristianII}. 

In the context of red novae, the two dominant mechanisms proposed in the literature to trigger the merger are Darwin instability and outer Roche lobe overflow. In the first case, in a tidally synchronized binary, the instability is triggered when the spin angular momentum of one star, typically an expanding post–MS component with a large moment of inertia, becomes equal to one-third of the total orbital angular momentum of the binary. Maintaining synchronization then requires angular momentum to be drawn from the orbit, leading to orbital decay and a runaway inspiral. Darwin instability is favored in systems with extreme mass ratios, typically $q \lesssim 0.1$. For example, for the red nova M31-LRN-2015, \citet{MacLeod2017} proposed rapid spin–orbit angular momentum exchange to explain the fast rise to peak luminosity observed in this transient. 

In the case of binaries experiencing increasingly high rates of mass transfer, material is likely to escape through the outer L2/L3 Lagrange points rather than being efficiently transferred (and accreted) between the stars. The outflowing gas carries away a substantial amount of orbital angular momentum, leading to efficient orbital tightening. This angular-momentum loss can drive a runaway inspiral, ultimately triggering the merger event.  For V1309 Sco, mass loss through the outer Lagrange points has been proposed to explain both the properties of the progenitor system and the rapid orbital decay observed during the pre-outburst phase \citep{Tylenda2011,Zhu2016V1309Sco,Pejcha2014,Pejcha2017,MacleodLoeb2020}.

A combination of both scenarios is also plausible. For example, while Darwin instability was pointed out as the leading mechanism for the initial orbit tightening in V1309 Sco, this eventually caused L2/L3 overflow, which quickly accelerated the system to the formation of a common envelope and final coalescence \citep{NandezV1309}. 


 The dynamical (total or partial) ejection of the envelope causes the electromagnetic red nova transient, and, depending on the binary's properties, the final remnant can be a tighter binary composed of the primary's core and the secondary star, or a stellar merger \citep[][]{SchneiderAnnRev}. Although ejecta mass estimates of red novae are usually well below the expected donor's envelope mass---supporting the merger scenario
\citep{Tylenda2011,Blagorodnova2021}---the literature is still inconclusive whether some of these events may instead result in a surviving binary \citep{Karambelkar2023ApJ,Howitt,Twum2026arXivRates}. 
Indeed, some studies have attributed red novae to non-terminal common-envelope ejection events \citep[e.g.,][]{IvanovaScience} and grazing envelope evolution \citep{SokerGEE,SokerGEE-ILOTS}. In both scenarios, a binary avoids a merger either through efficient dispersal of the common envelope or by avoiding entering the CE phase whatsoever, owing to jetted outflows \citep[e.g.,][]{RedNovaReqJets}. 

Red nova outbursts are primarily powered by gravitational energy, with possible additional contributions from recombination energy \citep{IvanovaRecomb,MacLeod2017,MetzgerPejcha2017,Reichardt2020,Henneco2024,ChenIvanova2024} and the prevalent role of shock heating \citep{MetzgerPejcha2017,Kirilov2025ApJ}. As discussed in \cite{TylendaCKfinal}, for the case of the red nova CK Vul, a merger involving an RGB star and a helium white dwarf could also lead to a temporary activation of thermonuclear helium burning, contributing some 20\% of the outburst luminosity. In certain cases, accreted mass may reach the primary's hydrogen-depleted core and ignite nuclear burning, which can potentially contribute extra luminosity to the outburst and even lead to explosive thermonuclear runaway \citep{Podsiadlowski2010MNRAS}.

\section{The outburst} \label{sect-outburst}

\begin{figure}[ht]
    \centering
    \includegraphics[width=1\textwidth]{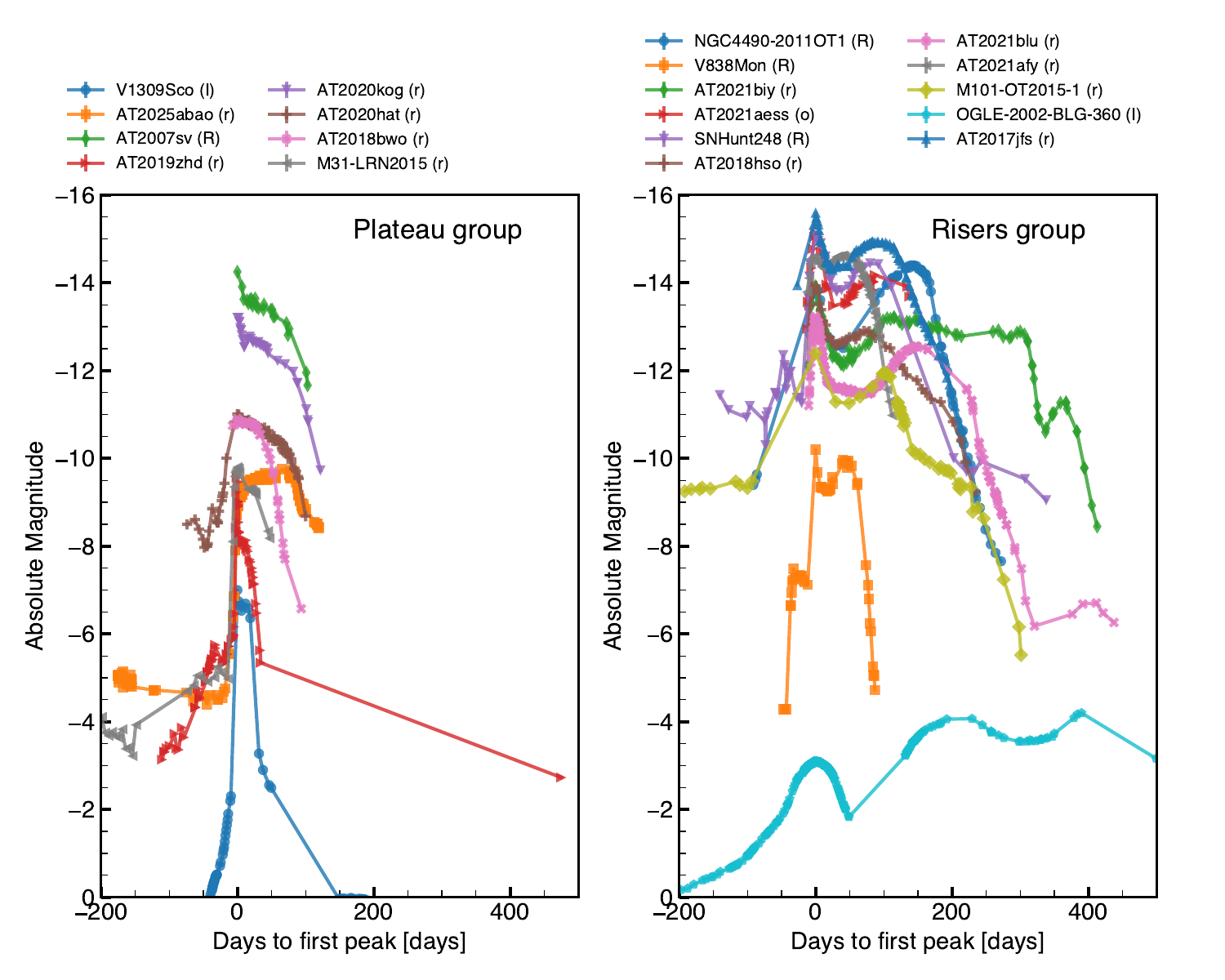}
    \caption{Light curves of red novae. The selected sample is divided into two groups. The `plateau group' (left panel) and the `risers group' (right panel). Each object is shown with a different marker and color, as specified in the legend. The light curve photometric band is indicated in parentheses. }
    \label{fig-lightcurves}
\end{figure}

\begin{sidewaystable}
\tiny
\caption{Red novae classified until 2026 and their basic parameters}\label{tab-main}
\begin{tabular*}{\textheight}{@{\extracolsep\fill}cccccccccccc}
\toprule%

Name & Distance & Host & Year of & LC & Plateau & $L_\mathrm{peak}$ & Peak abs. mag & FWHM & $M_\mathrm{dust}$ & Note & Main\\

 & (kpc,Mpc) & galaxy & main peak & type & (d) & (\Lsun) & Band Mag$\pm$Err & (\kms) & (\Msun) & & ref. \\
(1) &(2) &(3) &(4) &(5) &(6) &(7) &(8) &(9) &(10) &(11) &(12)\\
\toprule%

CK Vul & 2.6-3.5 & Milky Way & 1671 & yes R 3 & $\sim$600 & 6.2E6 & $M_V\;-12.3 \pm 0.6$ & 1420 & 1.0E-2 & $^{1}$ & [1] \\
V4332 Sgr & 4.2-5.5 & Milky Way & 1994 & yes P 4-5 & 200 & 4.5E4 & $M_V\;-7.9 \pm 0.6$ & 240 & 6.0E-4 & $^{2}$ & [2] \\
V838~Mon & 5.9 & Milky Way & 2002 & yes R 3-4 & 76 & 6.9E5 & $M_R\;-10.1 \pm 0.2$ & $<$250 & 9.4E-4 & $^{3}$ & [3] \\
BLG-360 & 4.1-5.6 & Milky Way & 2004 & yes R 3 & $>$1000 & 7.0E5 & $M_I\;-3.3 \pm 0.3$ & $\sim$100 & 1.2E-2 & $^{4}$ & [4] \\
V1309 Sco & 2-5.6 & Milky Way & 2008 & yes P 2 & 27 & 3.2E4 & $M_V\;-7.0 \pm 1.4$ & 300 & 1.0E-3 & $^{5}$ & [5] \\
ZTF SLRN-2020 & 4 & Milky Way & 2020 & yes? P? 2 & 0 & 34 & $M_r\;+2.2 \pm 1.0$ & 120 & 1.3E-6 & $^{6}$ & [6] \\[10pt]
M31-RV & 0.78 & M31 & 1988 & ? P? 1? & $\geq$60 & 7.5E5 & $M_{R_c}\;-9.9 \pm 0.2$ & 460 &  & $^{7}$ & [7] \\
M31-LRN-2015 & 0.78 & M31 & 2015 & yes P 1-2 & 41 & 8.7E5 & $M_r\;-9.8 \pm 0.4$ & 200 & 7.7E-4 &  & [8] \\
AT2019zhd & 0.78 & M31 & 2019 & yes P 2-3 & 26 & 7.0E5 & $M_r\;-9.6 \pm 0.1$ & 280 &  &  & [9] \\
AT2025abao & 0.78 & M31 & 2024 & ? R 2-3 & 100 & 9.1E5 & $M_r\;-9.8 \pm 0.1$ & 700 & 4.0E-7 & $^{8}$ & [10] \\[10pt]
AT1997bs & 9.2 & NGC 3627 & 1997 & ? U 2 & 140 & 1.6E7 & $M_R\;-13.7 \pm 0.2$ & 585 & 1.0E-3 & $^{9}$ & [11] \\
AT2007sv & 18.9 & UGC 5979 & $<$2007 & ? P 2 & 100 & 3.0E7 & $M_R\;-14.3 \pm 0.4$ & 800 &  & $^{10}$ & [12] \\
N4490-OT2011 & 9.6 & NGC 4490 & 2011 & yes R 2 & 210 & 3.0E7 & $M_R\;-14.4 \pm 0.3$ & 350 & 1.8E-3 & $^{11}$ & [13] \\
N3437-2011OT1 & 18.6 & NGC 3437 & 2011 & no R 2 & 147 & 1.0E7 & $M_R\;-13.9 \pm 0.4$ & 350 &  & $^{12}$ & [14] \\
UGC12307-2013-OT1 & 39.7 & UGC12307 & 2013 & no R $\geq$1 & 148 & 3.0E7 & $M_R\;-15.0 \pm 0.2$ & 400 &  & $^{13}$ & [15] \\
SNhunt248 & 22.5 & NGC 5806 & 2014 & yes R $>$2 & 154 & 3.5E7 & $M_R\;-15.1 \pm 0.4$ & 440 & (1:10)E-5 & $^{14}$ & [16] \\
AT2014ej & 20.6 & NGC7552 & 2014 & ? R 2 & 83 & 7.7E7 & $M_r\;-15.1 \pm 0.5$ & 900 &  &  & [17] \\
M101-2015OT1 & 6.4 & M101 & 2015 & yes R 2-3 & 230 & 6.0E6 & $M_R\;-12.4 \pm 0.5$ & 500 &  & $^{15}$ & [18] \\
AT2015fx & 23.8 & NGC 2748 & 2015 & ? P $\geq$2 & 120 & 3.3E8 & $M_r\;-13.6 \pm 0.2$ & 700 &  & $^{16}$ & [19] \\
AT2017jfs & 34.7 & NGC 4470 & 2017 & ? R 2 & 176 & 1.4E8 & $M_r\;-15.6 \pm 0.2$ & 450 &  & $^{17}$ & [20] \\
AT2018bwo & 6.6 & NGC 45 & 2018 & ? P 1? & 41 & 2.6E6 & $M_r\;-10.9 \pm 0.1$ & 580 & 7.5E-5 &  & [21] \\
AT2018gzz & 66.9 & NGC 271 & 2018 & ? R 2? & 115 & 4.9E7 & $M_r\;-14.5 \pm 0.2$ & 280 &  &  & [22] \\
AT2018hso & 20.2 & NGC 3729 & 2018 & yes? R 2 & 210 & 2.8E7 & $M_r\;-13.9 \pm 0.1$ & 400 &  & $^{18}$ & [23] \\
AT2020hat & 5.2 & NGC 5068 & 2020 & yes P 1-2 & 100 & 2.0E6 & $M_r\;-11.0 \pm 0.1$ & 250 &  & $^{19}$ & [24] \\
AT2020kog & 23.8 & NGC 6106 & 2020 & ? P 2? & 120 & 2.3E7 & $M_r\;-13.2 \pm 0.5$ & 380 &  &  & [25] \\
AT2021afy & 48.7 & UGC 10043 & 2021 & ? P $>$2 & 120 & 5.3E7 & $M_r\;-14.6 \pm 0.6$ & 410 &  &  & [26] \\
AT2021biy & 7.5 & NGC 4631 & 2021 & ? R 3 & 210 & 4.1E6 & $M_r\;-13.9 \pm 0.2$ & 430 & 3.0E-4 & $^{21}$ & [27] \\
AT2021blu & 8.6 & UGC 5829 & 2021 & ? R 2 & 242 & 1.8E7 & $M_r\;-13.3 \pm 0.2$ & 470 & 4.2E-5 & $^{20}$ & [28] \\
AT2021aess & 34.4 & NGC 1359 & 2021 & ? R $>$2 & 150 & 3.5E7 & $M_r\;-15.1 \pm 0.2$ & 470 &  & $^{22}$ & [29] \\
\hline
\end{tabular*}
\vspace{2pt}
\begin{minipage}{\textwidth}
{\bf Columns:} Col.(2) Distance is given in kpc for Milky Way objects and in Mpc otherwise. Col.(4) Year of the main peak, not necessarily of the first peak. Col.(5) The light curve type is coded by three aspects: whether the light curve has the precursor feature, light curve shape defined in text (P and R for `plateau group' and `risers group', respectively), and the number of distinct peaks in multiwavelength observations (arbitrary).  Col.(6) The time span of the plateau phase in the (visual) light curve. Col.(8) See original papers for error derivations. Col.(9) Line FWHM as a proxy of the main ejecta velocity. When possible, measurements for metal lines were used, avoiding Lorentzian hydrogen emission wings. Col.(10) Highest reported dust mass (may include pre-eruption dust). Col.(12) Main reference, esp. justifying red nova classification.  
\end{minipage}
\begin{minipage}{\textwidth}
{\bf Comments:} $^{1}$misclassified as a hibernating classical nova; 
$^{2}$misclassified as classical nova; 
$^{3}$misclassified as classical novae; 
$^{4}$misclassified as a lensing event; no spectrum taken in outburst; 
$^{5}$misclassified as classical nova; 
$^{6}$YSO?,a.k.a. ZTF 20aazusyv; 
$^{7}$no linewidth measurement; remnant not recovered; 
$^{8}$still in outburst; a.k.a WNTR23BZDIQ; 
$^{9}$misclassified as SN impostor; 
$^{10}$misclassified as SN impostor; 
$^{11}$a.k.a. AT2011kp; 
$^{12}$a.k.a. SNhunt31; 
$^{13}$a.k.a. AT2013lw; 
$^{14}$probably not a RN; a.k.a. AT2014ib; 
$^{15}$a.k.a. AT2015dl; 
$^{16}$misclassified as SN impostor; 
$^{17}$misclassified as SN IIn; 
$^{18}$misclassified as ILRT; 
$^{19}$infrared precursor; 
$^{20}$misclassified as LBV; 
$^{21}$GEE?;
$^{22}$a.k.a. ZTF21acpkzcc; 
\end{minipage}
\begin{minipage}{\textwidth}
{\bf References:} [1] \cite{Kato2003}; [2] \cite{Martini1999}; [3] \cite{Tylenda2005}; [4] \cite{TylendaBLG}; [5] \cite{Tylenda2011}; [6] \cite{DeNature}; [7] \cite{2004A&A...418..869B}; [8] \cite{Williams2015}; [9]\cite{Pastorello2021_2019zhd}; [10] \cite{Karambelkar2025ApJ}; 
[11] \cite{Pastorello2019review}; [12] \cite{ReguittiIR}; [13] \cite{Smith2016} [14] \cite{Pastorello2019review}; [15] \cite{Pastorello2019review}; [16] \cite{Kankare2015AA}; [17] \cite{Stritzinger2020AA}; [18] \cite{Blagorodnova2017}; 
[19] \cite{ReguittiIR}; [20] \cite{Pastorello2019review}; [21] \cite{Blagorodnova2021};
[22] \cite{Karambelkar2023ApJ}; [23] \cite{Cai2019}; [24] \cite{Pastorello2021AA}; [25] \cite{Pastorello2021AA}; [26] \cite{Pastorello2023}; [27] \cite{Cai-2021biy}; [28] \cite{Pastorello2023};  [29] \cite{Karambelkar2023ApJ}
\end{minipage}
\end{sidewaystable}
 
\begin{figure}[h!]
    \centering
    \includegraphics[page=1,width=0.85\textwidth]{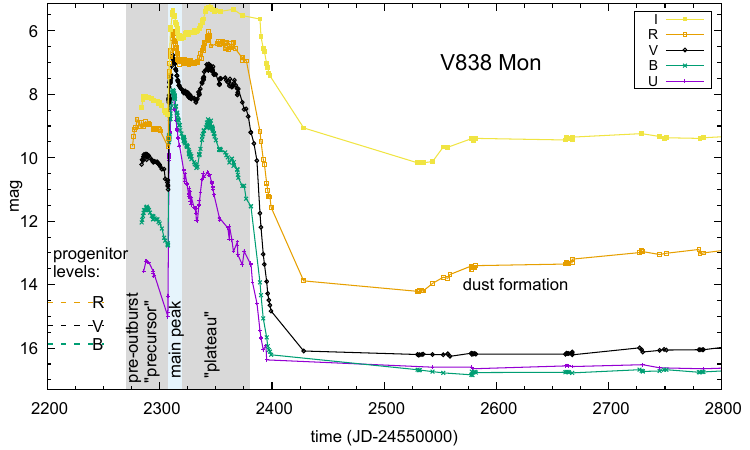}
    \includegraphics[page=2,width=0.85\textwidth]{figures/plotv838multicolor.pdf}
    \caption{Multi-color light curves of V838~Mon in the optical (upper) and NIR (lower. The main phases of a typical red nova outburst discussed in the text are highlighted in gray and cyan. Observed data points are not corrected for extinction. The progenitor pre-outburst magnitudes for optical bands are shown on the left (the corresponding NIR progenitor magnitudes are well below the scale shown in the bottom panel). There are no measurements directly preceding the phase marked here as pre-outburst. The observations were compiled from \citet{Munari2002,Goranskij2020,banerjee2002,Loebman,Crause2003,Crause2005,Watson2005,KamiALMA,Olivier,woodward,Liimets}. Solid lines simply connect the observed points except for $N'$-band where we applied arbitrary interpolation.}
    \label{fig-lightcurve-v838multicolor}
\end{figure}

\subsection{Light curves}\label{sect-lightcurve}
Light curves and evolving SEDs of red novae are unique among astrophysical transients. Despite spanning a wide range of peak luminosities and light-curve shapes, they can be broadly grouped into two main morphological classes illustrated in Fig.\,\ref{fig-lightcurves}: the ones showing a slow brightness evolution, which we call the `plateau group,' and the `risers group' showing a re-brightening to a second major peak. Most, but not all, members of the `plateau group' show an initial $\sim$2-week-long peak, followed by a plateau lasting up to 100 days, and then a rapid decline. The members of the `risers group' also show a fast initial peak, but in this case, the decline from the peak is followed by a second re-brightening, which sometimes reaches or even exceeds the brightness of the initial peak. The timescales of this second peak are longer, extending the overall duration of these transients from 80 to over 400 days if we consider a third or even a fourth peak in these light curves. 

Given this morphological diversity, color evolution becomes a critical diagnostic to distinguish red novae from other transients. After the initial rise or peak, red novae quickly evolve to lower temperatures, which can be traced in their progressively reddened SEDs. Compared to other explosive transients like classical novae and SNe, the red nova ejecta cools down rapidly via adiabatic expansion. A well-observed example is V838~Mon, whose light curves across different bands are shown in Fig.\,\ref{fig-lightcurve-v838multicolor}. While the blue bands quickly fade after the first peak, emission in IR and mid-IR (MIR) bands increases with time. Red nova light curves typically evolve through the following phases:

\paragraph{(I) Pre-outburst or precursor brightening }
Pre-outburst or precursor brightening can occur months to years before the main outburst, and is characterized by a much lower amplitude, often 1--3 mag above the quiescent progenitor level, than the main outburst. 
Detecting this phase in archival data is challenging. The signal is often too faint or the transient position too uncertain for recovery in existing surveys.
As a result, only the nearest red novae and those with brighter progenitors have been observed during this stage. The most notable example showing precursor brightening was V1309 Sco, where OGLE captured the steady increase in the eclipsing binary's light curve, starting at about five years before its optical peak (see highlight D. in Fig.~\ref{fig-v1309lc}). 

Among extragalactic systems, the best example of a precursor brightening is AT2025abao in M31 \citep[WNTR23bzdiq; see][]{Karambelkar2025ApJ}, with archival photometry spanning 20 years before its outburst. The data reveal that the brightening in this system began about 10 years before the main outburst and the colors during this phase were comparable to those of the progenitor star.
Other red novae with detected precursor emission are SNHunt248, M101-OT2015-1, M31-2015-LRN, and AT2019zhd \citep{Kankare2015AA,Blagorodnova2017,Blagorodnova2020,Pastorello2021_2019zhd}. They show a 1--5\,mag brightening starting $\lesssim$5 years before the outburst. 

Notably, precursors often end with a short-lived dimming episode, hereafter referred to as `the dip,' occurring just before the fast rise. For instance, it was observed in V1309 Sco (see first half of the highlight B. in Fig.~\ref{fig-v1309lc}), where it was assigned to increased obscuration from a dusty disk. Due to the limited availability of follow-up observations before the main peak, color evolution during the dip phase has been constrained for only a small number of objects. For V838~Mon and AT2025abao, the observed colors show only a small reddening compared to the rest of the precursor phase \citep[$\leq$0.3\,mag;][]{Reguitti2026_AT2026abao}, arguing against significant dust obscuration.

\paragraph{(II) The main outburst}

The main outburst begins with a rapid rise in brightness, typically at a rate of 0.4--1\,mag\,day$^{-1}$. For objects without a detected precursor, this stage marks the first identification of the system as a transient in time-domain surveys. During this phase, light curves reach their highest temperatures, typically in the range 4000–-9000\,K, and it is therefore sometimes referred to as the `blue peak.' A few transients reached even higher temperatures, of $\sim$11\,000\,K, resulting in detected ultraviolet (UV) emission \citep[AT2020kog and AT2021biy;][]{Pastorello2021AA,Cai-2021biy}. However, among the plateau group, some red novae seem to skip the blue peak overall, directly transitioning to the next phase (e.g., AT2020hat and AT2025abao). In many cases, the main peak corresponds to the maximum light of the event, especially at shorter visual wavelengths. The post-peak decline is comparable to, but generally slower than the rise, and is associated with a rapid cooling of the photosphere.

\begin{figure}[ht]
    \centering
    \includegraphics[trim=0 25 0 5, width=1.0\textwidth]{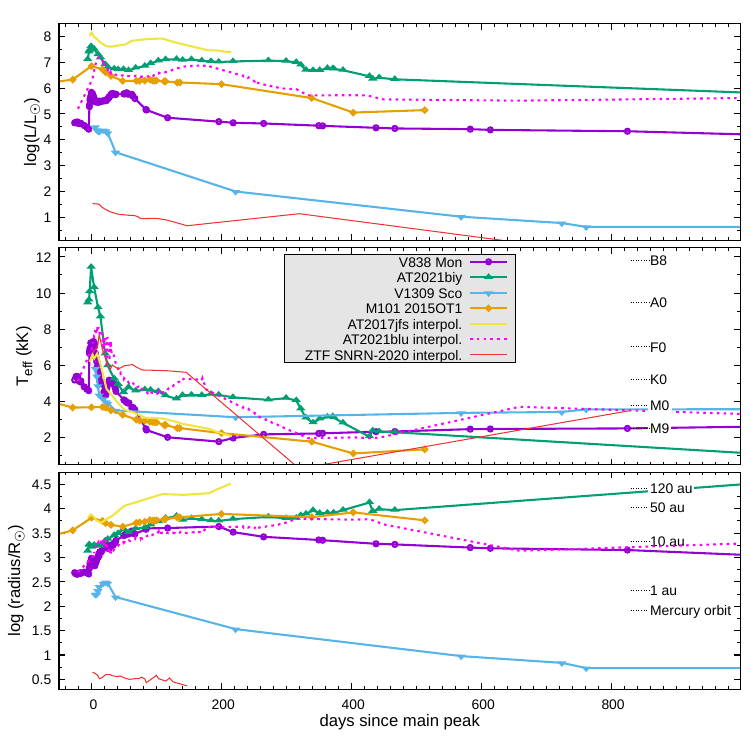}
    \caption{Evolution of basic (pseudo-)photospheric parameters of red novae. The shown parameters were derived mainly from black-body fits to the observed SEDs. Data are from \citet{Tylenda2005, Cai-2021biy, Tylenda2011, TylendaSED, Blagorodnova2017, Pastorello2019_AT2017jfs, KarambelkarJWST}, Wavasseur et al. (in. prep.), \cite{DeNature,LauSLRN}; the last 3 curves show interpolation between data points for better clarity. See the original papers for error bars which are not included here. Distance to V1309 Sco is uncertain.}
    \label{fig-RTeffL}    
\end{figure}

\paragraph{(III) Plateau or red peak} 
Depending on the morphological class, the fast decline from the main peak may either flatten into a slower plateau phase or give way to a secondary re-brightening to a similar or modestly fainter maximum ($\Delta m < 1$), initiating the `red peak'. The plateau phase may comprise a series of small-amplitude peaks and dips rather than a smooth decline.
Colors at this stage are redder than at earlier phases (see Fig.~\ref{fig-RTeffL}), but for some objects from the risers group, a clear second increase in temperature is observed \citep[UGC12307-2013OT1 and SNhunt248;][]{Pastorello2019review}. Because of its longer duration, the plateau phase often dominates the total radiated energy, and for some red novae, maximum light in the red optical and the IR bands is reached during this stage (e.g., BLG360 or V838~Mon observed in $I_C$, or NGC4490-2011OT1). Additionally, in several red novae classified as `risers', an additional late-time re‑brightening is seen, producing a lower luminosity third peak (e.g., BLG360, SNHunt248, AT2021blu) or, in extreme cases, even a fourth peak, as in AT2021biy.


\paragraph{(IV) The decay}
During the decay phase, the plateau (or red peak) ends, and the light curve starts to decline much faster (around 0.1--0.2\,mag\,day$^{-1}$). Optical to MIR photometry taken at this stage shows SEDs with clear IR excesses, which have been associated with signatures of dust at temperatures of 100--1500\,K \citep[][Wavasseur in prep.]{Tylenda2011,Blagorodnova2020}. Although the post-outburst bolometric luminosity remains above the progenitor level for decades \citep{KamiALMA,ForesToribioKochanek2026}, the post-plateau evolution at visual wavelengths often falls well below the progenitor level due to extinction from freshly condensed dust. The lower temperatures in the remnant and its surrounding dense gas contribute to the nucleation of dust grains, which reprocess the central hotter emission of the coalesced star into longer wavelengths. Contrary to the fast decay of optical light, the IR flux for some red novae keeps increasing for months to years after the end of the main optical peak \citep{KarambelkarJWST,ReguittiIR} (see V838~Mon in Fig.\,\ref{fig-lightcurve-v838multicolor}) and perhaps some red novae can only be observed through this late IR signature without an optical counterpart \citep{Jencson2019}. Late post-outburst photometric evolution and relaxation phase of red nova remnants is described in Sect.\,\ref{section-remnants}. 

\medskip
The duration of each phase (I--IV) varies widely among red novae (see Fig.\,\ref{fig-lightcurves}), possibly due to different progenitor masses, different evolutionary stages of the progenitor stars, different binary configurations at the onset of the event (see Sect.\,\ref{sect-relations}), and different orbital instability mechanisms at play. For example, in the relatively fast red nova V838~Mon in the `risers group', the precursor phase lasted $\gtrsim$35 days, the main peak lasted 10 days, and the plateau stretched for over 70 days. For comparison, M31-LRN-2015 in the `plateau group' and with similar peak brightness was even faster, with a plateau length of only $\sim$40\,days. For the very slow red nova BLG-360, the timescales were 470, 240, and $>$660 days for precursor, main peak, and plateau, respectively, illustrating the diversity of red novae light curves. 


Despite their heterogeneity, the characteristic multi-peak, quickly reddening light curves remain an important tool for identifying new transients of this type and are often readily distinguishable from other classes of outbursts. For example, in classical novae, the ejecta expansion leads to increasing effective temperature as deeper, hotter layers become visible (a small subsample exhibits secondary variations or `jitters' in their light curves). This produces a systematic evolution toward bluer colors with time, which is the opposite of what happens in red novae. Gap transients such as ILOTs do not have a plateau and show a continuous decline from the peak \citep{PastorelloFraser2019,Jencson2019ApJ,Valerin2025AA_ILRTphot} with little color evolution. The LBV outbursts are characterized by longer timescales than those of red novae and possibly display previous activity \citep{Pastorello2010_SN2000ch,FailedSNBeasor}. It is, however, still often the case that red nova light curves have been confused with other types of transients or even microlensing events (see notes in Table \ref{tab-main} and most recent reclassifications in \citealp{ReguittiIR}). Due to their unique circumstances, some historic red novae, such as CK Vul or BLG360, were classified solely based on their optical or IR light curves and SED evolution, as no spectra were taken during their outbursts. 



Multi-wavelength light curves of red novae have often been used to compare the colors of the transient with low-gravity stars to estimate their spectral classification. In addition, studies have also built multi-epoch SEDs to trace the changes in the physical parameters of the expanding object, most often under the assumption of a black body surrounded by a spherical dust shell. This approach provides estimates of semi-bolometric luminosity and parameters such as effective temperature, photospheric radius, expansion velocity, and dust properties \citep[e.g.,][]{Tylenda2005,Pastorello2019review,Blagorodnova2020,ReguittiIR} (see Fig.~\ref{fig-RTeffL}). Such analyses commonly assume spherical symmetry and are based on sparse photometric data, rarely extending beyond the near-IR (NIR). This can lead to an underestimation of dust-related IR emission and makes the results especially incomplete at late times, when the luminosity is increasingly dominated by emission at longer wavelengths.

The growing sample of well-covered light curves has shown a consistent picture of their properties during the outburst. The main peak is characterized by the highest effective temperatures, typically 7000--12\,000\,K. The corresponding effective radii are in the range of 100--8000\,\Rsun, in extreme cases reaching $\approx$40\,au. During the plateau phase, the radii grow by a factor of a few, sometimes reaching even 12\,000\,\Rsun, and the effective temperature drops to 4000--4500 K. After the decline phase (years after the main outburst), in some objects it was observed that the expanding material becomes progressively geometrically thinner, resulting in the shrinkage of the effective radius; the thinning eventually exposes the inner coalesced core, which is a bloated star of spectral type M with effective temperature of 2000--3500 K \citep[e.g.,][]{TylendaEngulf}. Nevertheless, the majority of central stars remain obscured for much longer due to a thicker dust ejecta \citep{Steinmetz,KarambelkarJWST}. Their effective radii grow at a rate consistent with the observed gas expansion velocity known from spectroscopy. So far, optically thick ejecta with effective radii as large as $\approx$200 au have been reported \citep{KarambelkarJWST,Mauerhan2018MNRAS}. The size evolution in red nova outbursts has been investigated mainly through indirect methods because direct size measurements, e.g., using IR interferometry, would be challenging even for the closest objects identified to date \citep[cf.][]{MobeenII}. We also caution here that in late phases of the evolution of some red novae, their photometric fluxes in the optical to NIR are grossly dominated by emission features of atoms and molecules \citep[e.g.,][]{2004ApJ...604L..57B,KamiMason}, not continuum, and the type of analysis we discussed here leads to erroneous results \citep[as discussed in][]{Steinmetz}.  

Ultraviolet observations of red novae are sparse, during outburst \citep{Rauch, Smith2016, Pastorello2023} and in the post-outburst phase \citep{Mauerhan2018MNRAS}. No red nova in outburst has been detected in X-ray or low-frequency radio emission, not even the nearest bright sources \citep{Antonini}. Under normal circumstances and for non-compact star mergers, such emission is not expected \citep[but see][]{SokerTylenda2007MNRAS}, as thermal emission at these energy regimes must be weak. To date, red novae outbursts have not been detected in gravitational waves (but see Sect.~\ref{sect-future})

Light echoes are occasionally associated with red nova outbursts, but such cases are rare, as the appearance of an extended light echo requires that the eruption take place near dusty interstellar medium, and so it is more likely for young or very massive progenitors. Up to date, V838~Mon light echo remains the iconic example of this phenomenon observed both in scattered optical light \citep{BondEcho} and in thermal IR \citep{BanerjeeEcho}. Another possible light echo was postulated for N4490-OT2011 \citep{ReguittiIR}, who proposed this scenario to explain the lack of MIR color evolution in the remnant source during the first six years after the red nova event. 

\subsection{Spectral characteristics}\label{sect-spectral}

\begin{figure}[ht]
    \centering
    \includegraphics[trim=30 30 10 0,width=\textwidth]{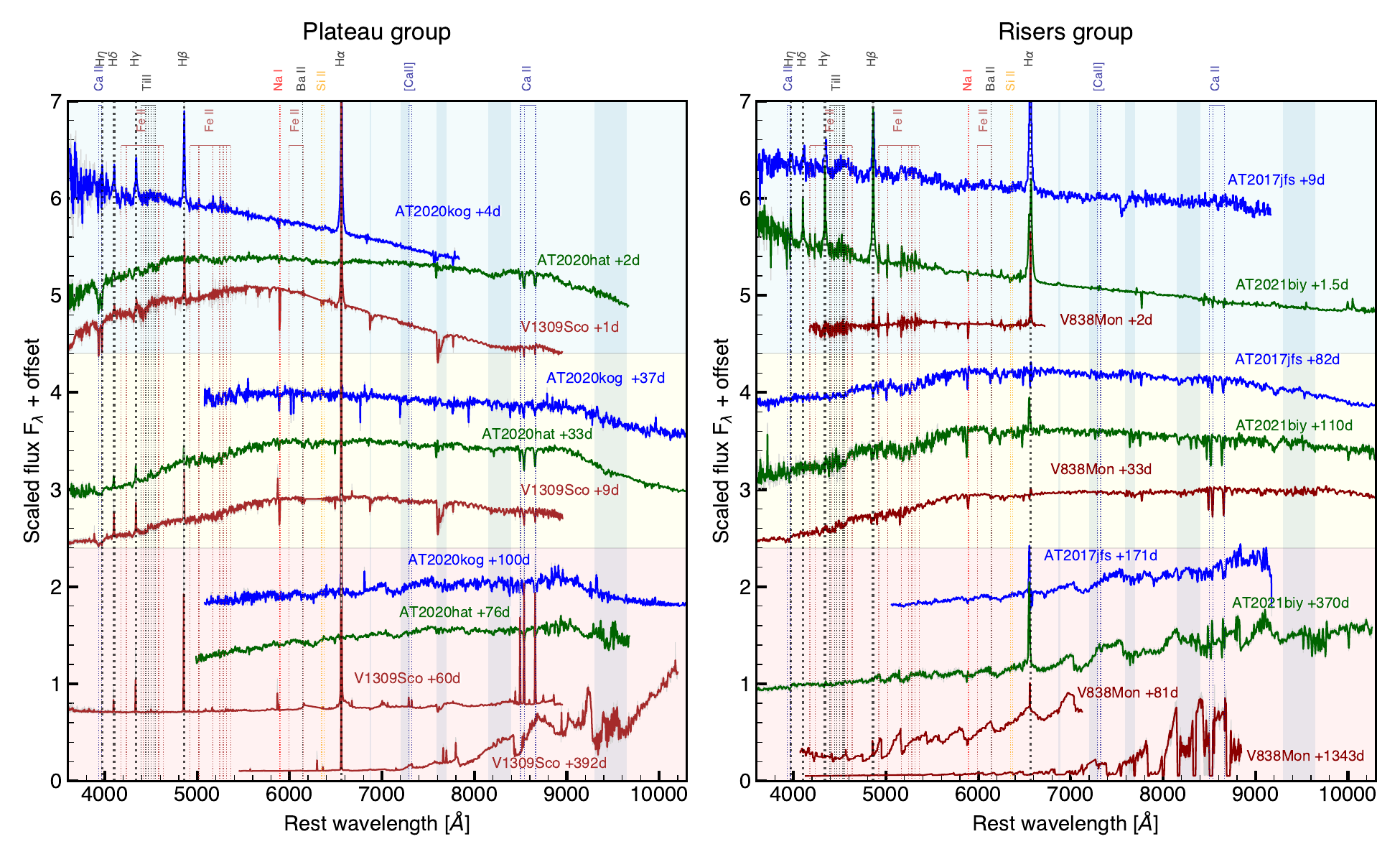}
    \caption{Spectral evolution for the `plateau group' (left) and `risers group' (right) of red novae (cf. Fig.~\ref{fig-lightcurves}). The reference epoch is given in reference to the date of the optical peak. Spectra acquired around the main (blue) peak have a blue background; spectra taken during the plateau (red peak) phase have a yellow background; spectra taken during the decay phase have a red background. Each group comprises three objects of different peak brightness. The data have been obtained from \citet[][AT2020kog and AT2020hat]{Pastorello2021AA}, \citet[][AT2017jfs]{Pastorello2019_AT2017jfs}, \citet[][V1309Sco]{Mason2010}, \citet[][AT2021biy]{Cai-2021biy}, and \citet[][V838Mon]{BondEcho,KamiKeck,Munari2002Proceedings}. The spectra have been corrected for Galactic extinction and redshift.}
    \label{fig:specsequence}
\end{figure}

The spectral evolution of red novae discussed in this section mostly focuses on optical wavelengths, which have been used extensively to identify red novae. Grouped by their light curve morphology, a sample of optical spectra is shown in Fig.~\ref{fig:specsequence} and corresponds to high-, medium-, and low-luminosity events shown in Fig. \ref{fig-lightcurves}. The background color indicates the phases during which the spectra were taken and which are discussed below.

\paragraph{(I) Pre-outburst}
Because red nova precursors are most often identified retrospectively in archival data, spectroscopic observations during this phase are rare. So far, spectroscopic characterization during precursor brightening only exists for AT2025abao \citep{Karambelkar2025ApJ}, which was identified as an object potentially undergoing common envelope ejection from an AGB star. The earliest spectra were taken at optical and NIR wavelengths $\sim$2\,yr before the outburst discovery in October 2025. The authors noted a discrepancy in the typing of the stellar type. While the optical emission matched a hotter M1-type star, the identified features of TiO, CO molecular absorption, as well as water vapor, better matched a cooler M8--M9-type star, resulting in a mixed temperature spectrum which persisted during the whole observational period until January 2025. This phenomenon was previously observed in other red novae, such as V838~Mon and the planet-merger ZTF-SLRN-2020. The emission pattern was consistent with a hotter star surrounded by cooler dense shells \citep{Lynch2004,Lynch2007}. While the shorter wavelengths originated deeper in the envelope, longer wavelengths traced the emission from the cooler, outer layers. This is commonly observed in YSOs.

\begin{figure}[ht]
    \centering
        \includegraphics[width=0.49\textwidth]{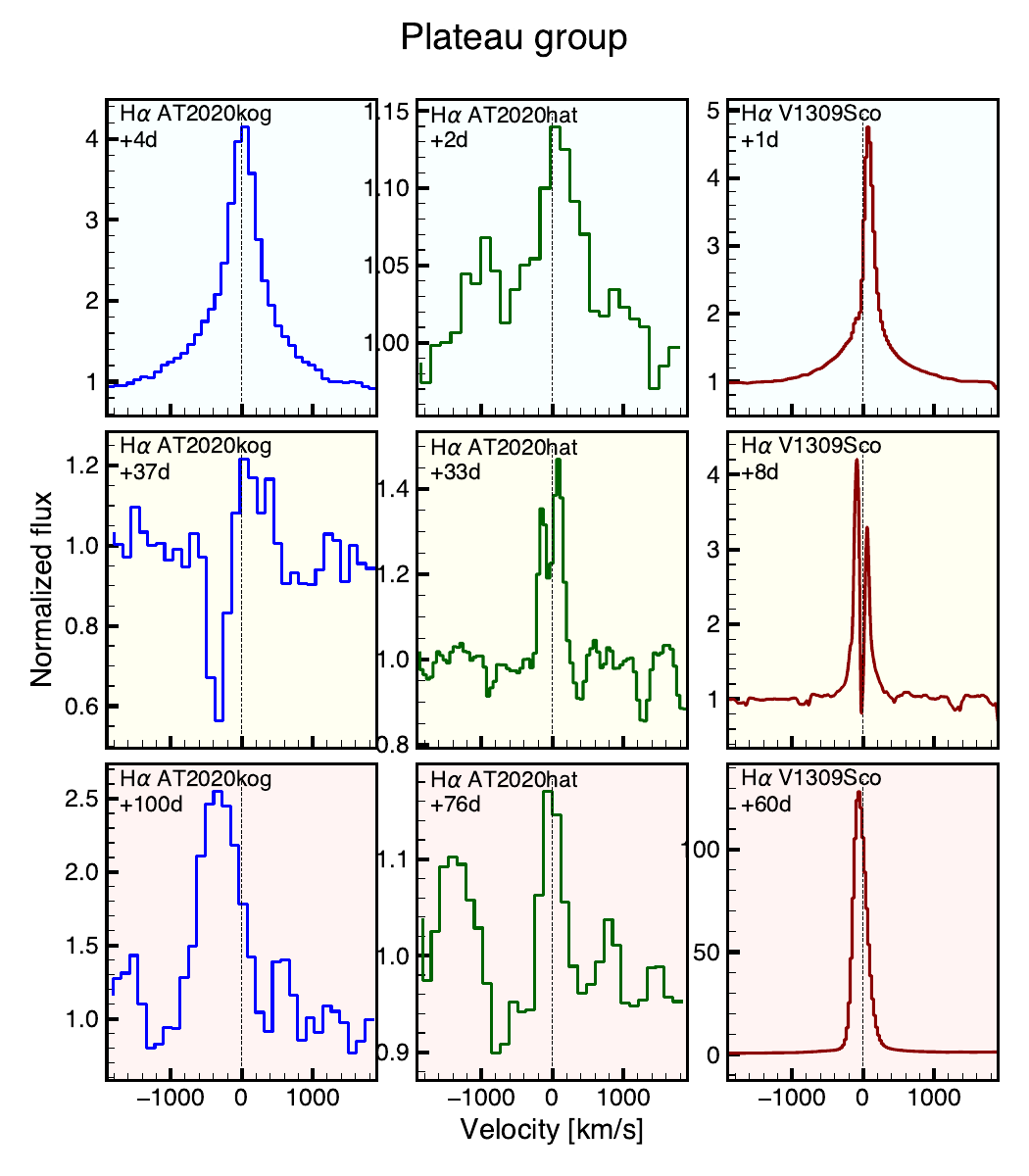}
        \includegraphics[width=0.49\textwidth]{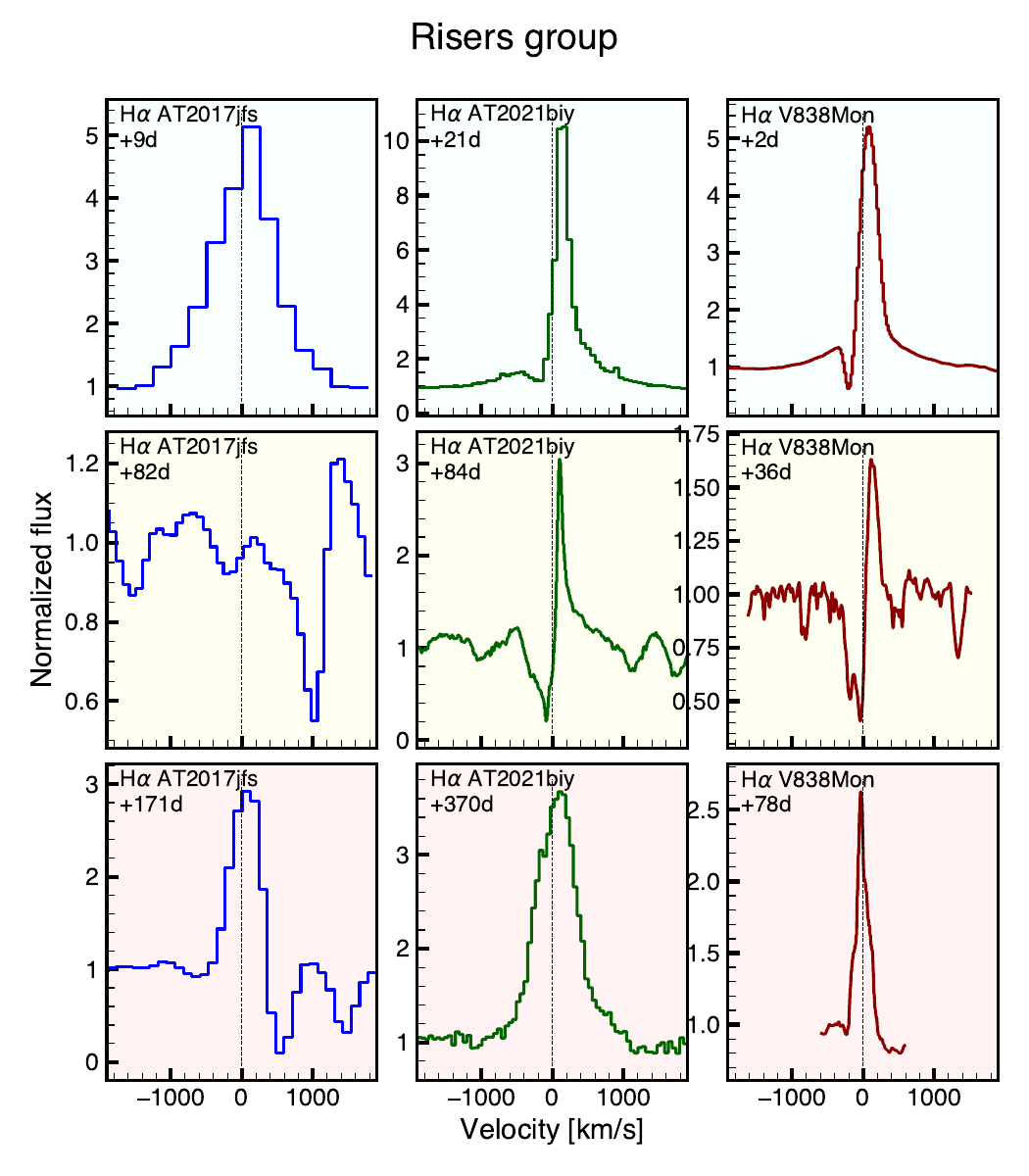}
    \caption{Spectral evolution of \halpha for objects in Fig.~\ref{fig:specsequence}.  Blue, yellow, and red backgrounds indicate the respective phases as in Fig.~\ref{fig:specsequence} (tom to bottom): blue peak, plateau/red peak, and decay. The names of the objects and their epoch relative to the date of the optical main peak are shown in each panel. The spectra are scaled to the continuum and smoothed with a Gaussian 0.3\,\AA\ filter for display purposes. Data obtained from \citet[][AT2020kog and AT2020hat]{Pastorello2021AA}, \citet[][AT2017jfs]{Pastorello2019_AT2017jfs}, \citet[][V1309 Sco]{Mason2010}, \citet[][AT2021biy]{Cai-2021biy} and also Wavasseur et al. in prep., \citet[][V838~Mon]{BondEcho,MunariConf,WisniewskiSpecpol}.}
    \label{fig:specsequence_halpha}
\end{figure}

\paragraph{(II) The main outburst}
During the main peak, red novae show a hot continuum and strong, narrow\footnote{Narrow, as compared to other transients, especially SNe.} ($<$2000 \kms; typically much less) Balmer emission, reminiscent of other eruptive transients such as SNe IIn or LBV eruptions. At this stage, the peak luminosity provides a useful first indicator for classifying an event as a supernova, but it is not sufficient on its own, leading to frequent misclassifications. As shown in the \halpha spectra in Fig.~\ref{fig:specsequence_halpha}, during the main peak, the hydrogen lines display Lorentzian profiles with extended ($>$1000 \kms) electron-scattered wings \citep[cf.][]{WisniewskiSpecpol}. For some luminous events \citep[e.g., AT2021blu and AT2021afy,][]{Pastorello2023}, the reported Balmer \halpha to \hbeta\ flux ratio was $\approx$2, somewhat smaller than expected from the nominal value of $\simeq$2.85 corresponding to Case B recombination of gas at 10\,000\,K \citep{Osterbrock1989_book}. This effect, observed in cataclysmic variables and SNe\,IIn \citep{Elitzur1983ApJ,Chugai2004MNRAS}, suggests that either the Balmer lines become optically thick and are reabsorbed, or that internal collisional processes enhance the population of the upper levels of the Balmer transitions, effectively boosting \hbeta\ emission and flattening the Balmer decrement toward smaller ratios, or both. The hydrogen emission indicates that dense ($>10^8$ cm$^{-3}$) matter undergoes ionization and recombination in regions above the expanding photosphere.

The \halpha profile often has a narrow absorption component, overlapping with the blue wing of the much stronger emission component \citep[e.g.,][]{Martini1999,WisniewskiSpecpol,Mason2010}, and it may quickly change location or temporarily disappear \citep[e.g.,][]{Cai-2021biy,Pastorello2021AA}. The absorption may be interpreted as a signature of the circumstellar material that is on the line of sight but located far from the eruption so that it is not significantly affected by the ionization front. At least in some cases, it may be the material lost by the system long before the red nova outburst \citep{Martini1999, MasonShore}. Several objects of the `risers' group may also show a broad absorption component, indicative of an ongoing fast outflow, forming a classical P\,Cygni-type profile. In terms of absolute fluxes and compared to the local continuum, emission in \halpha is most intense during the main outburst and fades away during the plateau phase. 


Spectra obtained with a sufficient spectral resolution during the outburst display absorption lines for elements other than hydrogen.  These are often interpreted as features arising in the photosphere of the optically thick expanding ejecta. The absorption spectrum is thus used for spectral typing, complementary to temperature constraints obtained with SED analysis (cf. Sect. \ref{sect-lightcurve}). Examples of features often observed this way are the H\&K doublet and IR triplet of \io{Ca}{ii}, visual lines of \io{Ba}{ii}, \io{Sc}{ii}, \io{Ti}{ii}, and, most commonly, \io{Fe}{ii} (see line identification in Fig.~\ref{fig:specsequence}). This pseudo-photospheric or indeed photospheric component matches spectral types from A--G in the early phases, and M supergiants (luminosity class I) in the latest phases of the eruption.  In some extreme cases, for instance, during the main peak of NGC4490-2011OT1, the effective temperature read from spectra was as high as 18\,000 K (corresponding to spectral type B4, see \citealp{Pastorello2019review}), but such hot photospheres are rare among red novae, especially among those with low-mass progenitors. It should be kept in mind that strict or accurate spectral typing is not possible for these eruptive objects, as the spectrum does not arise in hydrodynamic or thermodynamic equilibrium and line profiles are contaminated by non-photospheric components. Regardless, the evolution through spectral types is often very rapid, especially in the final phases of the optical outburst, when a change of one spectral type over one night is not uncommon.

Effective temperatures of 7000--11\,000 K near the bolometric peak (or at the main peak) inferred from spectral typing are relevant for assessing the role of recombination energy in powering the outburst. The role of the associated release of internal energy on the system evolution is hotly debated in the literature (cf. Sect.~\ref{sect-LCinterpretation}) and may depend on the nature of the binary undergoing the red-nova event \citep[cf.][]{Reichardt2020}.


Emission in lines other than in the Balmer series is comparatively weak during the main peak, but not unmet (e.g., \io{Na}{i} and \io{Ba}{ii} lines in V838~Mon, \citealp{Iijima}; \io{Fe}{ii}, \io{Ti}{ii}, \io{Ca}{ii} and \io{O}{i} in NGC4490-2011OT1, \citealp{Pastorello2019review}; \io{Na}{i}, \io{Fe}{i}--\textsc{ii}, \io{Mg}{i}, and [\io{O}{i}] in V4332 Sgr, \citealp{Martini1999}). The permitted lines generally show a P-Cygni profile, with lower expansion velocities than the ones estimated from the \halpha line. In contrast to some other gap transients, the firmly identified red novae do not show [\io{Ca}{ii}] $\lambda\lambda$7291,7323 during the main peak, but the lines have been detected in the plateau phase \citep[e.g.,][]{Mason2010}. 

\paragraph{(III) Plateau or red peak}

After the main peak, the continuum reddens considerably. Balmer emission lines also dramatically weaken, with low or undetectable \hbeta and only marginal \halpha. Moderate-resolution spectra of the \halpha region also reveal the presence of low-velocity narrow absorption lines, likely associated with a cooler shell or disk surrounding the recent merger event. For plateau spectra, observers often report `a forest' of metal lines, especially pronounced near 5000\,\AA.  During this phase, metallic lines previously showing P-Cygni profiles now transition to pure absorption, and the \io{Ca}{ii} triplet becomes stronger. In some objects, H$\alpha$ next regains its flux in the later phases of the plateau (see Fig. \ref{fig:specsequence_halpha}). Late-time spectra, taken hundreds of days after the main peak, show that the Balmer emission can eventually disappear when the spectrum turns into an M-type star. 

\paragraph{(IV) Decay}

In the latest phases of the fast fading from the plateau, the absorption spectrum shifts from yellow towards progressively more red photospheres, eventually displaying a spectrum mimicking M-type giants and supergiants. Most characteristic for this phase are absorption bands of TiO and VO (the latter only in the red part of the optical spectra). The latest spectral type reported in this phase is later than M10 \citep{evans}, but it should be noted that these dimming photospheres especially are embedded in cool circumstellar molecular gas, which strongly affects the absorption features used for spectral classification. This may lead to the assignment of spectral types later than actually characterizing the stellar photosphere. While the molecular bands at the end of the outburst of a new transient are generally a smoking gun of a red nova, there have been cases when objects miss any detectable features \citep[e.g., BLG360;][]{SteinmetzBLG} or the object was still too hot for the molecular bands to be apparent in the spectra \citep[e.g., NGC4490-2011OT1;][]{Smith2016}. In the plateau and later phases, the photospheric absorption spectrum and that of circumstellar molecules are combined with the emission spectrum of hydrogen and of species \io{Na}{i}, \io{K}{i}, [\io{Ca}{i}], [\io{Ca}{ii}], \io{Ca}{ii}, [\io{O}{i}], and \io{Mn}{i} \citep[e.g.,][]{Martini1999}.

As described in more detail in Sect.~\ref{section-remnants}, early post-outburst spectra of red novae indicate an environment of low temperatures. When the stellar-remnant photosphere is observed, which is a rare case even for the Galactic red novae, it shows a giant or supergiant photosphere of spectral type M \citep[e.g.,][]{Kami-specpol}. The post-outburst system is best studied through spectral features of the circumstellar medium, often exclusively through emission lines, including gas and dust probes at IR and submillimeter wavelengths. Some spectral appearances of red-nova remnants can mimic young stellar objects (YSOs; for instance, those with molecular emission bands, \citealp{GuoZ}). A surprisingly common spectral feature in the early post-outburst phases of red novae is the presence of molecular bands of AlO \citep{Banerjee2015}. Bands of CrO have been observed exclusively in these remnants \citep{KamiMason,Steinmetz}. 

\paragraph{Kinematics over the entire outburst}
Emission and absorption lines have been used to derive the merger ejecta velocities. However, those measured from the widths of the hydrogen recombination lines in emission are almost always higher (even 2000 \kms\ FWHM) than those measured with other probes. Scattering of electrons in the recombining plasma can often broaden the lines and give them a Lorentzian shape. This is true especially in the early parts of the main peak, when the ionization degree is highest (see top row of Fig.~\ref{fig:specsequence_halpha}). Therefore, outflow velocities derived from other lines, especially from the optical \io{Na}{i}, resonance doublet, are generally a more reliable measure of the ejecta kinematics (if the interstellar component can be accounted for). These lines trace very well the cool matter that is associated with all red novae. The FWHM velocities measured this way in spectra of red novae in outburst are typically 100--500 \kms\ and correlate with other properties of the system (see Sect.~\ref{sect-relations}). The continuum definition is often a significant source of uncertainty and can affect the measured velocities. 

The peak or central positions of emission and absorption lines can shift over the duration of the outburst, but it is unclear what dominant factor is causing it. Asymmetric and variable line profiles, when unresolved with poor spectral resolution, may be a plausible explanation. Studies of the red nova remnants have demonstrated that the central line velocities observed during the outburst are generally not a good measure of the systemic velocity \citep{KamiSubmm}. Moreover, the line widths change over the course of the outburst, and those indicate real changes in the ejecta velocities. The highest ejecta velocities are found near maximum brightness and often exceed the escape velocities of the presumed progenitors. Matter ejected in later phases of the outburst is slower. These variable ejecta velocities are also well evidenced in the kinematic structure of the remnants observed years, decades, and centuries after the outburst.

\subsection{Non-optical spectra}
Red novae in outburst have been mainly studied at optical wavelengths, and little is known about their spectral characteristics at shorter (UV) and longer (MIR--FIR--submm) wavelengths during the eruption. Red and NIR spectra of V1309 Sco acquired during its main peak showed \io{Fe}{ii} emission lines and the Paschen and Brackett series mostly in absorption, except for H$\beta$ to $\gamma$ Paschen transitions, which showed only emission \citep{Rudy2025}. Later in the main peak, absorption bands of CO and H$_2$O could have been present \citep{Rudy2,Rudy1,Rudy2025}. The 2.3--2.4 $\mu$m bands of CO were observed throughout the entire outburst of V838~Mon \citep{Geballe} as well in late-time spectra for some extragalactic red novae \citep{Blagorodnova2021,Cai-2021biy,Karambelkar2023ApJ}. As soon as dust condensation occurs (plateau or later), IR dust features show up. Most characteristically for the galactic and extragalactic red novae, a few years after the main outburst, a broad `10 $\mu$m' feature is observed in absorption \citep{nicholls,KarambelkarJWST}. The feature is caused by freshly condensed dust consisting of silicates and, less often, of alumina or iron grains \citep[e.g.,][]{Banerjee2015,SteinmetzBLG}.

\subsection{Polarimetry and spectropolarimetry}\label{sect-specpol}
Spectropolarimetry experiments are rare for red novae, but a few observations have been performed during or shortly after the outburst \citep{Desidera,WisniewskiSpecpol,Cai-2021biy} and several years later, during the post-outburst phases \citep{Kami-specpol, MobeenIII}. When corrected for interstellar extinction, the continuum polarization of the luminous red novae is very low, at a level of a few \%.
In the case of V838 Mon, polarimetric observations revealed that the main luminosity peak coincided with the maximum polarization observed of this object (intrinsic polarization of $0.7-1.0\%$), indicating that the emitting region was not spherical \citep{WisniewskiSpecpol,Desidera}. At shorter wavelengths, the object appeared to be more polarized, which suggested the existence of additional mechanisms besides Thompson scattering. Freshly formed dust, or older dust predating the explosion, could also contribute to scattering at this low level. However, the changes in polarization in V838 Mon were just of the order of days, which were at odds with longer dust formation timescales.

Studies of polarization in spectral features are even more scant. During and shortly after its outburst, V838~Mon showed only small depolarization effects across H$\alpha$. Due to the very modest signal-to-noise ratio (S/N), these observations were not very informative. For the low-mass red nova remnant V4332 Sgr, however, continuum polarization as high as 17\%\ was observed and interpreted as scattering off the dusty disk and bipolar outflows surrounding the star \citep[][see also Sect. \ref{sect-geometryCSM-lowmass}]{Kami-specpol}. The presence of pronounced depolarization in atomic lines and molecular bands revealed a stratified circumstellar structure, in which the gas producing fluorescent emission is decoupled from the asymmetric continuum-scattering geometry. Such observations can thus be used to investigate the structure of the objects in outburst or during the remnant phase. A major limitation is that such observations require both high S/N and high spectral resolution; as a result, detailed spectropolarimetric studies are restricted to the brightest events and have proven challenging even for the most luminous red novae in the Milky Way.

\subsection{Interpretations of the outbursts}\label{sect-LCinterpretation}

The interpretation of the observed outbursts, especially the multiple peaks in the light curves, is still debated. In the early days of the red nova research, it was proposed that coalescence occurs through a series of grazing collisions, with each periastron passage producing a separate peak in the light curve \citep[cf.][]{TylendaSoker2006,TylendaBLG}. Another hypothesis linked each peak to the engulfment of a planet \citep{Retter}. However, these early hypotheses did not receive much support in recent studies. In particular, it would be difficult to explain why we often observe 2--3 peaks in systems of very different progenitor types and orbital parameters. Simulations of common-envelope evolution often relate to the red nova mergers but encompass a much broader range of phenomena than red novae; we therefore refer the reader to dedicated reviews \citep{Ivanova2013Rev,Ivanova2020ceebook,RoepkeDeMarco2023,SchneiderAnnRev}, and focus here on current efforts to reproduce the observed light curves of red novae, specifically.

Recent studies encompass 1D \citep{Lipunov, ChenIvanova2024,MetzgerPejcha2017, MatsumotoMetzger, ZhuoChen2026}, 2D \citep{Kirilov2025ApJ}, and 3D \citep[e.g.][Mu in prep.]{HatfullIvanova2021} (radiation-) hydrodynamical models, with some success in explaining the observations. A major advance was the inclusion of radiation transport and radiation hydrodynamics \citep{ChenIvanova2024,HatfullIvanova2021,Kirilov2025ApJ}, but fully self-consistent models are still missing. Below, we briefly review a selection of recent ideas proposed to explain the observed characteristics of red nova outbursts. However, it may still be premature to generalize the physical mechanisms, as factors such as viewing angle, orbital configuration, mass ratio, and evolutionary stage can introduce significant diversity, and both observations and models remain limited. 

\paragraph{Precursor brightening} 

Only the closest and brightest objects were confidently detected during the gradual precursor rise months to years before the main outburst, with widely varying timescales and amplitudes (see Table \ref{tab-main}; Sect.~\ref{sect-lightcurve}). For example, in M31-LRN-2015 the precursor began tens of orbits before peak \citep{MacLeod2017}, whereas in V1309 Sco it started $\sim$1300 orbits earlier \citep{Tylenda2011,Pejcha2014,Pejcha2017}. For the more extended AGB progenitor of AT2025abao, the precursor emission initiated $\sim$10\,years before its transient event \citep{Karambelkar2025ApJ}. The connection to progenitor properties and cause of orbital instability, though, remains unclear. 

While the precursor rise has rarely been modeled in the radiation-hydrodynamic studies of the light curves, there is generally a consensus \citep[but see discussion in][]{NandezV1309} that this phase is related to a runaway buildup of circumbinary matter, mainly via the outer Lagrange L2 point. Runaway mass loss from the system accelerates orbital decay, ultimately driving the binary toward merger and producing the characteristic `concave-up' light-curve evolution observed in precursor emission \citep{Pejcha2017,MacLeod2018b,MacleodLoeb2020}. The evolution of binaries experiencing extreme mass-transfer rates is an active area of research, providing new constraints on the pre-merger (or pre-supernova) mass-loss phase through detailed treatment of momentum exchange and the formation and structure of mass outflows \citep[e.g.][]{Lu2023,Scherbak2025}. Here, we mostly focus on the connection with red nova events.

\cite{Pejcha2017} \citep[see also][]{Pejcha2016,Pejcha2016b} specifically investigated the precursor evolution of red novae, with a primary focus on reproducing the observed properties of the V1309 Sco progenitor system. They demonstrated that the rise in brightness was due to collisions and shocks within dense material accumulated by the binary in the orbital plane. In later phases of the precursor (e.g., near 2007 in Fig.~\ref{fig-v1309lc}), the shock luminosity approaches the binary's luminosity itself, fueled by increasing disk mass and higher temperature of the gas feeding the outflow. \citet{MacLeod2017} proposed that the precursor peak is related to tidal heating of the primary, which is conceptually consistent with the `velocity shear' model of \cite{Pejcha2017}. The mass accumulated during the precursor phase may be comparable to that ejected during the dynamical interaction, and thus strongly influences later evolution, calling for more observational identifications of precursors. Observationally, this picture is consistent with the bipolar morphologies and disks detected in low-mass mergers like V1309 Sco and V4332 Sgr \citep{KamiSubmm,Steinmetz}. 



Observations of the light curve of V1309 Sco show that at the end of the precursor phase the eclipsing binary modulation disappears (see highlight B. in Fig.~\ref{fig-v1309lc}), i.e., when the binary becomes embedded in the optically thick material. Some linked it to the onset of the common envelope \citep[e.g.,][]{Pastorello2019review}.


\paragraph{Dip} 
The dip may be a consequence of the terminating precursor phase, and it can be related to the viewing angle. It was best studied in V1309 Sco, where the decrease in $I$-band flux (Fig.~\ref{fig-v1309lc}) was interpreted as caused by the buildup of circumstellar material obscuring the central binary at visual wavelengths in our line of sight \citep{Pejcha2017}. This model naturally explains the appearance of the dip at the latest phases of the precursor and the disappearance of the eclipses. Alternatively, \cite{Tylenda2011} assigned the feature to an enhanced column of dust in the line of sight to the coalescing stars, but the timescales for this dust to be formed appear to be too short to explain the fast light curve dimming.

\paragraph{Main (blue) peak}
The rise to the main peak is sometimes interpreted as the first contact between the merging stars (the plunge-in phase), during which dense material is compressed and strong (hot) shocks form in the primary's envelope \citep[e.g.,][]{SokerTylenda2006,NandezV1309}. For instance, hydrodynamic calculations suggest an increase in ejecta velocities at this phase due to loss of co-rotation as the orbit of the merging pair tightens \citep{MacLeod2018a}, initiating the dynamical phase of the merger \citep{Ivanova2013Rev}. In the classical common envelope evolution picture, the orbital energy of the binary is used to unbind (part of) the envelope, which then expands almost freely at velocities above the local escape velocity \citep{ChenIvanova2024}. Most authors agree that the main peak is due to the release of thermal energy of this unbound, expanding, shock-heated gas. During this stage, the ejecta is mostly ionized and optically thick, forming a likely asymmetric pseudo-photosphere (see Sect.~\ref{sect-specpol}) that expands at velocities of a few hundred \kms. This phase resembles other eruptive events associated with strong winds, which may explain the characteristic main-peak temperatures near 7500\,K \citep{DavidsonEruptions, Quataert} (cf. Fig.~\ref{fig-RTeffL}) and strong hydrogen emission. The interaction models of \citet{Kirilov2025ApJ} show that early shocks between merger ejecta and circumbinary material can also partially contribute to the total luminosity, but this effect does not dominate at peak light and appears sensitive to the initial conditions.


Soker et al. \citep[e.g.,][ and references therein]{RedNovaReqJets} advocate an alternative interpretation of the rapid luminosity rise during the main peak. In their \emph{grazing envelope evolution} model, which avoids a merger, the rapid rise is caused by jets forming near the more compact companion star and slamming into the surrounding material. 
%


\paragraph{Plateau (red peak)} 

Early work by \citet{IvanovaScience} adapted the framework of hydrogen recombination cooling, originally developed for type IIP SNe by \citet{Popov1993}, to account for the observed properties of red nova light curves. Within this scenario, several subsequent studies \citep[e.g.,][]{ChenIvanova2024,MatsumotoMetzger} have emphasized hydrogen recombination as the dominant power source of the extended plateau phases seen in luminous red novae. In these models, recombination energy is required to power the plateau because most of the initial thermal energy of the ejecta is lost to adiabatic expansion rather than being radiated away. \cite{ChenIvanova2024} and \cite{Chen2026ApJ} also proposed that during the plateau, the ejecta transitions from a radiation-dominated to matter-dominated regime, and that the formation of molecular hydrogen can power an additional, late-time lower luminosity plateau. So far, recombination models have explained reasonably well red novae signatures for fainter transients such as V1309 Sco or AT2019zhd. However, the most luminous events seem to require ejecta masses significantly larger than those inferred from observations, or they struggle to reproduce the observed temperature evolution \citep[cf.][]{MatsumotoMetzger,RedNovaReqJets}.

The longer, redder peaks observed in the light curves of the `risers' group were explained as a result of the interaction of the merger ejecta with the pre-existing circumbinary material. While the scenario was first introduced by \cite{MetzgerPejcha2017}, \cite{Kirilov2025ApJ} investigated the interaction in detail and found that the structure and geometry of the pre-merger circumbinary material influence the amplitude and length of the plateau phase, analogous to circumstellar medium-interacting SNe \citep[cf.][]{Dessart2016}. The ejecta can strongly energize the circumbinary matter so that it can remain bright even after the shock wave is gone. Substructures in the light curves (i.e., extra peaks or dips) can be directly linked to the structure of the circumbinary material (likely a torus) penetrated by the radiative shocks. In the scenario considered by \cite{Kirilov2025ApJ} \citep[see also][]{MacLeod2017}, there is no special role of recombination energy in this phase of the outburst: hydrogen recombination does happen but does not increase the luminosity of the plateau and only slightly prolongs its duration. One possibility is that among red novae, different populations may require different mechanisms driving the plateau (red peak) phase. For instance, the `plateau' and `risers' groups (Sect. \ref{sect-lightcurve}) may be physically different in that context. 

To date, most simulations of red nova outbursts have assumed intrinsically spherical merger ejecta \citep[but see][]{NandezV1309}. Nevertheless, asymmetries are expected to develop during the plateau phase in many merger models. In more recent hydrodynamical simulations, initially spherical ejecta interact with a dense circumbinary torus or disk, which redirects the flow and suppresses equatorial expansion. As a result, the ejecta expand preferentially along the polar directions, producing a bipolar outflow structure \citep{Pejcha2017,MetzgerPejcha2017}. An alternative interpretation is proposed by \citet{RedNovaReqJets} within a non-terminal grazing envelope evolution (GEE) scenario. In this framework, the ejecta consist of multiple components with different opening angles, ranging from highly collimated jets to wide-angle outflows. Their interaction with the surrounding medium generates a sequence of shocks and mass-ejection episodes that combine to form the observed plateau. However, detailed multidimensional simulations of the GEE scenario are still lacking. 

The role of dust formation in shaping the light curve is an active topic of research, as dust condensation is notoriously difficult to implement in simulations of eruptive events \citep[cf.][Mu in prep.]{GonzalezBolivar2024,BermudezBustamente2024}. While some simulations of red nova outbursts find that dust formation does not strongly affect the light curve evolution up to the end of the plateau phase \citep[e.g.,][]{ChenIvanova2024}, dust formation in the remnant is an essential requirement to closely reproduce the observed light curve shapes \citep{HatfullIvanova2021,Kirilov2025ApJ}, color evolution, and the dust masses measured in young red nova remnants.

\paragraph{Post-plateau decay} 
The immediate post-plateau evolution is related to the dissipation of energy stored in the pre-merger torus and merger ejecta, but at much lower densities and temperatures than in the earlier phases \citep{Kirilov2025ApJ}. On longer time-scales, the post-plateau relaxation phase is typically interpreted as partial gravitational contraction of the remnant and geometrical dispersal of the ejecta \citep{TylendaSoker2006,Schneider2016MNRAS}. Further modifications of the transient's brightness can be due to infall activity \citep{GoranskijATelV4332}. Low-amplitude changes years and decades after the eruption have been observed in many Galactic objects, but their origin is unclear. See Sect.~\ref{section-remnants} for discussions on the post-outburst relaxation in red nova remnants.

\section{The progenitors}\label{sect-progenitors}
Pre-outburst photometry of red novae comes from surveys that serendipitously covered the position of the progenitor years before the red nova outburst. The parameters of these systems are listed in Table \ref{tab-progenitors}, and mainly correspond to the primary, which in most cases is assumed to dominate the progenitor's flux. The properties of the secondary components are much less constrained, and only in the case of V1309 Sco was the secondary directly observed before the red nova event. Table \ref{tab-progenitors} covers a wide range of the primary masses, from $\approx$1 to 40 \Msun. The progenitors also span a wide range of evolutionary stages, and include pre-MS stars, MS stars, subgiants and yellow supergiants, RGB stars, AGB stars, and even a white dwarf and a planet. The red nova progenitors as a population are related to the red nova rates discussed in Sect.~\ref{sect-stats}. Below, we first discuss the progenitor data for Galactic red novae before introducing the larger sample of known extragalactic progenitors. We next describe efforts to identify objects evolving into a red nova and emphasize the role of triples among red nova progenitors.

\begin{sidewaystable}
\caption{Properties of red nova progenitors}\label{tab-progenitors}
\footnotesize
\begin{tabular}{l l l | l l l l | c}
\toprule
Object & Merged binary & Evolutionary & \multicolumn{4}{c|}{Primary's properties} & Main \\
     & ($M_\odot$) (if known) &   stage or type & $M$ ($M_\odot$) & $L$ ($L_\odot$) & $T_\mathrm{eff}$ (K) & $R$ ($R_\odot$) & refs.\\
\hline
  CK Vul & 2+0.3 & RGB+HeWD &  &  &  &  & [1] \\[1pt]
  V4332 Sgr & $\sim$1 & sub-giants? & $1^{+1}_{-0.3}$ &  &  &  & [2] \\[1pt]
  V838~Mon & (5-10)+$\sim$0.4 & MS+pre-MS & $7.5^{+2.5}_{-2.5}$ & $900 \pm 200$ & $17000 \pm 1000$ &  & [3] \\[1pt]
  BLG-360 &  & RGB+? & $\sim1$ &  &  &  & [4] \\[1pt]
  V1309 Sco & 1.5+0.2 & contact sub-giants & $1.5^{+0.5}_{-0.5}$ & $5.8 \pm 2.8$ & $4500 \pm 200$ &  $\sim 1.6$& [5] \\[1pt]
  ZTF SLRN-2020 & (0.8-1.5)+planet & MS+super-Jupiter &  $1^{+0.5}_{-0.2}$& $\sim$0.3 &  & $\sim$1 & [6] \\[1pt]
  M31-LRN-2015 & (3-5)+(0.1-0.6) & subgiant & $4^{+1.5}_{-1}$ & $600 \pm 200$ & $5250 \pm 1250$ & $30 \pm 10$ & [7] \\[1pt]
  AT2019zhd &  & &  &  &  &  & [8] \\[1pt]
  AT2025abao &  & M-type AGB & $7^{+2}_{-2}$ & $16\,000 \pm 4100$ & $3500 \pm 200$ & $350 \pm 50$ & [9] \\[1pt]
  AT1997bs &  &  & $40^{+20}_{-20}$ & $(1.6 \pm 0.9) \times 10^{5}$ & $13500 \pm 6500$ &  & [10] \\[1pt]
  N4490-OT2011 &  & LBV? YSG? & $30^{+10}_{-10}$ & $(1.3 \pm 1.2) \times 10^{5}$ & $14000 \pm 6500$ &  & [11] \\[1pt]
  N3437-2011OT1 &  & YSG? & $>10$ &  &  &  & [12] \\[1pt]
  UGC12307-2013-OT1 &  & YSG? & $>10$ &  &  &  & [13] \\[1pt]
  SNhunt248 &  & YHG & $30^{+2}_{-2}$ & $(4 \pm 1)\times 10^{5}$ & $6500 \pm 300$ & $\sim500$ & [14] \\[1pt]
  M101-2015OT1 &  & YSG & $18^{+1}_{-1}$ & $88\,000 \pm 8000$ & $6600 \pm 300$ & $220 \pm 25$ & [15] \\[1pt]
  AT2015fx &  & YSG & $17.5^{+2.5}_{-2.5}$ & $1 \times 10^{5}$ & $7250$ &  & [16] \\[1pt]
  AT2018bwo &  & YSG & $13^{+3}_{-2}$ & $18\,000 \pm 1300$ & $6280 \pm 340$ & $114 \pm 9$ & [17] \\[1pt]
  AT2018hso &  &  &  & $(4.7 \pm 3.0) \times 10^{5}$ & $13500 \pm 4000$ & $125 \pm 50$ & [18] \\[1pt]
  AT2020hat &  & YSG (mid-K) & $\lesssim8$ & $3700 \pm 600$ & $4800 \pm 500$ &  & [19] \\[1pt]
  AT2021blu &  & YSG & $15.5^{+2.5}_{-2.5}$ & $41\,000 \pm 6000$ & $6800 \pm 300$ & $144 \pm 14$ & [20] \\[1pt]
  AT2021biy & (17-24)+5? & YSG &  & $(10\pm 1) \times 10^{4}$ & $5900 \pm 200$ &  & [21] \\[1pt]
\toprule
\end{tabular}
\footnotetext{{\bf References:}[1] \cite{TylendaCKfinal} [2] \cite{KamiSubmm} [3] \cite{TylendaProgenitor} [4] \cite{TylendaBLG} [6] \cite{Stepien} [7] \cite{DeNature} [8] \cite{Pastorello2021_2019zhd} [9] \cite{Karambelkar2025ApJ} [10] \cite{VanDyk1999_1997bs_progenitor,VanDyk2000_1997bs,Adams2016MNRAS} [11] \cite{Smith2016} [12] \cite{Pastorello2019review} [13] \cite{Pastorello2019review} [14] \cite{Mauerhan2015} [15] \cite{Blagorodnova2017} [16] \cite{Tartaglia2016ApJ_AT20215fx} [17] \cite{Blagorodnova2021} [18] \cite{Cai2019} [19] \cite{Pastorello2021AA} [20] \cite{Pastorello2023} [21] \cite{Cai-2021biy}}
\end{sidewaystable}

\subsection{Galactic red nova progenitors}\label{sect-progenitors-gal}
Characterizing the progenitors is challenging. The available archival pre-outburst photometry is often non-simultaneous and limited to a few bands, rarely reaching the IR regime. Additional problems are related to extinction correction (i.e., reddening) and field contamination. The extinction problem is especially complex because, in addition to the interstellar component, many of the progenitor stars have circumstellar dust, which may add an extinction component of a different characteristic than the canonical interstellar dust. Another challenge is the potential presence of emission lines in systems evolving towards red nova. V1309 Sco remains the only object whose progenitor was observed with a few-day cadence before the red nova outburst \citep{Tylenda2011} (Fig.~\ref{fig-v1309lc}). Even in this case, though, the observations were made quasi-simultaneously in only two photometric bands, at best. None of the Galactic red nova progenitors has been observed spectroscopically.

An additional complication, more relevant for Galactic objects than for the extragalactic ones, is the problem of their distances. Distances are required to constrain the source luminosity and, consequently, the evolutionary state. The only Galactic red nova with a well-known distance is V838~Mon. Its distance of 5.9$\pm$0.4 kpc has been constrained with several methods, including polarimetric light echo observations \citep{sparks} and interferometric maser parallax \citep{Ortiz-Leon2020}. Distances to other objects are often based on their reddening, ISM features, galactic radial velocity, or kinematical expansion of the ejecta. These methods carry substantial uncertainties, leading to luminosity and mass estimates that can vary by factors of a few. 

To illustrate the challenges and the impact of uncertainties in characterizing progenitors, we revisit the case of V838~Mon, which erupted in 2002. \citet{TylendaProgenitor} summarized the available progenitor data, including photometric measurements in six bands spanning from $B$ to $K_s$. In addition, Goranskij et al. (\citeyear{goran2004}) examined a large collection of photographic plates from 1928--1994 and found no variability in the $B$ and $V$ bands, justifying the comparison of data obtained at widely separated epochs. This archival effort, later updated in \cite{GoranskijHistorical,Goranskij2020}, provides one of the longest--albeit irregular--historical records for a red nova progenitor.

Early post-outburst spectra revealed that V838~Mon has a distant companion whose spectral features correspond to a B3 or B4 MS star (luminosity class V). Tylenda et al. compared the observed fluxes with standard stellar colors compiled from the literature (largely based on now outdated calibrations from the 1980s). Because the interstellar extinction toward the object was poorly constrained, they considered two values of $E(B-V)$, 0.9 and 0.71 mag. In their analysis, they assumed that the low-mass secondary consumed in the merger did not contribute significantly to the spectral energy distribution (SED), and modeled the system as the combined light of the merger primary and the distant blue companion. They found that the combinations B3V+B1.5V and B4V+A0.5-protostar best reproduced the observations for the two assumed reddening values, respectively.

For comparison, using a more modern calibration of stellar colors and spectral types from \citet{Mamajek2013} and adopting $E(B-V)=0.9$ mag, one obtains a B3V+B3V configuration \citep[cf.][]{TylendaKaminskiEcho}. The inferred stellar masses remain highly uncertain: masses for B3V stars span $\sim$5--10 \Msun\ in the literature, and the range becomes even broader when other spectral types are considered (down to $\sim$2 \Msun\ in the A0.5 protostar scenario). The V838~Mon system is a member of an open cluster of B-type stars \citep{AfsarBond}, yet the progenitor and its companion are $\sim$1.3 mag fainter than other cluster members. This discrepancy has been investigated \citep{TylendaKaminskiEcho,Goranskij2020}, including the possibility of patchy extinction across the cluster, but no satisfactory explanation has been found. Possible interpretations include circumstellar or circumbinary extinction, rotational deformation, or a higher luminosity class (IV) for the components. Interestingly, a somewhat more evolved (and dimmer) evolutionary state for the progenitor has also been proposed on theoretical grounds, where most of the pre-outburst contribution would be attributed to the distant tertiary star \citep{Temmink2025}.

The case of V838~Mon highlights how difficult it is to reliably identify progenitor properties, even when relatively extensive historical data and a well-constrained distance are available. The mass of the engulfed companion was estimated to be 0.1--0.5 \Msun\ \citep{TylendaSoker2006}, representing the minimum accreted mass required to power the observed radiative and kinetic output of the outburst. These estimates assume a primary mass of 8 \Msun\ and a radius of 7.5 \Rsun. A rigorous propagation of uncertainties shows that most parameters describing the progenitor are uncertain to within at least an order of magnitude. Consequently, stellar mass estimates for the objects listed in Table \ref{tab-progenitors} should be treated with caution: at best, they allow a classification into broad categories such as low-, intermediate-, or high-mass stars, rather than precise determinations. The most robust parameter that can currently be inferred is the evolutionary stage of the progenitor system.

Tracing back the evolution of the identified progenitor binaries is not an easy task either. As mentioned, V1309 Sco's progenitor was the only system evolving into a merger with both components observed in the photometric data. The observations showed a low-mass binary in deep contact, similar to W\,UMa systems but with a slightly longer orbital period of 1.43 days and in a more advanced evolutionary stage. Several studies attempted to model the evolution of V1309 Sco leading to the state in which it was observed in the 2001--2008 period, just before the red nova event. That of \citet{Stepien} was most comprehensive because it attempted to reproduce the evolution starting at zero-age main sequence (ZAMS). In his best model, the object starts as a 1.2+0.5 \Msun\ binary with an orbital period of 2.8 days. Magnetic braking causes loss of angular momentum; the orbit shrinks, and the radius of the primary increases as it evolves to the subgiant branch. After 5.4 Gyr, the system undergoes the first Roche-lobe overflow (RLOF) mass transfer phase and a successive common envelope phase, which produces an Algol-like system with inverted masses of 0.51+1.22 \Msun\ and a significantly shorter orbital period of 1.2 d. The mass transfer continues but at a much lower rate, and the now more massive star evolves towards the RGB, building up a helium core. This continues until Darwin instability triggers the merger. At the age of 7.85 Gyr, the system has the properties of the observed progenitor of V1309 Sco: a contact binary with a mass of 0.16+1.52 \Msun, orbital period of 1.44 d, and a surface temperature of $\sim$5000 K. Both components have helium cores of $\approx$0.15 \Msun\ at the moment of the red nova event. This detailed model is very fine-tuned to V1309 Sco and should not be generalized, especially to W\,UMa systems. Despite the many uncertainties---including mass loss, strength of magnetic fields, assumed internal structure of the stars--- the model of \citet{Stepien} remains the most sophisticated attempt to model a red nova progenitor.

\subsection{Extragalactic red nova progenitors}\label{sect-progenitors-ex}

Detections of extragalactic red nova progenitor systems were usually possible using archival images from space-based \citep[Hubble Space Telescope, Spitzer Space Telescope; see][]{Smith2016,Blagorodnova2021} and deep ground-based images \citep[e.g.,][]{Blagorodnova2017,Pastorello2021AA}. Because of the limiting magnitudes of these datasets, usually only the brighter stellar progenitors (or the closest ones) were detected. 
Currently, as shown in Table \ref{tab-progenitors}, there are 15 extragalactic systems (out of 23 listed in Table \ref{tab-main}) with a direct archival detection of a progenitor. 

\begin{figure}[ht]
\begin{center}
 \includegraphics[width=0.75\textwidth]{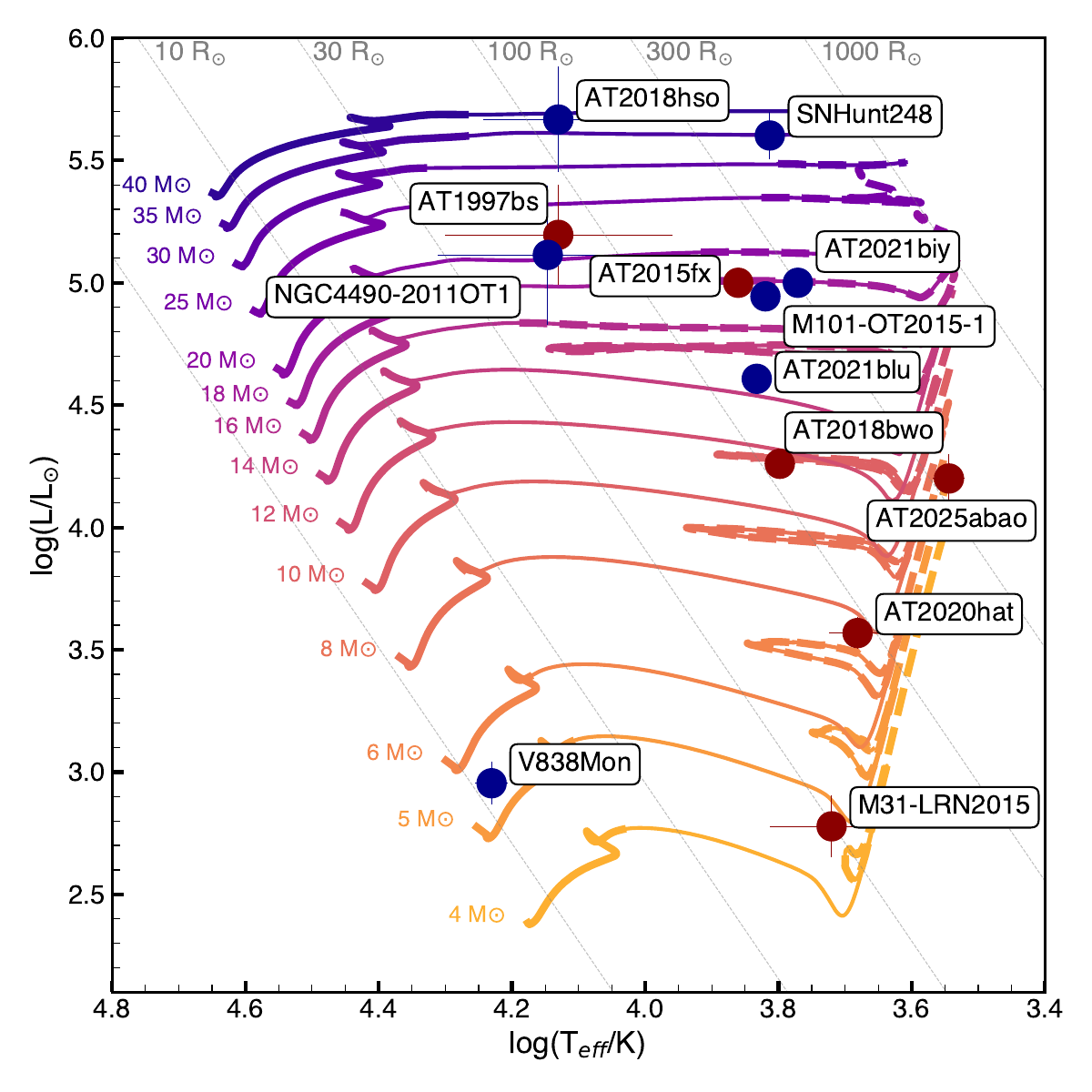} 
\end{center}
\caption{Luminosities and effective temperatures for selected red nova progenitors. Drawn lines show MESA MIST models \citep{Dotter2016_MIST,Choi2016ApJ_MIST} for initial masses between 4$-$40\,\Msun at solar metallicity and 0.4 critical rotation. The main sequence is marked with thick solid lines. The transition to core He burning is represented by thin lines. The He core burning and later phases are represented by dashed lines.}
\label{fig:progenitors} 
\end{figure}

The majority of archival images were obtained in at least two photometric bands. It was possible to derive the color and intrinsic luminosity of the more evolved and massive primary star, which was assumed to dominate the light of the whole system. To determine the mass range and evolutionary state of the primary star, some studies compared the observed magnitudes in different filters with synthetic photometry provided by grids of stellar evolution tracks \citep{Cai-2021biy}. Other studies derived the stellar SEDs using blackbody, modified blackbody, or model-atmosphere synthetic spectra, and then compared them with model grids of single and binary evolutionary tracks \citep{Blagorodnova2017,MacLeod2017,Pastorello2021AA}. Figure \ref{fig:progenitors} places these sources in the Hertzsprung-Russell (HR) diagram. 

The dominant spectral types for red nova progenitors appear to be yellow supergiants, which are stars in fast expansion after the end of their hydrogen core burning (i.e., in the Hertzsprung gap). This dominance of a single type of star in the observed population is striking and contrasts with the broader range of progenitor types of the less-luminous red novae known from the Milky Way. Some less massive progenitors that might break this extragalactic monopoly are AT2020hat (a K-type supergiant) and AT2025abao (an AGB star), and possibly a few luminous stars that share similarities with LBVs. The apparent lack of evolved donors, especially in systems with $<5$\,\Msun, has been attributed to dust obscuration: prolonged, non-conservative mass transfer can cause gas to accumulate, cool, and form dust around the binary, hiding the final ejection event \citep{MacLeod2022}.

A few studies developed detailed binary evolution models, often using the stellar evolution code MESA \citep[Modules for Experiments in Stellar Astrophysics;][]{Paxton2011,Paxton2013,Paxton2015,Paxton2018,Paxton2019,Jermyn2023}, to characterize the pre-merger systems during quiescence. These included more comprehensive descriptions of physical properties of the binary (donor's mass and evolutionary stage, mass ratio, and period), mass-transfer rates \citep{Blagorodnova2021}, and different orbital instability criteria (Wavasseur in prep.). These model-supported studies converge on the conclusion that the extragalactic progenitors observed as yellow supergiants likely underwent a delayed dynamical instability \citep[DDI;][]{HjellmingWebbink1987ApJ}. In this evolutionary channel, the Roche-lobe overflow begins as stable mass transfer but progressively accelerates as the donor’s structure evolves, ultimately triggering dynamical instability and leading to a merger on timescales of a few hundred years. During this phase, the donor star experiences a substantial decrease in luminosity by up to an order of magnitude \citep[cf.][]{Temmink2025}, as energy from the inner regions is diverted into stellar expansion rather than transported outward as radiative luminosity. Observationally, this implies that donors undergoing high mass-transfer rates may appear anomalously faint, and can therefore be misidentified as lower-mass stars when interpreted using single-star evolutionary models.

\subsection{Searches for systems directly evolving into red novae}


The identification of a change in period in the eclipsing binary progenitor of V1309 Sco has fueled multiple attempts to predict the next red nova. For example, \cite{SokerTylenda2006} proposed searching for changes in orbital periods of W\,UMa contact binary systems with low mass ratios. Indeed, W\,UMa binaries with critically low mass ratios, $q<0.09$, have long been believed to be tidally unstable and particularly prone to Darwin instability \citep[][]{Rasio1995ApJ,PestaPejcha2023}. Recent studies present a more complex view on which binaries merge and on the necessary time scales \citep[e.g.,][ and references therein]{Wadhwa2025,Jiang2026}. 
Current W\,UMa catalogs seem to include a few candidates within the instability parameter space \citep[e.g.,][]{Wadhwa2025SerAJ,Jiang2026,Poro2026}, but problems with deriving precise stellar parameters and excluding the `third light' (or other tertiary companion effects) have hampered definitive identifications \citep{Kurtenkov2017,Wadhwa2025}. The search continues with multiple observing campaigns \citep[e.g.,][]{Gazeas,kobulnicky}. Since W\,UMa stars are relatively frequent (0.2\% in the solar neighborhood, \citealp{Rucinski}), the scarcity of the likely merger candidates suggests that not many red nova progenitors can be identified this way without considerable improvement in the sensitivity of our observations, which may soon become feasible with Rubin-LSST \citep{LSSTroadmap}. We also note that the progenitor of V1309 Sco was already in an advanced evolutionary stage when first observed, as it was a sub-giant or a giant \citep{Stepien}, while most W\,UMa stars are considered to be on the main sequence. \cite{kobulnicky} found that only W\,UMa systems evolving off the main sequence have a chance to develop the Darwin instability and merge.

Stimulated by the V1309 Sco's progenitor properties, there have also been more general searches of binary systems with rapidly decreasing orbital periods. \citet{Pietrukowicz} browsed the vast OGLE repository, and among over 22\,000 systems with short periods, they found 56 binaries with some period decrease.  The maximal decay rate found is of --1.94$\times$10$^{-4}$ d\,yr$^{-1}$. In the same sample, however, Pietrukowicz et al. found 56 systems with increasing periods, which led them to conclude that all the observed period changes, positive and negative, are most likely due to the presence of a third body (or due to stellar spots). One system with a decreasing period caught special attention, when \citet{MolnarRednova} identified a W\,UMa system, KIC 9832227, in the \textit{Kepler} field. Multiepoch observations acquired with different telescopes suggested that the object had a decaying orbit, somewhat similar to V1309 Sco. In 2017, the authors predicted that the system would merge in 2022. However, a few years before the awaited merger, an error in the analysis was found \citep{Socia}, and the system's variable period was explained by a third body (\citealp{Kovacs}; but see \citealp{Salinas23}). KIC 9832227 did not become a red nova. It seems, therefore, that finding red nova progenitors based solely on orbital changes is very difficult and often misleading. In fact, based on theoretical considerations linking orbital decay to unstable mass loss, \citet{MacleodLoeb2020} showed that the first time derivative of the orbital period alone is insufficient to identify systems approaching merger, and that the second derivative provides a more robust diagnostic \citep[see also][]{Pejcha2014,Pejcha2017}.

Yet, another potential method of identifying red nova progenitors relies on a slow and steady increase in brightness of the progenitor system, i.e., during the precursor phase (see Sect.~\ref{sect-lightcurve}) linked directly to the inspiraling phase just before the merger \citep[Sect.~\ref{sect-LCinterpretation}; e.g.,][]{MacleodLoeb2020,Pejcha2017}. 
The slow rise has a much lower amplitude than the outburst, typically $\lesssim$ 3 mag. Based on the notion that many of the extragalactic luminous red novae had precursors with properties consistent with yellow supergiants, \citet{Addison} and \citet{Tranin} searched photometric archives of the Milky Way and the Local Group galaxies for yellow supergiants that mildly increased their brightness. Additionally, \citet{Garcia-Moreno2026AA} attempted to identify emission line stars in the Hertzsprung gap that showed variability. Since yellow supergiants are rapidly evolving, stringent selection criteria could be applied to find them in monitored galaxies. It was found that M31, M33, and M101 are the galaxies where spotting a red nova precursor is most likely and that these precursors come from primaries with masses close to either 5 or 10 M$_{\sun}$. Several Galactic and twelve extragalactic candidates for luminous red nova precursors were proposed and await verification. The searches of \citet{Tranin} and \citet{Addison} for yellow supergiant precursors are probably the most comprehensive (volume-wise, considering candidates in hundreds of galaxies) studies to date. They are, however, somewhat biased towards a certain subtype of red novae progenitor and source selection was based on sometimes uncertain extinction corrections for the Milky Way and nearby galaxies. This approach will turn out to be especially potent for finding red nova progenitors in the era of the Rubin-LSST survey \citep{LSSTroadmap}. 


\subsection{Mergers in triple systems}\label{sec-triples}
Although this topic remains relatively unexplored in the literature, there is growing evidence to suspect that a significant fraction of red nova progenitor systems are triples or higher-order multiples. V838~Mon remnant certainly has a surviving companion on a wide orbit of a few 100\,au, so, within the merger hypothesis, the collision took place in a hierarchical triple \citep{DesideraMunari,KamiALMA}. Additionally, the structure and kinematics of the molecular outflows of CK Vul observed some 300 yr after its eruption suggest that they arise from ejecta interaction with a high-gravity (e.g., a dwarf) companion \citep{KamiCKalma2}. This implies a triple system as the progenitor. Moreover, the majority of progenitors of the extragalactic red novae belong to the 10--30 \Msun\ population (Table \ref{tab-progenitors}) for which the fraction of triples is 60--70\% \citep{Multiplicity1,Multiplicity2}.

Triggering an instability that could lead to a merger is dynamically easier in hierarchical triples than in binaries (e.g., through the Kozai--Lidov cycles or as in \citealp{GlanzTriples}), and therefore searching for triple progenitors among red novae is fully justified \citep[e.g.,][]{Toonen2022,triplemergers,Kummer,HagaiFabrycky,Perets2026}, as much as it is hard in practice. The potential presence of a surviving companion in red nova remnants, however, puts extra uncertainties on our characterization of the remnants, as the companion would be an extra heat source that needs to be accounted for in SED analysis and may trigger shocks through interactions of the passing ejecta \citep[cf.][Sect. \ref{section-remnants}]{KamiALMA}.

\section{Red novae as a population}\label{sect-population}


\subsection{Scaling relations}\label{sect-relations}

\begin{figure}[ht]
\begin{center}
 \includegraphics[width=\textwidth]{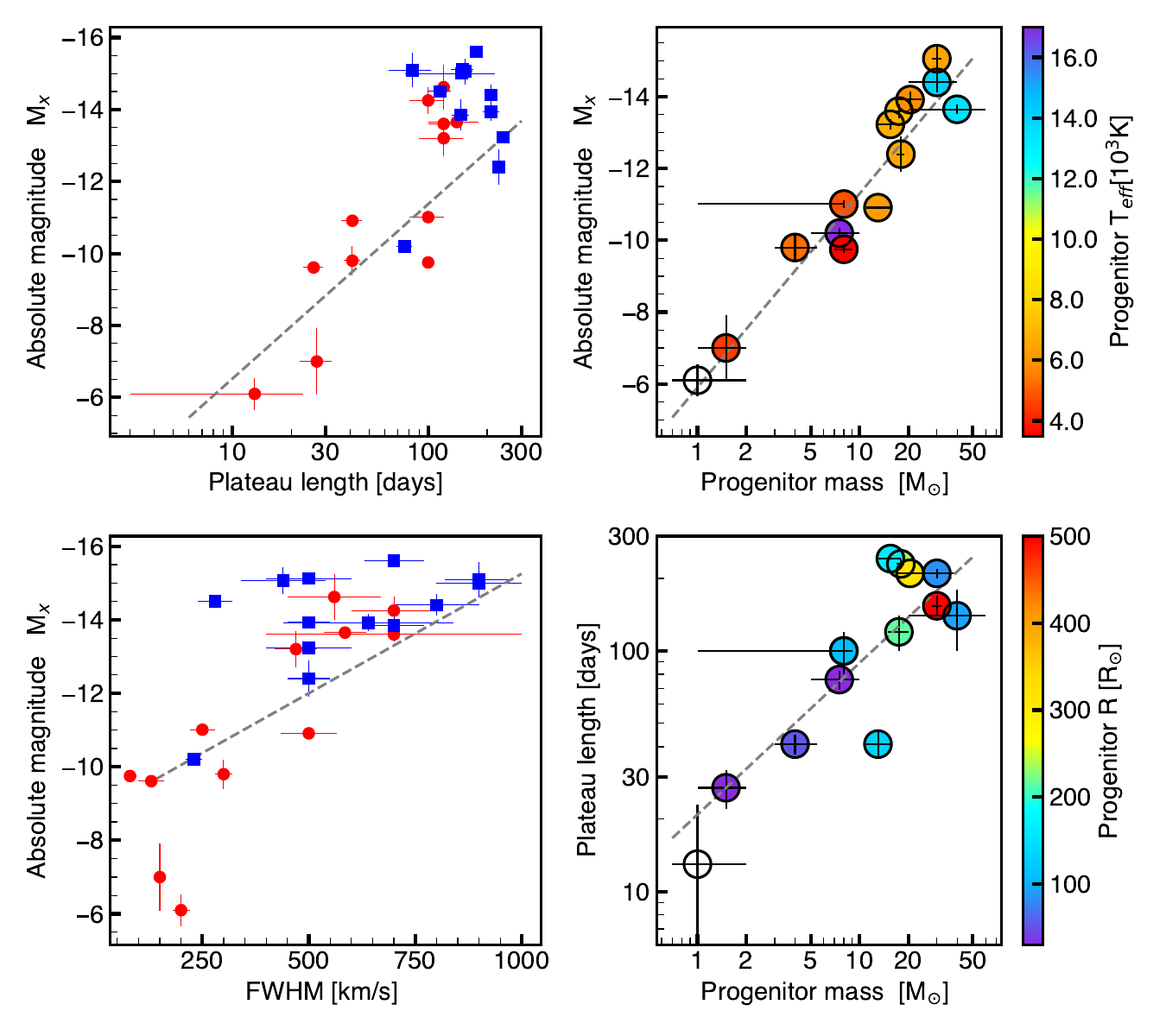} 
\end{center}
\caption{Scaling relations for red novae in Table \ref{tab-main}. Left panels: Comparison of observable properties relative to the outburst. Blue squares show the population of the `risers' group, while the red circles represent the `plateau' group (cf. Sect.~\ref{sect-lightcurve}). Absolute magnitudes refer to bands as in Table \ref{tab-main}, where \textit{x} refers to \textit{V/R/r} band. The FWHM represents typical linewidths during the eruption, as in Table \ref{tab-main}. Right panels: Comparison between the properties of the primary progenitor star and the outburst properties of red novae. Points are color-coded by the progenitor's temperature and radius (cf. Sect.~\ref{sect-progenitors} and Table \ref{tab-progenitors}; a white color indicates a lack of information.)}
\label{fig:scaling-relations} 
\end{figure}

Red novae represent a highly heterogeneous group. Although they are considered `intermediate luminosity transients', the faintest objects, such as V1309 Sco or V4332 Sgr, clearly overlap with the average luminosity of classical novae and the brightest red nova events are comparable to faint SNe (cf. Fig.~\ref{fig-diagram}). The spread in visual absolute magnitudes of nearly 10 mag suggests a range of physical mechanisms driving the red nova outburst. The red nova diversity is also emphasized by the wide variety of light curves, including their time spans and shapes (see Sect.~\ref{sect-lightcurve}). Among this apparent variety, recent studies explored possible correlations between the properties of red novae and of their progenitor systems that could be generalized for this entire transient family. Some of these correlations are shown in Fig.~\ref{fig:scaling-relations}, and our best-fitting relations applied to data in Tables \ref{tab-main} and \ref{tab-progenitors} are listed in Table \ref{tab-relations}.

\begin{table}[ht]
\caption{Best-fitting linear relations.}\label{tab-relations}
\begin{tabular}{@{}ll@{}}
\toprule
Relations from \cite{kochanek} & \\ 
$M_V =-7.2(\pm1.5) \times \log_{10}(M_\mathrm{prog} [M_{\odot}])-3.6(\pm 0.8)$ & \\
$M_I =-5.1(\pm0.7) \times \log_{10}(M_\mathrm{prog} [M_{\odot}])-5.6(\pm 0.4)$ & \\ \hline

Relations from \cite{Cai-2021biy} & \\ 
$M_V =-5.56(\pm 0.69) \times \log_{10}(M_\mathrm{prog} [M_{\odot}])-4.91(\pm 0.26)$  &\\\hline

 Relations* for data in Tables \ref{tab-main} and \ref{tab-progenitors} & \\ 
$M_X =-4.85(\pm2.06) \times \log_{10} (t_\mathrm{plateau}[d]) - 1.67(\pm4.22)$ & $\chi^2_r$=0.20\\
$M_X =-5.4(\pm0.6) \times \log_{10}(M_\mathrm{prog} [M_{\odot}])-5.9(\pm 0.7)$ &$\chi^2_r$=0.23\\
$M_X =-0.0065(\pm0.0040) \times \mathrm{FWHM}[\mathrm{km\,s}^{-1}] - 8.75(\pm1.98)$ & $\chi^2_r$=10.0\\
%
$\log_{10} (t_\mathrm{plateau}[d])=0.63(\pm0.16) \times \log_{10}(M_\mathrm{prog} ([M_{\odot}]) +1.32(\pm 0.18)$ &  $\chi^2_r$=0.04 \\
\botrule
\end{tabular}
\footnotetext{*Error-weighted fits to data using the nonlinear least-squares Marquardt-Levenberg algorithm and the corresponding reduced chi-square values ($\chi^2_r$). $M_X$=$M_r$, $M_R$, or $M_V$ as in Table \ref{tab-main}.}
\end{table}

Early studies proposed different parameterizations to uncover scaling relations in red novae. Based on a handful of events known at that time, \cite{Kulkarni2007Natur} proposed a relation between the peak magnitude and the duration of the outburst on a log scale. Figure \ref{fig:scaling-relations} shows that for the updated sample of red novae, these two properties are somewhat related (although with some scatter), especially for the members of the `plateau' group; the `risers' generally occupy the high-luminosity and high-duration region of the diagram. This correlation was also shown in the theoretical work of \cite{IvanovaScience}, which interpreted red novae as recombination-powered transients. In their model, the plateau luminosities and durations of the light curves are linked to the amount of unbound mass ejected during the merger, its kinetic energy, and the initial launch radius of the outflow. This framework naturally accounts for the observed scatter: for progenitor stars of equal mass, the plateau luminosity is expected to be broadly similar, while the plateau duration can increase up to an order of magnitude for more evolved systems. 

In a complementary approach, \cite{Kashi2010arXiv} and \cite{Soker2011arXiv} considered that all intermediate luminosity transients are powered by mass accretion from a companion onto an MS star. These authors introduced an Energy Time Diagram, which displays the total energy of the transients (radiated plus kinetic) as a function of the duration of the eruption, defined as a 3-mag decline in the $V$-band. In this diagram, intermediate‑luminosity transients occupy a diagonal band, revealing a nearly linear correlation between total energy and outburst duration. The lowest energy events in this sample correspond to the planet–brown dwarf merger scenario \citep{Bear2011MNRAS, Metzger2012MNRAS}, with characteristic timescales of only a few days. The central part of the diagram is dominated by red novae, with durations between a month and a year, and the highest energy events correspond to the long-lasting LBV eruptions analogous to $\eta$ Car's Great Eruption. This representation links red novae to the broader population of intermediate-luminosity transients.

The detection and characterization of the progenitor stars has also made it possible to relate the properties of the quiescent system with the resulting outburst. The early work of \cite{kochanek}, based on four Galactic red novae with progenitors known at that time, proposed the existence of a correlation between the progenitor's mass and peak luminosity of the transient as $L \propto M^{2.9\pm0.6}$ (see Table \ref{tab-relations}). More massive progenitors, such as that of V838~Mon, generated brighter events than low-mass progenitor systems like V4332 Sgr or V1309 Sco. This correlation, depicted in Fig.~\ref{fig:scaling-relations}, has now been expanded to the more luminous and massive extragalactic red novae \citep[e.g.][]{Blagorodnova2021,Cai-2021biy} and the mass-luminosity relation was further revised to $L \propto M^{2.22 \pm 0.28}$ (see also Table \ref{tab-relations}). \cite{Pastorello2023} has also shown that the progenitor magnitude (proxy for mass) not only correlates with the luminosity of the main blue peak, but also the magnitude of the second red peak, and the magnitude of the precursor. Notably, their sample shows that among events with durations shorter than 100 days (mostly members of the `plateau group' defined above), the duration is shorter when the first blue peak is substantially more luminous than the red peak. In contrast, for events lasting longer than 100\,days (the `risers group'), no clear correlation is observed between the blue‑to‑red peak luminosity ratio and the event duration.

The theoretical model of \cite{Pejcha2016b} suggested that the luminosity of red transients is proportional to the asymptotic outflow velocity measured for these events, as L2 mass loss and inter-shocks would be the main power mechanism for red nova emission. 
Although their study included objects since then removed from the red nova category (e.g., SN2008S or NGC-300OT),
Fig.~\ref{fig:scaling-relations} shows that this correlation holds, albeit with a large scatter, which increases for events brighter than $-$10 in absolute magnitude. It is important to note, however, that velocities are hard to measure, and a significant fraction of the observed scatter likely reflects methodological differences in how and at what phase the velocities are measured. 

While velocity measurements inferred from narrow absorption lines may be a more precise way to derive the ejecta speed, the lack of medium- and high-resolution data for the observational sample complicates this approach. Consequently, the width of the usually broader and stronger \halpha line has been commonly used as a proxy for the bulk ejecta velocity. The drawback is that this line has been shown to exhibit electron scattering wings around the main peak, and often displays multiple and evolving absorption components that are challenging to model. The overall FWHM has also been shown to vary with time \citep[see Fig. 13 from][]{Pastorello2023}, with the highest values reported during the first optical peak, so the lack of a standardized reference epoch to report FWHM measurements further complicates comparisons across different studies. In some cases, spectra could only be obtained after peak luminosity, or were taken at insufficient spectral resolution, introducing additional systematic uncertainty into the inferred velocities.

Interestingly, \citet{Pastorello2023} showed that the \halpha line luminosity measured at +7 days post-peak exhibits significantly less scatter than the corresponding H$\alpha$ FWHM values reported at the same epoch, and correlates more tightly with the peak bolometric luminosity. This suggests that while the total energy radiated in the line is relatively stable, variations in ejecta geometry and line-forming conditions can introduce substantial additional scatter in FWHM-based velocity estimates.

\subsection{Event statistics}\label{sect-stats}
Constraining the total number of red nova events is a very difficult task. Estimates based on the observed rates in the Milky Way and nearby galaxies face the problem of completeness and selection bias, whereas population synthesis models must include the largely unconstrained physics of mass and angular-momentum transfer at different evolutionary phases of the binary, including the still poorly understood common envelope evolution. Nevertheless, notable attempts to calculate red nova rates have been made. 

From an observational perspective, \cite{SokerTylenda2006} estimated that bright transient merger events (`mergebursts') occur in a Milky Way–type galaxy once every 10–-50\,yr. This estimate was derived by scaling the formation rate of blue stragglers inferred by \cite{Ciardullo2005} for old stellar populations, and then adjusting this rate for a much narrower and lower-mass range of secondaries expected for stellar mergers. The prediction applies to the most violent and luminous merger events, such as V838 Mon–like eruptions, and does not address the occurrence rate of fainter, lower-energy red nova events. Their rate is in agreement with the observational lower limit of  0.019\,yr$^{-1}$ for a Milky Way type galaxy \citep{Ofek2008ApJ}

\citet{kochanek} performed a more systematic study of red nova rates. In their Bayesian analysis anchored in the detection of three Galactic events (V4332 Sgr, V838 Mon, and V1309 Sco), they found that the number of red novae scales with luminosity as $dN/dL \propto L^{-1.4\pm0.3}$. They expect 1.3 events per 10\,yr for the transients of the luminosity of V1309 Sco or brighter, and expect only one event of the magnitude of V838~Mon every 30\,yr in a galaxy like ours. 
These estimates were in good agreement with rates in the population synthesis results of \citet{Belczynski}. Kochanek et al. noted that the red nova population should be dominated by MS stars merging with evolved stars, with only a small fraction (20\%) being  MS-MS pairs. The study also pointed to a strong relation between the peak luminosity and the mass of the progenitor for the known events of $L\propto M^{2.5\pm0.5}$ (see Sect.~\ref{sect-relations} and Table \ref{tab-relations}). 

Utilizing the binary population synthesis, \citet{Howitt} and \citet{Twum2026arXivRates} simulated a large population of binaries and estimated the light curve properties for those deemed to be unstable with recombination emission models \citep{IvanovaScience,MacLeod2017}. As a novelty, \citet{Howitt} parametrized different assumptions about the kinetic energy of the unbound mass released during the merger, which represented an additional energy sink to fully unbind the envelope. Their approach predicted an intrinsic Galactic event rate of 0.1\,yr$^{-1}$, and a volumetric rate of 8$\times 10^{-4}$ Mpc$^{-3}$ yr$^{-1}$. \citet{Twum2026arXivRates} replaced the commonly used tabulated core–envelope binding energy parameter $\lambda$ with detailed stellar structure profiles computed using MESA for single-star donors, enabling a more consistent estimate of the ejecta mass across a range of binary configurations. Although no specific rate was provided by this latter study, both works predicted a bi-modal distribution of plateau luminosities in their populations, corresponding to fainter merger events and brighter full common-envelope ejections (leaving a compact surviving binary). In practice, detections from magnitude-limited surveys such as Zwicky Transient Facility (ZTF) and LSST are expected to be dominated by the brighter class.

Whereas most earlier studies derived red nova rates primarily from Milky Way events, \citet{Karambelkar2023ApJ} instead based their estimates on transient detections in nearby galaxies ($<150$ Mpc) from the Census of the Local Universe (CLU) survey conducted with the ZTF. After applying completeness corrections to their identification of six (plus five tentative) red novae, they derived a volumetric rate of red novae of $\approx 8 \times 10^{-5}$ Mpc$^{-3}$ yr$^{-1}$ for absolute $r$-band magnitudes between --16 and --11 (i.e., comparable in brightness to V838~Mon or brighter). Since their experiment was sensitive to brighter events than those known from the Milky Way, they were able to demonstrate that the rate is a steep function of luminosity, $\propto L^{-2.5}$, for absolute magnitudes $\lesssim$--10. It is therefore steeper than proposed by \cite{kochanek} for the Milky Way events (broadly reproducing the initial mass function), but is shallower than the binary population synthesis of \citet{Howitt}. The broken power law for the rates as a function of luminosity is interesting as it may manifest different outburst mechanisms. In line with previous theoretical work, \citet{Karambelkar2023ApJ} speculated that the lower-luminosity events are caused by coalescing stars, while their high-luminosity analogs are a result of non-terminal common-envelope ejections in massive binaries. 

The estimates of \citet{Howitt} and \citet{Karambelkar2023ApJ} of the observational rates of the most luminous (thus most massive) red novae seem to be comparable to compact mergers that are potentially detectable through contemporary gravitational-wave observatories. This notion may initially imply that some of the most luminous transients are indeed related to progenitors of the merging black hole and neutron star systems \citep[see also][]{Twum2026arXivRates}. Their common-envelope phase is hypothesized to be necessary to shrink the orbits of these systems before they spiral into a merger. Nevertheless, the binary population synthesis-based analysis of \cite{Jain2025LRN_GW} found that the fraction of red novae associated with an observed gravitational-wave signal (from a binary neutron star or a neutron star-black hole) would be as small as $\sim 10^{-3}$, implying that the vast majority of transients will lead to mergers, and not to compact systems.

The predicted rates are currently (as of 2026) challenged by the lack of new identifications of a Milky Way red nova since 2008 (although three red novae were already identified in our neighbor M31 from 2015 to 2026). We should have seen at least one Galactic event since the outburst of V1309 Sco. Perhaps our estimates were overly optimistic. This discrepancy may reflect statistical fluctuations, poor knowledge about the patchy Galactic extinction, or some sort of selection bias.
For example, the mentioned studies of red nova rates predict a higher frequency for low-luminosity events that are supposedly missed in the contemporary data. Note that most of the studies were performed before the detection of the planet-induced (subluminous) red nova of \citet{DeNature}.

Because red nova rates are so strongly coupled to the crucial aspects of binary star evolution, there is a hope that future sensitive sky surveys will be able to provide more robust rates in the Milky Way and the Local Group. For instance, it is expected that the optical photometric survey of LSST-Rubin \citep{LSSTroadmap} will be able to observe 20--1500 red novae per year \citep{Howitt,Karambelkar2023ApJ}. Assuming the duration of the LSST-Rubin project of 10 yr, one could expect a significant increase in the number of red novae in the Milky Way and nearby galaxies, but it remains uncertain how many will be robustly identified and followed up spectroscopically. There are also expectations that, in addition to merger and common-envelope transients, LSST-Rubin can provide another example of a close binary spiraling into a merger like V1309 Sco \citep{LSSTroadmap}.

Most of the rate estimates performed so far have focused on optical detections of these transients. Since some red novae are expected to be dust-obscured even before the main outburst, some events will be intrinsically infrared \citep{Jencson2019,MacLeod2022}. Comprehensive IR sky surveys are highly desired to establish the census of merging non-compact stars, a quest that should be feasible for the Milky Way population. Such surveys covering a substantial part of the sky in the IR have already started to operate \citep[e.g.,][]{GattiniDe2020PASP,WINTER,Sumi2025AJ_PRIME} or are expected to be carried out by upcoming space observatories (see Sect. \ref{sect-future}).

\section{Remnants}\label{section-remnants}

\subsection{The star}\label{sect-remnants-star}
\begin{figure}[ht]\centering
    \includegraphics[scale=0.93]{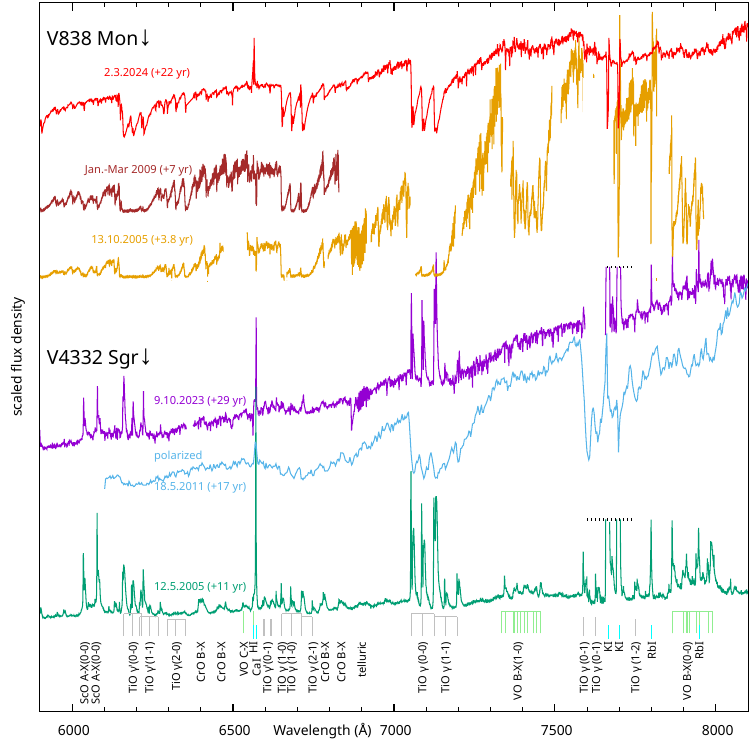}
    \caption{Late evolution of red nova spectra. {\bf The top three spectra} are of V838~Mon which erupted in 2002 and represents a high-/intermediete-mass red nova remnant \cite[adapted from][Kamiński, in prep.]{KamiKeck,Tylenda-V4332UVES}. Spectra a few years after the eruption show very deep molecular absorption bands, often saturated, which become shallower with time (e.g., TiO $\gamma\prime$ 1--0 near 6700 \AA). Emission lines have been relatively sparse in V838~Mon. Uncontaminated photospheric lines appeared only recently. {\bf The bottom three spectra} are of V4332 Sgr, which erupted in 1994 and is used here as a template for spectral evolution of low-mass remnants \citep[spectra from][Kamiński, in prep.]{Tylenda-V4332UVES,Kami-specpol}. The 2005 and 2023 spectra are dominated by circumstellar emission in neutral atoms and molecules, and only recently have a weak photospheric component. Strong emission in the \io{K}{I} doublet and severe telluric features were blanked for clarity. The 2011 spectrum shows polarized flux, representing the much purer photospheric spectrum of spectral type $\sim$M5, similar to that of V838 Mon. Main spectral features are labeled at the bottom of the plot (cyan markers for atomic features, green for VO, and grey for TiO).}
    \label{fig-remnantsSpec}
\end{figure}

While the circumstellar remnants have been observed extensively, photospheric spectra of only two central stars--V838~Mon and V4332 Sgr--have been obtained so far. Their most recent evolution is shown in Fig.~\ref{fig-remnantsSpec}. The photospheric spectrum of V838~Mon, even though it was significantly contaminated by circumstellar features, became recognizable shortly after the eruption and has been studied extensively at optical and NIR wavelengths. In the case of V4332 Sgr, whose central star is strongly blocked by a dusty disk, the spectrum is visible due to scattering on dust above and below the plane of the disk, and is most clearly seen in polarized light \citep[Fig.~\ref{fig-remnantsSpec};][]{Kami-specpol}. The \io{Ca}{ii} triplet in absorption is likely the first strong photospheric feature to reappear after the eruption. At later times, the spectra of the stellar remnants show cool, $T_\mathrm{eff}$=3200--3700 K, photospheres (spectral types M3--M5) of low gravity. These objects are therefore close to the Hayashi limit. While temperatures are relatively easy to determine, owing to the presence of molecular bands of TiO and VO, the spectra acquired so far have been of insufficient quality for a direct measurement of $\log g$. With well-constrained temperature and distance to V838~Mon, the radius of the star is estimated at 464 R$_{\sun}$ \citep{KamiALMA}. Recent observations show that, at least in the case of V838~Mon and V4332 Sgr, the obscuring material clears out from the line of sight, and deep high-resolution observations should be possible in the near future, offering the prospect of improved characterization of both photospheres. 

SED modeling of the remnants also provides constraints on the effective temperatures of the stars, but is generally much less accurate than direct spectral typing (as it is degenerate with respect to reddening and assumed dust properties). A few years after the outburst, central remnants are found at temperatures from 2200 K \citep[e.g., AT2021blu and AT2021biy;][]{KarambelkarJWST} to 4000 K \citep{SteinmetzBLG,ReguittiIR}. Merger remnants are typically M-type stars. 


Because of their very extended envelopes, freshly coalesced stars are expected to have convective atmospheres typical of ordinary red (super-)giants. Convective cells on the surface can explain their erratic post-outburst photometric variability \citep[e.g.,][]{Liimets} and relatively large macroturbulence velocities (priv. comm.). \citet{KamiLit} advocated that a strong signature of lithium in the spectrum of V838~Mon is a consequence of strong convective mixing in the envelope of this star, but it remains unclear whether this is a common feature among the red nova remnants.  

It is worth noting that the products of mergers in the extremely massive red novae, with progenitor masses above, say, 50 \Msun, may be much hotter stars. In the case of SN Hunt 248 with a yellow hypergiant progenitor of 60 \Msun, the remnant is thought to be of spectral type B \citep{Mauerhan2018MNRAS}. The identification of the object as a red nova is, however, still questionable.


\paragraph{Winds} Having gigantic sizes and cool photospheres, the stellar remnants of red novae should be prone to developing dust-driven winds similar to those of red supergiants and of AGB stars. Because the higher-mass remnants (like V838~Mon and many extragalactic red novae) remain very luminous even after the eruption has ended, there is a possibility that additional momentum is absorbed by gas through broad molecular absorption bands. A long-lasting wind is certainly observed in V838~Mon through optical absorption features \citep{KamiLit} and was spatially resolved at millimeter wavelengths \citep{KamiALMA}. The wind is clumpy and has the mass-loss rate of roughly 10$^{-4}$ M$_{\sun}$ yr$^{-1}$ \citep{TylendaEngulf}, rivaling the most extreme cases of mass loss in red supergiants. The similarity of the V838~Mon wind to that of extreme red supergiants is striking, particularly in that the wind powers asymmetric SiO masers at 43 and 86 GHz \citep{Deguchi2005,Ortiz-Leon2020}. Stellar winds may be one way the coalesced star gets rid of excess angular momentum and may significantly shorten the relaxation phase.

\paragraph{Spin and magnetic fields} Unfortunately, no magnetic field or rotation measurements are available for red novae or their remnants. The obscuration by dust and veiling of the spectra by circumstellar gas makes it difficult to measure these properties using well-known techniques, which are difficult to apply even to standard field giants \citep[cf.][]{Ma2024Beteleguse}. Note that rotation and surface B fields are challenging to measure once the star is expanded to hundreds of solar radii and has strong convective motions at the surface. A general expectation and theoretical models seem to agree that young merger products should be rapid rotators and have enhanced magnetic fields for a wide range of masses of the progenitor components and a wide range of collision parameters \citep[][see also Sect. \ref{sec-remnant-analogs}]{Wrotation,Frost,ScheniderNature2019}. 
Rotation and magnetic fields are also expected to strongly influence the evolution of the remnant after the thermal relaxation phase \citep{Sills2001}.  

\paragraph{Pulsations} Immediate merger products are expected to be prone to pulsation instabilities. V838~Mon, the brightest of the known remnants, shows photometric variability at time scales comparable to its dynamical timescale \citep[][Kamiński in prep.]{Liimets}, but the period does not seem to be stable and cannot be taken as a signature of pulsations. Presently, no solid evidence of pulsations in red nova remnants exists. Some theoretical work has been done to analyze pulsations in fully relaxed merger products \citep{RuiFuller,HennecoPulsationsI,HennecoPulsationsII}, but not during the relaxation phase. There is a common belief that the merger event should trigger non-radial pulsations in the coalesced remnant that can distinguish a merger product from stars after mass transfer or accretion \citep{WeltyRamsey1994}. 

\paragraph{Post-outburst relaxation phase of the star} 
In the merger interpretation of red novae, the extended cool remnants are a natural consequence of the fast accretion and energy deposition within the primary \citep[cf.][]{Tylenda2005,BlueRing,Lipunov}. The accumulated gravitational energy is an extra luminosity source that makes the stars brighter than their progenitors. The relaxation to a stable configuration is expected to proceed on the thermal timescale, but detailed calculations for individual cases are missing. However, the post-outburst evolution of red nova stellar products on the currently accessible timescales of decades to centuries is expected to be slow. Indeed, while dynamical changes are observed in the circumstellar environments of red nova remnants, no drastic changes are observed in the stars \citep[e.g.,][]{MobeenII,Geballe2025}. Only slow, gradual increases in flux are observed in V4332 Sgr and V838~Mon \citep[][Kamiński in prep.]{Tylenda-V4332UVES,Liimets}. Occasional sudden dips in the light curve have been reported, and are thought to arise from circumstellar material moving across the line of sight, infalling gas, or possibly shock interactions with the surrounding medium \citep[][see also Sect.~\ref{sect-geometryCSM}]{Tylenda-V4332UVES,LauSLRN,Cai-2021biy,GoranskijATelV4332, Goranskij2020}. 

While the stellar photosphere appears to be changing slowly, IR observations of the continuum source often show a steady expansion at velocities of 100--300 \kms. This is most likely due to the expanding dusty ejecta, which remains optically thick at IR wavelengths even over a decade past the eruption \citep{SteinmetzBLG, KarambelkarJWST}, reaching radii of the order of $10^{16}$ cm (1000 au). With time, however, this dusty photosphere also becomes transparent.

Eventually \citep[e.g., after the thermal timescale of $\sim$500 yr in the case of V838~Mon, assuming its current parameters in][]{KamiALMA}, the coalesced star should achieve a new stable configuration that appears younger than the progenitor primary. For instance, in the context of the cluster surrounding V838~Mon and given its most likely mass of 8 \Msun, the object should eventually become a blue straggler. V1309 Sco, in turn, is expected to evolve into a red giant \citep{Stepien}, rather than a blue straggler. Earlier suggestions of an evolution towards a blue-straggler were based on a misinterpretation of the available data (see the discussion in \citealt{Steinmetz}). An interesting analog of red nova descendants on a time scale of thousands of years may be the suspected merger products, the Blue Ring Nebula \citep{BlueRing} and HP Tau \citep{Reipurth}. This topic is discussed further in Sect.~\ref{sec-remnant-analogs}. 

\subsection{Geometry of the circumstellar medium}\label{sect-geometryCSM}
The circumstellar structure of red nova remnants is highly complex. Observations and merger simulations show material that was dispersed before the merger, during the merger burst, and, in some cases, after the collision during the `remnant' phase (e.g., winds from the coalesced star or fallback material). These components are thought to interact, often giving rise to shocks. The circumstellar components can take the form of disks (tori or spiral arms), structured bipolar outflows (including jets), and spherical or irregular ejecta. For most known red nova remnants, these components remain spatially unresolved. Imaging observations in which these structures are partially resolved are limited to 2--5 nearby red novae. Most of what we know about the remnants' architecture thus comes from unresolved observations. These studies demonstrate that there is no single blueprint that explains observations of all known remnants, but some commonalities can be identified. Below, we present two predominant types of architectures discussed in the literature and based on observations of the remnants decades after the red nova event. Merger simulations, which we do not cover here, explore an even wider range of cases but are reviewed elsewhere \citep[e.g.,][]{Schneider}.    

\subsubsection{Low-mass remnants}\label{sect-geometryCSM-lowmass}
\begin{figure}[ht]
\begin{center}
 \includegraphics[trim=2 10 12 10,clip=true,width=0.85\textwidth]{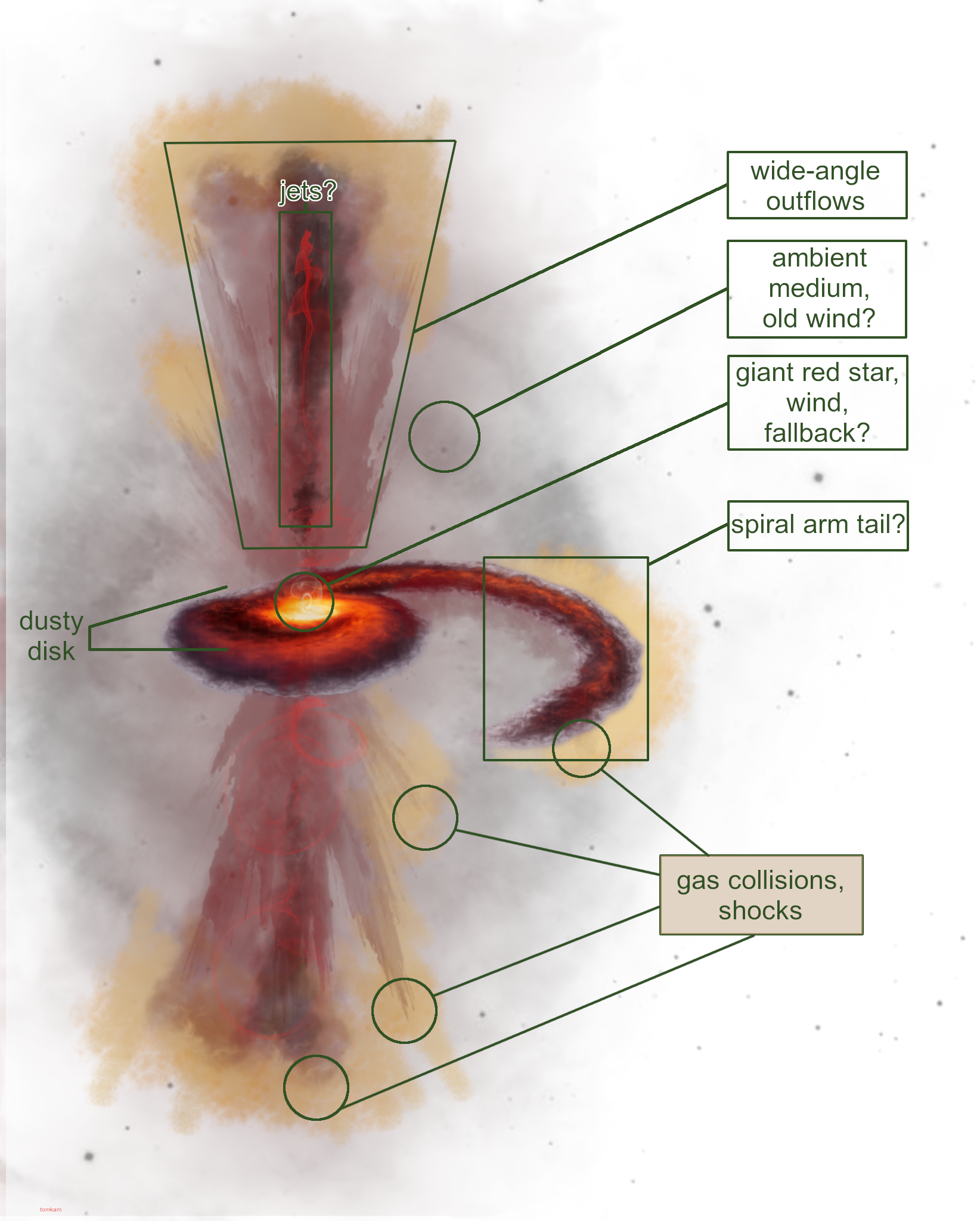} 
\end{center}
\caption{Schematic representation of a low-mass red nova remnant with a disk. The depicted features are not to scale. The scheme reproduces observations of V1309 Sco and V4332 Sgr.}
\label{fig-remnantStructure-lowmass} 
\end{figure}

The first type of remnant structure we describe is based on observations of the Galactic red nova remnants of V1309 Sco and V4332 Sgr (and to some degree on the CK Vul case). They are both thought to be products of mergers of solar mass stars; V1309 Sco was certainly a contact binary of two subgiants whose orbit was orthogonal to the sky plane \citep{Stepien,Pejcha2017}. These properties are likely to determine the final remnant structure. Our schematic representation of this type of remnant is shown in Fig.\,\ref{fig-remnantStructure-lowmass}. Since distances to these objects are poorly constrained, the spatial scales (and masses) we provide below should be regarded as very rough first estimates.

\paragraph{Disk} The bloated, coalesced star (Sect.~\ref{sect-remnants-star}) is surrounded by a dusty disk with dust temperatures of 200--950 K \citep{Kami-V4332-disk}; the temperatures might reach the dust sublimation temperature ($\approx$1700 K) at the inner rim \citep{Banerjee2015}. This range of temperatures is typical for protostellar disks with radial extents of roughly 7--300 au, assuming a central source temperature of 2000 K and stellar radius of 300 \Rsun. The presence of the dusty disk in red nova remnants is evidenced by detailed analysis of the SEDs of the three mentioned remnants and by other more subtle arguments \citep[e.g.,][]{Kami-V4332-disk,Kami-specpol}. In all three objects, the disk is observed nearly edge-on. Presently, it is uncertain if these are the same disks that were formed during the precursor phase---i.e., a few months to years before the mergeburst, as observed in the light curve V1309 Sco \citep[][see E in Fig.~\ref{fig-v1309lc}]{Tylenda2011,Pejcha2017}---or whether they formed even earlier. 

The dusty disks in red novae must be relatively dense, as that of V4332 Sgr shows deep silicate absorption features and a water-ice band near 3 $\mu$m \citep[e.g.,][]{nicholls,BanerjeeIce}. The icy mantles normally form only in dense environments on timescales of Myr. The disk (or parts of it) may therefore largely predate the merger and inspiral phases. \citet{Pejcha2017} calculated that the total mass (dust and gas) of the disk formed just before the mergeburst of V1309 Sco, a feature dubbed `the death spiral outflow', was 0.05 \Msun\ \citep[see also][]{ZhuDustform}. Based on data presented for V4332 Sgr in \citep{Kami-V4332-disk}, our estimates suggest a similar disk total mass of 0.06 \Msun. For the nearly edge-on viewing angles, the disks contribute strongly to polarized flux in the optical continuum (Sect.~\ref{sect-specpol}). 

While theoretical studies suggest the formation of spiral-like extensions or disk tails during the merger process (cf. Fig.~\ref{fig-remnantStructure-lowmass}; \citealt{NandezV1309,MacLeod2018a,MacLeod2018b}), there is currently limited observational evidence for such structures persisting in merger remnants decades after the event \citep[but see][]{MasonShore}. Numerous merger or common-envelope simulations predict the formation of a stable (i.e., bound) disk where most of the angular momentum of the former binary system is stored \citep[e.g.,][]{NandezV1309,MacLeodLoeb-preCE}. Unfortunately, no measurements of angular momentum have been possible in the observed disks. Some authors also suggested that a small portion of the dispersed but bound matter may be falling back on the central star long after the merger \citep[e.g.,][]{LauSLRN}, but solid observational support of this notion is still missing for stellar merger remnants. On a longer timescale, it was proposed that post-merger disks can eventually form Jupiter-mass planets \citep{MartinPlanets}.

\paragraph{Bipolar outflows}
The disk is associated with a pair of conical outflows expanding at speeds of 200--300 \kms. A spectropolarimetry experiment demonstrated that these lobes are composed of both dusty and cool, mostly neutral, gas \citep{Kami-specpol}. The dust grains scatter the stellar light from high above the disk plane at scattering angles close to 90$^{\circ}$, introducing strong polarization. The gas component is also seen in fluorescent emission from these regions owing to efficient scattering and includes strong optical resonance doublets of \io{K}{i} and \io{Na}{i}, and molecular bands such as TiO and VO \citep{Kami-specpol,Tylenda-V4332UVES}. The molecular bands suggest gas temperatures from 100 K to a few hundred K. Even cooler regions of the bipolar lobes (at 30--100 K) are seen in molecular emission at (sub-)millimeter wavelengths and have been partially resolved in V4332 Sgr by ALMA \citep{KamiALMA}. The ALMA interferometric observations revealed a bipolar outflow with maximal deprojected velocities of about 300 \kms. The velocity field was well reproduced, with gas speeds being proportional to the radial distance from the star. The radial extent of the lobes was about 600 au in 2016, consistent with the material being ejected in the 1994 eruption of V4332 Sgr. A similar circumstellar component, albeit spatially unresolved, is observed in V1309 Sco at optical-IR to mm wavelengths \citep{Steinmetz}. 

Most of the observed characteristics of the outflows are consistent with the scenario proposed by \citet{Pejcha2017} and \citet{MacLeod2018a}, whereby the expanding merger ejecta, originally in the form of a spherical or equatorial outflow, is deflected by a pre-existing spiral disk (or torus) into two opposing conical structures. On the one hand, future observations of the bipolar lobes should be easier as the material expands, but on the other hand, it will be more challenging to detect the outflows because the cooling material gets dimmer in emission lines. 

No highly collimated jets have been reported for the discussed remnant type \citep[except for CK Vul, possibly due to a companion, see][]{KamiCKalma2}, likely due to the limited spatial resolution of current observations. In spatially unresolved observations of V1309 Sco, shock tracers such as H$_2$ NIR emission or mm-wave emission of HCO$^+$ occur only in the blueshifted and redshifted components, respectively,  suggesting that the bipolar lobes are not identical. This is consistent with many merger and common-envelope simulations that highlight the rise of inhomogeneities in the dispersed matter \citep[e.g.,][]{NandezV1309}, including asymmetries in the bipolar component \citep[e.g.,][]{AFrank2018}.

\paragraph{Ambient medium} The mentioned shocks can also be interpreted as a signpost of an ambient medium into which the deflected merger ejecta expands. It is very likely that the medium was formed by relic mass loss from the progenitor occurring hundreds of years earlier, possibly by tenuous winds or some sort of precursor mass loss preceding the `death spiral outflow'. Some post-outburst observations of forbidden lines arising from these late interface shocks indicate densities of $10^{4\pm1}$ cm$^{-3}$ \citep{Steinmetz}, but much higher values have also been reported \citep[$10^8$ cm$^{-3}$;][]{MasonShore}. 

The ambient circumbinary medium may also be apparent during the outburst. Narrow absorption lines seen in Balmer H$\alpha$ and the resonance \io{Na}{i} doublet at fixed velocities may be a signature of this cool gas along the line of sight and expanding at terminal velocities that are lower than those characterizing the mergeburst ejecta \citep[e.g., 60 \kms\ in V1309 Sco,][see also Sect.~\ref{sect-spectral}]{MasonShore,McCollum}. This ambient medium is first affected only by the red nova radiation, triggering feedback fluorescent emission. It is later excited in the red-nova remnant by collisions with the ejecta. In V1309 Sco this ambient medium extends out to $<$3000\,au and may have an ellipsoidal distribution \citep{MasonShore}.

Observations indicate that there is an even more extended and thus older circumbinary component in the low-mass remnants. In studies which consider photometric fluxes extending to FIR and submm wavelengths, a cool ($\approx$30 K) dust component is identified at radii of a few thousand au \citep{TylendaSED,SteinmetzBLG}. These spatial scales are comparable to the Oort cloud. It remains unexplored how red-nova outbursts affect these systems' frontiers, but dust heating and even photo-desorption of ices should be considered \citep{KamiPhD}. 

\subsubsection{High-mass remnants}
\begin{figure}[ht]
\begin{center}
 \includegraphics[trim=0 0 0 00,clip=true,width=0.6\textwidth]{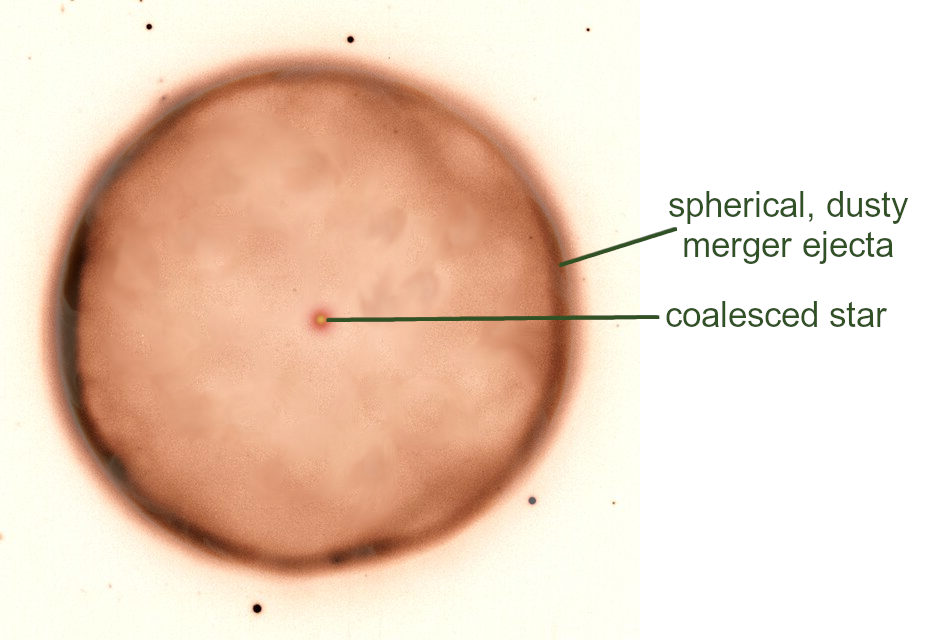}
  \includegraphics[trim=0 61 0 0,clip=true,width=0.6\textwidth]{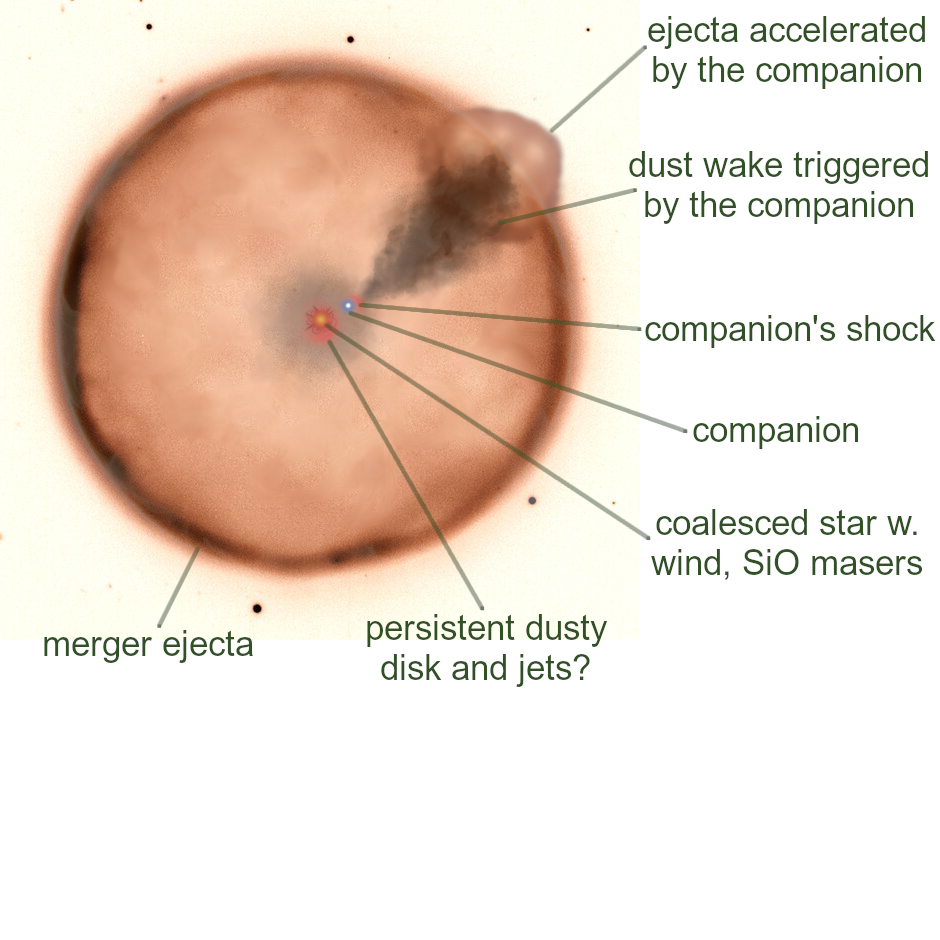}
\end{center}
\caption{Schematic representations of `high-mass red nova remnants`. The top panel shows a remnant with no companion, while the bottom presents a case with a companion, as in V838~Mon. The depicted features are not to scale.}
\label{fig-remnantStructure-highmass} 
\end{figure}

The extragalactic high-mass red novae, with predominantly YSG progenitors, are difficult to study after the outburst, and only a few objects have been observed years after the end of the eruption. The limited data, usually in the form of SEDs, are interpreted within the assumption that the remnant is dominated by spherically symmetric dusty ejecta \citep{KarambelkarJWST,ReguittiIR}. Our picture of these remnants is thus basic, as depicted in the top panel of Fig.~\ref{fig-remnantStructure-highmass}. The ejecta is assumed to be freely expanding with its terminal velocity, $v$, and with radius increasing as $vt$; it is thus reaching radii of a few hundred au in the first few years of the post-merger evolution and often remains optically thick at visual and IR wavelengths. Although late rebrightening of a remnant is interpreted as arising from circumstellar shocks \citep{Cai-2021biy}, it is unclear which components are involved. So far, no observations support the presence of disks or collimated outflows, such as jets, among the red novae with massive primaries.       

However, since the Galactic red nova V838~Mon has a progenitor with a mass of $\approx$8 \Msun\ and its luminosity ($10^6$ \Lsun) approached those of the extragalactic events, it may be treated as a nearby analog of these high-mass cases. Still, having a young (MS or pre-MS) progenitor, V838~Mon may be a special case among the analyzed red novae, but for illustrative purposes we will treat this well-studied object as a possible template for the extragalactic population. The remnant structure is schematically shown in the lower panel of Fig.~\ref{fig-remnantStructure-highmass}.  

The central object is a cool, low-gravity star whose luminosity exceeds that of the progenitor. This may result in the star developing a wind similar to that of red supergiants (cf. Sect.~\ref{sect-remnants-star}). The high luminosity and dusty wind are also prerequisites for activation of SiO masers. As the remnant evolves, the wind occupies an increasingly large volume, but its terminal velocity and mass-loss rates are much slower than the ejecta. For V838~Mon, the wind speed is about 50 \kms\ while the ejecta velocities were reaching 300 \kms\ \citep{KamiLit}; the current mass loss rate is at least three orders of magnitude lower than during the mergeburst \citep{TylendaSpec2009}.

In addition to the dusty wind, the coalesced star in V838~Mon is surrounded by a flattened, dusty structure, which is apparent in NIR-MIR interferometric data at radii of $\lesssim$10 au \citep{Mobeen23,MobeenII}. While this structure is persistent, its exact form and origin remain uncertain \citep{MobeenIII}. It is therefore unclear if we should expect its analogs in extragalactic remnants.

Interferometric imaging of the ejecta in V838~Mon by ALMA showed that the bulk of the gas lost during the merger has the form of a spherical shell \citep{KamiALMA}. It is only distorted by the presence of a companion, which interacted with the passing ejecta and the post-outburst wind. As argued in Sect.~\ref{sec-triples}, mergers in triples should not be uncommon among the extragalactic population of red novae. The influence of the companion is twofold: on the shape of the ejecta and its composition. Ejecta passing the companion can be accelerated within a narrow cone. Radiation from the companion can photoionize the passing matter. Additionally, a shock may form upstream of the companion, changing the molecular composition of the matter and triggering more efficient dust formation, similarly to the case in binary AGB stars \citep{Danilovich25}. This mechanism can significantly increase the dust production of red novae. All the effects are illustrated in Fig.~\ref{fig-remnantStructure-highmass} and described in more detail in \citet{KamiALMA}.

Since extragalactic objects dominate the known sample of red novae, it is important to verify if any of the features known from V838~Mon are common and confront them with  theoretical models of these more massive red novae.


\subsection{Composition of the circumstellar matter}
Most red nova remnants with noncompact progenitors do not show any compositional anomalies that would be common for the entire class, except, perhaps, for enhanced lithium abundances in many of them (see below). In the merger interpretation of red novae, the composition of the remnant and dispersed matter is directly linked to the composition of the progenitor binary components. Some simulations of off-axis (parabolic) collisions show that the material from the secondary is well mixed with that of the primary in the outer parts of the coalesed object \citep{TylendaSoker2006,Sills2001}. The remnant's surface composition may thus be a mixture of both stars.

\paragraph{Elemental composition} Except for CK Vul, which is a special case discussed below, no systematic anomalies in the elemental composition of red novae have been identified. Early claims of overabundances of $s$-process elements in the outburst spectra of V838~Mon should be taken with caution, as they were based on methods tailored to chemical analysis of quiescent stars \citep{Lynch2004,KipperConf}. Most of the remnants seem to have slightly subsolar metallicity, consistent with an old age (e.g., 8 Gyr for V1309 Sco) or a location in the outer Galaxy (V838~Mon is in the Outer Arm; \citealp{KipperConf}). A persistent presence of the $\lambda6707$ optical resonance line of lithium in the spectra of three red novae led \citet{KamiLit} to speculate that the remnants are Li enhanced due to the activation of the Cameron-Fowler mechanisms in the convective exteriors or due to other mixing mechanisms. It is uncertain if the same would happen for mergers involving stars more massive than those of the Galactic red nova. Also, the intrinsic origin of Li in red novae remnants is not the only possible explanation of the observed line \cite[see][for more discussion]{KamiLit}.  

\paragraph{Molecules} Although molecules play an important role in identifying and characterizing red novae shortly after outburst and over the subsequent decades, only a few studies have attempted to derive molecular abundances in the atmospheres \citep{Lynch2004,Lynch2007} and circumstellar media \citep{KamiSubmm,KamiALMA,KarambelkarJWST} of these objects. Molecules observed at optical, IR, and submillimeter wavelengths include primarily\footnote{Some identifications made in the literature, e.g., CaH in V838~Mon's spectrum \citep{Loebman} were incorrect; suggestions of NIR features of CN in extragalactic red novae are ungrounded.}: oxides CO, CO$_2$, AlO, CrO, VO, TiO, SO, SO$_2$, NO, SiO, ScO, YO, OH, H$_2$O, and AlOH; and sulfur-bearing species--SH, SiS, and H$_2$S--all commonly observed in oxygen-rich stars, old and young. While TiO and VO are the most recognizable features of late red nova spectra in the optical, CO and H$_2$O dominate at longer wavelengths and can be identified even in low-resolution, low-S/N spectra \citep{banerjee2002}.  Reliable estimates of circumstellar gas column densities remain very sparse. While local thermodynamic equilibrium is assumed to calculate partition functions, the molecules observed at IR wavelengths are always excited by the star's radiation and by dust layers at different radiation temperatures, making such estimates unreliable. Absorption features and submillimeter-wave spectroscopy yield dependable results. 

Currently, there is no quantitative evidence for anomalous molecular abundances, but researchers have noticed the prevalence of optical and IR bands of AlO, CrO, or PH$_3$, which are rarely observed (AlO) or have not been observed elsewhere \citep[CrO;][]{Banerjee2003,KamiMason,Steinmetz,LauSLRN}. These rare spectral features may be caused by special temporary excitation conditions, and not necessarily by the overabundance of the species or their parent metals. However, this notion requires a quantitative verification. Finally, it should be noted that some molecular species observed near V838~Mon, including AlOH, SO$_2$, and H$_2$S, were formed in shocks due to the interaction of the merger ejecta with the companion \citep{KamiALMA}. Thus, the observed molecular inventories likely reflect a combination of environmental factors, including interactions with the outer tertiary star. 

Spectra of extragalactic red novae usually have modest S/N, allowing only the most prominent absorption bands to be studied: TiO and VO in the optical \citep[e.g.,][]{Pastorello2021_2019zhd}, CO in the NIR, and H$_2$O in the NIR and MIR \citep[][Gomez-Muñoz in prep.]{Blagorodnova2021,Cai-2021biy, Karambelkar2023ApJ,Karambelkar2025ApJ}. 
Water bands are often so pronounced that they need to be considered in SED modeling of late remnants  \citep{KarambelkarJWST, Reguitti2026_AT2026abao}. 
 
\paragraph{Isotopes}The presence of molecular features in red nova spectra has made it possible to place the first observational constraints on the isotopic composition of the material in the circumstellar environment of the remnant. The best studied in this context is V838~Mon, whose IR bands of CO imply a $^{12}$C/$^{13}$C ratio of  10--100 \citep{Geballe} and submillimeter-wave emission yields $^{12}$C/$^{13}$C$>$80, all consistent, within uncertainties, with the interstellar medium value. Furthermore, \citet{KamiKeck} analyzed the main TiO isotopologues in the optical spectrum of V838~Mon, that is $^{46}$Ti$^{16}$O, $^{47}$Ti$^{16}$O, and $^{48}$Ti$^{16}$O, but found no deviation from the solar titanium isotopic ratios. The only other red nova remnant where isotopic measurement was attempted is V4332 Sgr. \cite{BanerjeeV4332lightcurve} found $^{12}$CO/$^{13}$CO$\approx$3 which may imply merger-triggered mixing. \cite{Banerjee26Al} put an upper limit of 10 on the $^{27}$Al/$^{26}$Al ratio from optical electronic bands of AlO, consistent with no significant enhancement with the radioactive isotope of $^{26}$Al. The case of CK Vul is discussed in a separate section below.

\paragraph{Solids} Dust spectral features have been studied mainly in the remnants, although there are a few observations of IR dust excess in progenitors \citep{TylendaBLG} and in outbursting red novae \citep{McCollum,Tylenda2011,TylendaBLG,Blagorodnova2020}. The dominant presence of oxides in the majority of the remnants is associated with the predominance of inorganic (noncarbonaceous) dust in their circumstellar media. 

Most red novae were observed to start forming dust days to months after the outburst \citep[e.g.,][Wavasseur in prep.]{Tylenda2005, TylendaSED, McCollum, WisniewskiPhoto, Blagorodnova2020}, giving rise to a broad and often structured 10\,$\mu$m feature. It is most often observed in absorption, but in some cases the emission component may dominate even a decade after the outburst (e.g., in V838~Mon; \citealp{Lynch2004}; AT2018bwo; \citealp{KarambelkarJWST}). The 10\,$\mu$m is identified as arising from a mixture of silicates (e.g., pyroxene and olivine, or the generic `astronomical silicates'\footnote{Although astronomical silicates have an admixture of carbonaceous dust and some red nova studies use soot or graphite opacities in SED modelling \citep[e.g.,][]{Blagorodnova2021,ReguittiIR}, there is no compelling evidence for organic dust in the red novae environments except for CK Vul.}) \citep{nicholls,woodward,SteinmetzBLG} but some studies emphasize a significant contribution from amorphous alumina dust, especially early (years) after the eruption \citep{Banerjee2015,woodward,SteinmetzBLG}. In a few cases, admixture of glassy olivines was necessary to reproduce SEDs of the remnants \citep{SteinmetzBLG,KarambelkarJWST}. FeO and MgO dust and other solid oxides or pure iron grains have also been suggested \citep{Banerjee2007,Banerjee2015,ZhuDustform}. 

The dust in red novae is thought to evolve on the timescales of decades, with signs of progressing annealing and an increasing level of oxidation \citep{Banerjee2015}. The diminishing contribution of the more refractory alumina being associated with the increase of silicates \citep{Banerjee2015,woodward} seems to be in line with the classical scenario of inhomogeneous dust nucleation in oxygen-rich circumstellar media, where silicate dust grows on alumina condensation cores \citep{GailSedmayr}. 

It is generally predicted that the mass of dust increases with the age of the red nova remnant, but there are so far only sparse observational constraints supporting this hypothesis \citep[][Wavasseur in prep.]{Blagorodnova2020,KarambelkarJWST}. Solid water ice was found only in V4332 Sgr \citep{BanerjeeIce}, but its origin is unknown, as there is no direct indication that it formed after the eruption, and it may be a trace of pre-existing material \citep[cf.][]{McCollum,TylendaSED}. Since solids play a great role in shaping the spectral energy distribution of red novae in their post-outburst evolution (and perhaps shortly before it), dust generally plays a major role in understanding red novae and mergers of noncompact stars. A vast amount of theoretical work has been invested in understanding dust formation in the context of non-compact mergers \citep[e.g.,][]{MacLeod2022} and in common-envelope ejection events \citep[e.g.,][Mu in prep.]{Iaconi2020,BermudezBustamente2024}, but the dust detailed composition, depending on the binary stars and the condensation sequence preceding the outburst remains largely unknown. 

The terminal dust mass yields were derived for a handful of remnants that are bright enough in the IR to be studied years to decades after the eruption. They range from $10^{-5}$ to $10^{-2}$ \Msun\ (see Col. 10 in Table \ref{tab-main}) \citep{KarambelkarJWST,ReguittiIR}. The one well-known planet engulfment event resulted in $10^{-11}$ \Msun\ of dust \citep{DeNature}, the lowest value ever reported for a red nova. The observations confirm the prediction that slow mergers, involving stars with extended atmospheres, produce much dustier remnants \citep{MacLeod2022,SteinmetzBLG}. Still, there are also hints that the red novae with more massive progenitors produce more net dust mass than the low-mass counterparts. 

\citet{KarambelkarJWST} analyzed the dust production budget of red nova remnants and proposed that, eventually, they contribute at least 25\%\ of dust mass estimated for core-collapse supernovae, which makes them important dust polluters on Galactic scales. This result accounts for the more frequent low-mass progenitors, which were assumed to be less efficient at dust production, and the more massive analogs, in which late interactions between ejecta and the circumstellar medium can considerably enhance the dust mass. Since our inventory of red novae likely misses the most dusty--and thus IR-only--transients, the role of red novae on dust production is probably still underestimated. We also note that the mass dust estimates should distinguish the ejecta component that would eventually disperse into the ISM from a gravitationally bound component that may be present around the remnant in a disk or torus and thus never reach the ISM.




\paragraph{Exceptional CK Vul} Most of the above characteristics are not valid for the case of the CK Vul remnant. Compared with the solar elemental composition, it is significantly enriched in He, Li, and N, but is depleted in H and O \citep{KamiSingle,TylendaXshooter,HajdukLit}. Moreover, the ratio of O/C may change across the remnant, with the closest observable vicinity of the stellar remnant being oxygen-rich (O$>$C) \citep{KamiCKalma1}. The molecular composition, explored mainly through (sub-)millimeter transitions, is also very extraordinary, as it is a combination of species typically only observed in oxygen-rich (e.g., SO and SO$_2$) and carbon-rich environments (e.g., CCH and CS) \citep{KamiNat,KamiSingle,KamiCKalma1}. In addition, many of the observed species are surprisingly complex, including CH${_3}$OH and CH${_3}$NH$_2$ \citep{KamiSingle}. Shock-induced infrared emission of the simplest molecule, H$_2$, is also observed \citep{TylendaXshooter,KamiCKalma1}. 

The current inventory of CK Vul's molecules includes 28 items, making it the richest red nova remnant in that respect. Many of these species appear in their isotopic variations, indicating highly non-solar isotopic ratios, including $^{12}$C/$^{13}$C$\approx$4, $^{14}$N/$^{15}$N$\approx$20, $^{16}$O/$^{18}$O$\approx$ 36, and $^{27}$Al/$^{26}$Al$\approx$7. The ratio of the aluminum isotopes was derived from the unprecedented observations of the rotational transition of the AlF molecule containing the radioactive isotope of $^{26}$Al \citep{KamiNatAstr}. Mid-IR observations of the remnant revealed many spectral features usually assigned to polycyclic aromatic hydrocarbons \citep[PAHs;][]{Evans2016}, which are signatures of UV-irradiated carbonaceous solids, making CK Vul the only remnant with such a clear display of organic dust. The overall chemical composition of CK Vul is explained by the hypothesis that this red nova (Nova 1670) was caused by a merger between a $\sim$2 M$_{\sun}$ red giant branch star and a He white dwarf (WD). The compact WD was able to plunge deep into the interior of the giant, disturbing its own He core surrounded by a layer rich in $^{26}$Al \citep{KamiNatAstr}. The merger also ignited partial He burning \citep{KamiSingle}. The observed remnant is a mixture of the nucleosynthesis products of CNO hydrogen burning from the progenitor giant, the partial He burning, and core material dispersed by the violent collision \citep{TylendaCKfinal}. The major difference between CK Vul and most other red novae is related to the nature of the progenitor binary and the short ignition of helium burning caused directly by the merger. This may be a common characteristic of mergers involving white dwarfs \citep[][and references therein]{Merwe1}.

\subsection{Related objects and post-relaxation analogs of remnants}\label{sec-remnant-analogs}

The red nova remnants will evolve into classes of objects long suspected of being merger products. The expectation is that V838~Mon will become a blue straggler; V1309 Sco, V4332 Sgr, and BLG360 are likely to evolve into red giants, or perhaps FK Com stars, possibly with extraordinarily high rotation and magnetic fields \citep[][]{Stepien,TylendaBLG}. Finally, CK Vul is believed to become an early-R type carbon giant \citep{TylendaCKfinal}. The immediate future of extragalactic red novae that have YSG progenitors remains largely unexplored, but after some thousands of years, some may become blue supergiant progenitors of core-collapse SNe, like SN\,1987A \citep{Podsiadlowski1992ApJ}, or progenitors of circumstellar interaction SNe \citep{Chevalier2012ApJ}.  

Understanding what kinds of stars are produced in red nova bursts is one of the main drivers of the field. Detailed predictions of the long-term evolution of the well-characterized red novae are still missing. Much more theoretical work has focused on the evolution after thermal relaxation, i.e., after the coalesced star returns to hydrostatic configuration, as already discussed in Sect.~\ref{sect-remnants-star}.  Observationally, we are limited to a small sample of Galactic remnants, most of which are only decades old. There is therefore a strong need to identify more, possibly older, remnants to expand the sample and extend the accessible timescales.

Identifying red novae through their outburst requires photometric and spectroscopic observations. As pointed out in Sect.~\ref{sect-history}, several historic objects were identified as red novae even though their eruptions took place long before the group of transients was defined (i.e., shortly after 2002). These include M31\,RV (eruption 1998), V4332 Sgr (Nova 1994), and CK Vul (Nova 1670). With these successful cases, it was believed that many other overlooked historic red novae could be found, even in cases when the exact coordinates of the transient were not well known \citep{Kimeswenger2007}. One expectation based on these characteristics of the historic transients is that they leave red, dusty, or otherwise peculiar remnants that can be distinguished from field stars. One interesting example of a search for an overlooked red nova is Nova 1943 \citep{Mayall1949}, a.k.a. V1148 Sgr. It was reported to exhibit a late-type spectrum with strong \io{Ca}{ii} absorption lines and possibly absorption bands of TiO, unusual for classical novae. Despite meticulous searching, its remnant has not been unambiguously identified \citep{BondV1148Sgr}. 

\begin{figure}[t]
\begin{center}
 \includegraphics[trim=0 10 0 0,clip=true,width=0.89\textwidth]{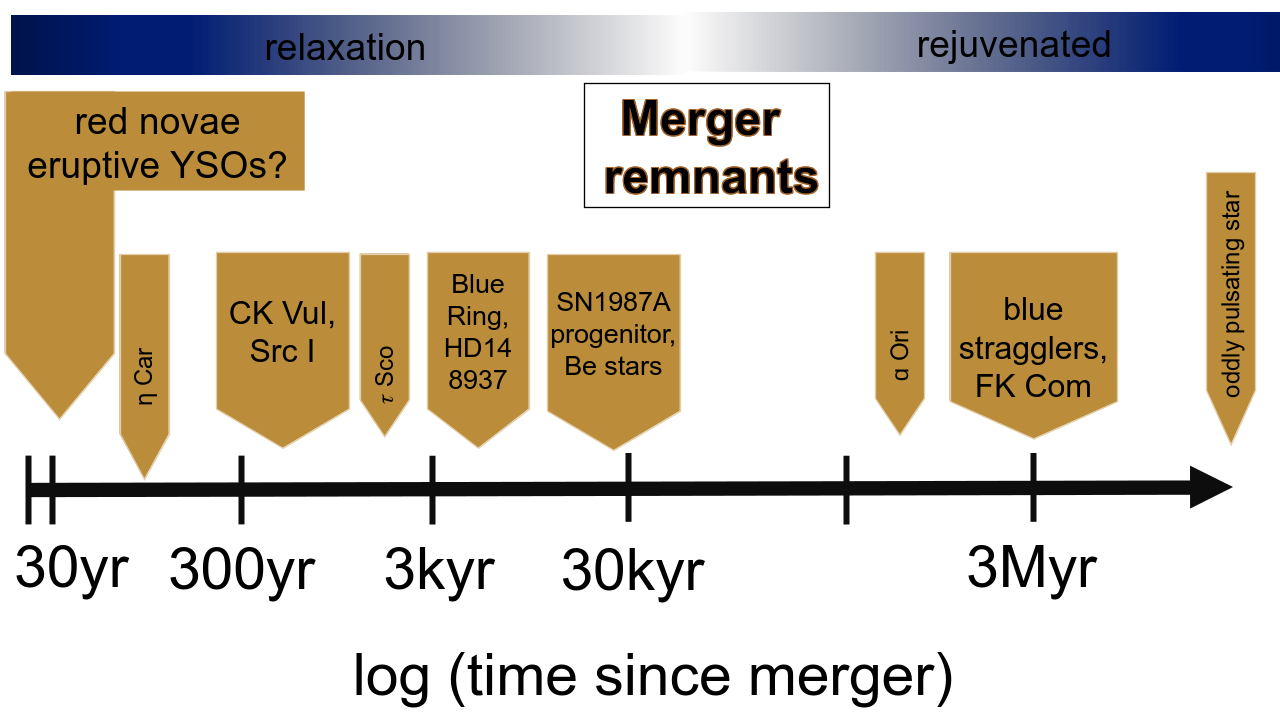}
\end{center}
\caption{Examples of postulated merger remnants on the time axis since the presumed merger. It is unclear into which objects red nova remnants will evolve on the presented timescales.}
\label{fig-oldremnants} 
\end{figure}

The search for remnants of historic outbursts continues \citep{KamiNovaeSurvey,BondV1148Sgr,Bond2018-M31RV}, but is realistically limited by observational records that extend back a few hundred years, at best \citep[e.g.,][]{Yang2025}. For much older red novae, we may not have a direct record of the outburst, which makes it difficult to formally classify them as red novae. In the astronomical literature, there are, however, hundreds of claims of objects being merger products from thousands to millions of years ago. Some examples are compared on the time axis in Fig.~\ref{fig-oldremnants}. Below, we mention only a few cases that seem most relevant to the red nova research.

One possible class of older merger remnants (relative to the currently observed sample of red novae) is represented by blue stragglers. V838~Mon, which is a product of a merger involving a B-type MS star, is predicted to contract into a rejuvenated B dwarf in the Kelvin-Helmholtz time scale of $\sim500$ yr \citep[using parameters from][]{KamiALMA}. As a member of a stellar cluster, V838~Mon is expected to evolve into a blue straggler. Most known blue straggler stars are likely to be significantly older. Only a fraction of blue stragglers are merger products of MS stars, and these are often difficult to distinguish from post-mass transfer systems \citep[for a recent review, see][]{BlueStgglerReview}. Ongoing studies of blue stragglers' rotation, surface magnetic fields, and elemental composition are a promising way to understand red novae and their post-outburst evolution to a hydrostatic configuration. Conversely, tracking the evolution of V838~Mon may be extremely informative for understanding the formation and evolution of blue stragglers. 

Merger products can also be identified as binary components that appear much younger than their companions. These are often cases of mergers in hierarchical triples (see Sect.~\ref{sec-triples}). The claimed merger products are likely to be puffed-up and highly magnetic stars with rapid rotation. However, angular momentum loss from winds and magnetic braking is expected to slow down the remnant on timescales of tens of Myr, making it a slow rotator compared to other cluster members \citep[see][]{Wang2022NatAs}. In some cases, this is associated with anomalies in the surface elemental composition. Examples include HD 148937 \citep{Frost}, $\theta$ Sco \citep{LewisThetaSco}, $\tau$ Sco \citep{Schneider,Keszthelyi2021MNRAS}, and some Cepheids \citep[e.g.,][]{espinoza}. On similar grounds, a merger scenario has been proposed to explain properties of B and B[e] supergiants \citep{KrausBeSG,Menon2024}, including FS CMa stars \citep{Dvorakova}. Perhaps a variation of these types of remnants is the future of V838~Mon remnant or some of the extragalactic red novae. 

Much younger are suspected merger remnants in star-forming regions. Here we provide two examples. On the high-mass end, {\it Source I} (or {\it SrcI}) in the Orion's Kleinmann--Low Nebula (Ori KL) is thought to be a possible merger product from $\approx$550 years ago \citep{Bally2020}. A dynamical interaction in a multiple system of several low- and high-mass protostars created the famous outflow of OMC1, and {\it Source I} is one of the walkaway stars in the complex. The candidate merger remnant is of the mass of 5--25 \Msun\ and possesses a large disk whose many properties are atypical for normal young stellar objects \citep{Testi2010,Wright2024}. 

Another merger candidate known from protostellar studies is HP Tau \citep[][and references therein]{Reipurth}. This object, of a mass of about 2 \Msun, is also a walkaway star moving away from a site where multiple protostars interacted dynamically 3000--5600 yr ago. Its current properties include high rotational speed, magnetic activity (at optical, radio, and X-ray regimes), and a rejuvenated appearance compared to other cluster members. These properties are remarkably reminiscent of longer-known suspected merger remnants, FK Com stars. This group of stars has been proposed to be products of mergers of contact W\,UMa stars \citep{Stepien2025}, that is, similar to the progenitor of the red nova V1309 Sco \citep{Stepien} (We note, however, that W\,UMa stars and V1309 Sco were much older than HP Tau progenitor). Interestingly, objects like HP Tau do not seem to have dusty disks. If they are direct descendants of systems like the V1309 Sco remnant, the disks must be eroded or dispersed on a timescale of thousands of years.      

There is also a group of star-forming regions that have been proposed to be more embedded merger sites than those discussed above, but still closely related to Ori KL. This peculiar group shares certain characteristics with circumstellar media of evolved stars and with star-forming regions. The class includes Ori KL, Sgr B2, W51N, G0.38+0.04, G19.61--0.23, G75.78+0.34, and IRAS 
19312+1950 (a.k.a. Deguchi-Nakashima object) \citep[][Gong et al. 2026, and references therein]{Deguchi2004PASJ,Cordiner2016}. All these regions contain components with active SiO masers, which otherwise are never observed in protostellar environments. Note that the red nova remnant V838~Mon is hosting an SiO maser, too. Kinematics of these regions indicate some past eruption, perhaps analogous to that in OMC1.  Whether it was caused by a red-nova-like merger is still being investigated. There are also unique objects seen toward the Galactic center that were proposed to be merger products similar to red nova remnants \citep{Ginsburh2024MUBLO,WitzelG2}.

The link of red novae to protostars may be even stronger. Accretion events, similar to FU Ori outbursts, share some characteristics with red novae: the raising slope toward the main peak, a kink in the rising part of the light curve reminiscent of the pre-eruption in red novae, the main-peak and plateau phase in the light curve, spectral appearance during outburst (including spectral type, Balmer lines, the \io{Ca}{ii} triplet, neutral and singly-ionized metals) and after (e.g., TiO and VO absorption), and linewidths \citep[e.g.,][]{Hillenbrand2025,FisherPPVII}. 

One difference is that classical FU Ori stars have lower outburst amplitudes (2--5 mag) than stellar red novae; they are typically also more embedded and remain longer in the active state, even decades to hundreds of years. The outburst mechanism is also, in principle, different since FU Ori eruptions are triggered by a disk instability, and accretion is fed from a large envelope. Nevertheless, both groups of objects are powered by accretion and shocks. The appearance of optical--IR molecular bands during or after the eruption, including rarely observed species like AlO, is especially intriguing. The commonalities between these two groups of eruptive variables are worth exploring \citep{Guo2025}. There are also related eruptions of young stars of shorter time span than classical FU Ori, which are often compared to red novae \citep[e.g.,][]{Lukas2020} and related to accretion of compact objects, such as protoplanets \citep{Wolf2024,Guo2025}. These objects may be good analogs of planet-engulfing red novae \citep{DeNature, KashiSokerPLanet1, KashiPlanet2}.

The famous `Great Eruption' of the luminous blue variable (LBV) $\eta$ Car is also often compared to red novae. This very massive ($\approx$120 \Msun) and young ($\lesssim$3 Myr) binary star dramatically increased its brightness by nearly 9 mag in 1843. This was followed by a prolonged 20-year plateau and 1--2 other peaks of increased brightness. This multi-peak character of the light curve, apart from its timescale, is similar to red novae. The main eruption was also as cool as those of red novae, as the spectrum of the eruption's light echo indicated a spectral type close to G2--G5 \citep{RestEcho}. Matter was lost at velocities of 100--600 \kms\ \citep{SmithEchoes}, consistent with what is typically observed in red novae. The remnant of the eruption is the bipolar Homunculus Nebula, whose inner regions are abundant in molecules and dust \citep{Gull}, just as in many bipolar red nova remnants. The interpretation of the eruption often leads to the suggestion it was triggered by a massive merger in a triple or higher-order system \citep[e.g.,][]{PortegiesZwart2016,SmithEchoes}. Despite these multiple similarities to red novae, $\eta$ Car is more classically included in the LBV family of variable stars and was omitted in our Table \ref{tab-main}. The red novae with the most massive progenitors, such as SN Hunt 248 or AT1997bs (a.k.a. SN 1994bs), may have even more in common with $\eta$ Car and P Cyg-like eruptions than with the low-mass red novae \citep{Mauerhan2018MNRAS}.

While the above examples are all relatively young remnants of presumed mergers at the main-sequence or before it, there are also more evolved suspected merger remnants relevant for red novae. \citet{Melis2020} reviewed the relation of dusty RGB stars, Phoenix Giants, to red novae and stellar mergers. These evolutionary-advanced Li-rich stars host dusty disks, which are speculated to be remains of planet engulfment from times when the star entered the first-ascent giant branch and significantly increased its size \citep{Jura2003,melis2009}. One such object, TYC 2597-735-1 associated with the Blue Ring Nebula, was proposed to be only 2000--5000 years old \citep{BlueRing}. Some of these merger candidates are puffed-up, active, magnetic, and rapidly rotating stars \citep{BlueRingXray}, and are also hypothesized to be evolving to the FK Com class. The relation of these Phoenix Giants to the red novae with red giant progenitors is still being explored \citep{SteinmetzBLG}.

Assuming that the mentioned objects are true remnants of a merger, it is interesting to ask how long one can observe the matter dispersed in a mergeburst or briefly before it. 
The nebula and disk in TYC 2597-735-1 and the bipolar outflow surrounding HP Tau suggest that the matter can be traced even after $\sim$5000\,yr  \citep{BlueRing,Reipurth}. In HD 148937, a bipolar cocoon is seen 75\,000 yr after the hypothesized merger \citep{Frost}. In SN 1987A, the hourglass nebula with the famous equatorial waist is seen 6650 years after the proposed merger of the progenitor \citep{Sn1987Aage}. No extended circumstellar matter has been reported for merger-born blue stragglers or FK Com stars, which may be even Myr after the coalescence. Although the survival of merger ejecta depends on its mass, velocity, and surrounding environment, it may remain detectable for up to $\sim$10\,000 years after the event. This longevity is encouraging for the future identification of merger products and remnants of red novae.


\section{Future observational prospects} \label{sect-future}

Progress in the study of red novae is likely to be driven along four main directions: the expansion of discovery samples through synoptic surveys (including IR facilities), earlier identification of progenitor systems, the development of more realistic multi-dimensional models of the outburst physics, and a shift from treating red novae as rare curiosities to using them as probes of binary evolution. Below, we outline several anticipated advances.

\subsection{Red novae in the 2030s} 

Future advances in the field of observational study of red novae will be related to progress on two major fronts: improvements in astronomical instrumentation and the release of large, multi-wavelength, multi-epoch catalogs from space missions and ground-based surveys. Large aperture telescopes such as the ESO Extremely Large Telescope \citep[ELT;][]{Gilmozzi2007_ELT}, Giant Magellan Telescope \citep[GMT;][]{Johns2012_GMT}, and, hopefully, Thirty Meter Telescope \citep[TMT;][]{Skidmore2015_TMT} will start operations in the early 2030s. The unmatched collecting power of such colossal apertures will enable scientists to obtain high-resolution spectroscopic and spectropolarimetric data on distant extragalactic red novae, allowing detailed comparisons with their Galactic counterparts and studies of the polarization signatures of individual lines. Diffraction-limited, milliarcsecond imaging for a 30\,m class optical-IR telescope will also resolve the closest environment to these objects, whether they are clusters of young stars or low-surface brightness nebulae. For red nova remnants, JWST and ALMA will continue to provide high-quality data on the molecular gas and dust that formed in recent and decades-old red-nova remnants. Multi-epoch observations of the evolving dusty content of red novae remnants will become a valuable resource to test rapid dust formation models and simulations, and the peculiar chemistry arising in these systems.

The future large time-domain catalog releases from optical and NIR surveys will be key not only in the identification of new transient events, but also in the characterization of their progenitor systems and past variability. In the 2030s, \textit{Gaia} will have final results released as DR5, containing its multi-epoch catalog for the whole duration of the mission. This will enable tracing back the evolution of nearby Galactic systems in high-precision photometry and low-resolution spectroscopy. It may also help to better constrain distances to the Galactic red nova remnants, benchmarking their physical properties. The final \textit{Euclid} data release, DR3, in the early 2030s will also provide a deep optical and IR baseline to retrospectively identify the progenitors of upcoming red nova outbursts. Slightly later, in the 2040s, millimeter wave facilities with surveying capabilities like AtLast and ALMA2040, will also provide a look at cool objects at longest wavelengths \citep{Atlast}.
Combined multi-epoch, multi-wavelength data on red nova progenitors and precursors offer great potential to shed light on how mass-transfer instability can be triggered in different binary systems, and create an observational benchmark for detailed theoretical studies and simulations.

Aided by the results of currently operating and future multiplexed spectroscopic surveys---providing accurate redshifts and characterization for millions of nearby galaxies and stars in the Milky Way---deep, wide-field photometric time-domain surveys will have a paved path for the discovery of large numbers of more distant extragalactic red novae, or the pre-outburst spectroscopic characterization of Galactic progenitor systems. Discoveries will come in different flavors: the early-time UV will be detected by ULTRASAT \citep{Sagiv2014_ULTRASAT}, optical data will be delivered by the ongoing Vera Rubin's LSST \citep{Ivezic2019ApJ_LSST}, and the cooler, highly dust-obscured transients will be detectable in the IR by the Galactic and High-latitude Wide Field surveys with the Roman Space Telescope \citep{Spergel2015arXivRoman}. This multi-wavelength coverage will be ideal to obtain a complete, long-term characterization of the evolution of thousands of red nova events. Similar to the progress observed for other, more common, transients, the large increase in well-characterized events has the potential to reveal different red nova families, associated with different progenitor populations, physical ejection mechanisms, and emission mechanisms. This will offer a unique opportunity to statistically assess the connection of red novae outbursts to common-envelope and merger physics, and observationally constrain the evolutionary pathways predicted by binary population models.

\subsection{Gravitational-wave studies of red novae}
Optimistically, the red nova research may go beyond the electromagnetic spectrum. Currently, red nova systems are unlikely sources of detectable gravitational wave (GW) signals for the now operating detectors LIGO-Virgo-Kagra, with peak sensitivity at $\sim$200\,Hz. This is because the radii of the non-degenerate stellar cores are much larger than those of neutron stars or black holes, and the signal would stop at much lower frequencies. Upcoming international space-based gravitational-wave observatories like LISA \citep[Laser Interferometer Space Antenna;][]{Danzmann1997,LISA2017arXiv}, with peak sensitivities in the 0.1--100\,mHz range, could identify a limited number of these events under strict conditions: the sources must be closer than 1\,kpc, and the signal should correspond to the merging compact cores after a failed CEE (rather than a tidal disruption of the companion at a larger radius) \citep{MoranFraile2023_GW,LISAupperlimitsCEE}. 

Currently, the most promising landscape for the study of red novae through GWs comes from missions operating in the frequency range between LIGO and LISA, i.e., at 0.1--10 Hz. One example is the Japanese DECIGO \citep[DECi-hertz Interferometer Gravitational wave Observatory;][]{Seto2001,Kawamura2011}, with its pathfinder mission B-DECIGO planned to launch in the 2030s, or the Advanced Laser Interferometer Antenna (ALIA) and the Big Bang Observer (BBO) \citep{CrowderCornish2005PhRvD}. Detections of stellar mergers and common-envelope events through gravitational waves will help to uncover the fraction of unseen, likely obscured population of interacting binaries, providing new constraints on current theoretical models. Looking ahead to the post-2030 era, multi-messenger observations of red novae are expected to play a key role in advancing our understanding of binary evolution.

\bmhead{Acknowledgements}
TK is grateful to T. Tylenda for initiating the field of red novae and inspiring many branches of research discussed in this review. We thank M. MacLeod and O. Pejcha for their comments on the manuscript. T.K. acknowledges funding from grant SONATA BIS no 2018/30/E/ST9/00398
from the Polish National Science Center (PI T. Kamiński).
N.B. acknowledges being funded by the European Union (ERC, CET-3PO, 101042610). Views and opinions expressed are, however, those of the author(s) only and do not necessarily reflect those of the European Union or the European Research Council Executive Agency. Neither the European Union nor the granting authority can be held responsible for them.
N.B. acknowledges financial support from grant CEX2024-001451-M funded by MICIU/AEI/10.13039/501100011033.
Research project PID2024-155585NA-I00 funded by MICIU/AEI /10.13039/501100011033.

\section*{Declarations}

\bmhead{Conflict of interest} The authors declare no conflict of interest.

\phantomsection
\addcontentsline{toc}{section}{References}
\bibliography{0bib.bib}


\end{document}